\documentclass[a4paper,fleqn,usenatbib]{mnras}

\usepackage{newtxtext,newtxmath}
 
\usepackage[T1]{fontenc}
\usepackage{ae,aecompl}

\usepackage{graphicx}	% Including figure files
\usepackage{amsmath}	% Advanced maths commands
\usepackage{amssymb}	% Extra maths symbols
\usepackage{multirow}

\usepackage[utf8]{inputenc}
\usepackage[export]{adjustbox}
\usepackage{wrapfig}
\usepackage{dutchcal}
\usepackage{braket}

\usepackage{ulem}
\graphicspath{ {images/} }

\newcommand{\lya}{\mbox{$\rmn{Ly}\alpha$}}

\newcommand{\wlyaO}{$\rm \lambda_{Ly\alpha}^{Obs}$}
\newcommand{\wlya}{$\rm \lambda_{Ly\alpha}$}

\newcommand{\llya}{\mbox{$L_{{\rm Ly}\alpha}$}}
\newcommand{\galform}{\texttt{GALFORM}}
\newcommand{\pmill}{\texttt{P-Millennium}}
\newcommand{\flareon}{\texttt{FLaREON}}
\newcommand{\lyart}{\texttt{LyaRT}}

\newcommand{\colorThin}{green}
\newcommand{\colorWind}{blue}

\newcommand{\ThinShell}{Thin Shell}
\newcommand{\GalacticWind}{Galactic Wind}

\newcommand{\GM}{GM}
\newcommand{\IC}{IC}
\newcommand{\VER}{GM-F}
\newcommand{\uVER}{uGM-F}

\newcommand{\NNU}{NN:Uniform}
\newcommand{\NNB}{NN:Bright}

\title [Systemic redshift of Lyman-\mbox{\boldmath$\alpha$} emitters]{Determining the systemic redshift of Lyman-\mbox{\boldmath$\alpha$} emitters with neural networks and improving the measured large-scale clustering.}
\author[S. Gurung-L\'opez. et al.]{
Siddhartha Gurung-L\'opez,$^{1,2}$\thanks{E-mail: sidgurung@cefca.es}
Shun Saito,$^{1,3}$
Carlton M. Baugh,$^{4}$
Silvia Bonoli,$^{5}$ 
\newauthor
Cedric G. Lacey,$^{4}$
and
\'Alvaro A. Orsi$^{6,2}$
.
\\
$^{1}$ Institute for Multi-messenger Astrophysics and Cosmology, Department of Physics\\
Missouri University of Science and Technology, 
1315 N. Pine St., Rolla MO 65409, USA\\
$^{2}$ Centro de Estudios de F\'isica del Cosmos de Arag\'on, Plaza San Juan 1, piso 2, Teruel, 44001, Spain. \\
$^{3}$Kavli Institute for the Physics and Mathematics of the Universe (WPI), Todai Institutes for Advanced Study,\\
the University of Tokyo, Kashiwanoha, Kashiwa, Chiba 277-8583, Japan\\
$^{4}$ Institute for Computational Cosmology, Durham University.\\
$^{5}$ DIPC, Manuel Lardizabal Ibilbidea, 4, 20018 San Sebastian, Spain. \\
$^{6}$ PlantTech Research Institute Limited. South British House, 4th Floor, 35 Grey Street, Tauranga 3110, New Zealand \\
}

\date{Accepted XXX. Received YYY; in original form ZZZ}

\pubyear{2020}

\begin{document}
\label{firstpage}
\pagerange{\pageref{firstpage}--\pageref{lastpage}}
\maketitle

\begin{abstract}

We explore how to mitigate the clustering distortions in Lyman-$\alpha$ emitters (LAEs) samples caused by the miss-identification of the Lyman-$\alpha$ (\lya ) wavelength in their \lya\ line profiles. We use the \lya\ line profiles from our previous LAE theoretical model that includes radiative transfer in the interstellar and intergalactic mediums. We introduce a novel approach to measure the systemic redshift of LAEs from their \lya\ line using neural networks. In detail, we assume that, for a fraction of the whole LAE population their systemic redshift is determined precisely through other spectral features. We then use this subset to train a neural network that predicts the \lya\ wavelength given a \lya\ line profile. We test two different training sets: i) the LAEs are selected homogeneously and ii) only the brightest LAE are selected. In comparison with previous approaches in the literature, our methodology improves significantly the  accuracy in determining the \lya\ wavelength. In fact, after applying our algorithm in ideal \lya\ line profiles, we recover the clustering unperturbed down to $1{\rm cMpc}/h$. Then, we test the performance of our methodology in realistic \lya\ line profiles by downgrading their quality. The machine learning technique  using the uniform sampling works well even if the \lya\ line profile quality is decreased considerably. We conclude that LAE surveys such as HETDEX would benefit from determining with high accuracy the systemic redshift of a subpopulation and applying our methodology to estimate the systemic redshift of the rest of the galaxy sample.  

\end{abstract}

\begin{keywords}
Radiative transfer -- Intergalactic medium -- ISM -- High-redshift -- Emission lines
\end{keywords}

%%%%%%%%%%%%%%%%%%%%%%%%%%%%%%%%%%%%%%%%%%%%%%%%%%

%%%%%%%%%%%%%%%%% BODY OF PAPER %%%%%%%%%%%%%%%%%%

\section{Introduction}\label{sec:intro}

Since the detection of the first galaxies emitting Lyman-$\alpha$ radiation more than twenty years ago \citep[e.g.][]{steidel96, hu98,rhoads00, malhotra02}, \lya\ radiation (with wavelength $\sim 1215.68 $\AA{} in rest frame) has been used as a successful tracer in the local \citep{Orlitova_2018,Henry_2018} and of the high redshift Universe \citep{ouchi08,steidel10,steidel11,Jones_2012,Chonis_2013,Erb_2014,Trainor_2015,oyarzun17,Matthee_2017_boot,Guaita_2017,Caruana_2018},  detecting galaxies even at the epoch of reionization \citep{Sobral_2015_CR7, Ouchi2018a, Shibuya_2018}. Ongoing cosmological galaxy surveys, such as the Hobby-Eberly Telescope Dark Energy Experiment~\citep[HETDEX]{Adams2011,Hill2008} and the Javalambre Physics of the Accelerating Universe Astrophysical Survey \citep{benitez14,Bonoli_2020}, aim at  unveiling the nature of the Dark Energy using LAEs at the high redshift Universe. One of the most useful tools to extract cosmological information from galaxy surveys is the galaxy clustering \citep[e.g.,][]{Shoji_2009}. 
Therefore, understanding the spatial distribution of LAEs has become more important than ever before.

The complexity in understanding the LAE clustering resides in the radiative transfer of \lya\ photons inside neutral hydrogen \cite{harrington73,neufeld90}. In first place, \lya\ photons are emitted in the HII regions around OB-{\it type} stars. Then they have to cross the interstellar medium (ISM), then the circumgalactic medium (CGM) and the intergalactic medium (IGM) until they finally reach our observatories. In all these three mediums there is neutral hydrogen, and therefore, they are optically thick to \lya\ radiation. Inside galaxies, it is commonly thought that \lya\ photons escape through outflow that modify the \lya\ flux and line profile \citep[e.g.,][]{ahn00, zheng02, ahn03, verhamme06, orsi12, Gronke_2016, Gurung_2018b}. Then, in the CGM the \lya\ radiation is spread around the galaxy creating the so-called \lya\ halos \citep[e.g.,][]{zheng10,Leclercq_2017,Behrens2017}. Finally, the \lya\ radiation enters inside the IGM, where, to a first approximation, the radiation bluewards the \lya\ wavelength is absorbed \citep[e.g.,][]{zheng11,laursen11,Byrohl_2019,GurungLopez_2020}. 

Furthermore, the clustering property of LAEs can be sensitive to the selection function which is typically determined by the flux threshold. For example, \lya\ radiaition is very sensitive to dust. Therefore, galaxies with low metalicities are preferentially observed as LAEs \citep{Sobral_2018a}. This translate into a lower clustering amplitude \citep{gurung18a}, as galaxies with lower metallicity exhibit are hosted in smaller dark matter halos \citep[e.g.]{lacey16}.  Moreover, it has been pointed out that the large scale properties of the IGM might play a role on the selection function of LAEs \citep{zheng10,zheng11}.  This could distort the clustering of LAEs and reduce the accuracy of LAE cosmological surveys \citep{Wyithe_2011}.  However, there is still debate in the community whether if there is a large scale IGM coupling with the observed \lya\ luminosity \citep{GurungLopez_2020} or not \citep{Behrens2017}.  These facts contribute to the complexity of understanding the radiative transfer of the \lya\ radiation and its impact on the clustering statistics.

A key challenge lies in determining the systemic redshift of an LAE only from their \lya\ line profile.  Due to the radiative transfer, the peak of the \lya\ line rarely matches the \lya\ wavelength.  In \cite{Verhamme:2018aa}, authors probed that taking into account the radiative transfer in the ISM was crucial to improve the determination of the systemic redshift of LAEs.  In their work, they used a relation between the shift of the line peak from \lya\ and the width of the line to correct for the redshift due to the radiative transfer into the ISM.   However, this recipe assumed the commonly used \ThinShell\ toy model for the outflow geometry.  In fact the relation between the width of the line and peak shift depends strongly on the outflow geometry used \citep{Gurung_2018b}. More recently, \cite{Muzahid_2019}, linked the LAEs star formation rate to the line peak offset, also, getting better results than when assuming that the peak of the \lya\ line is the \lya\ wavelength.   
    
Along a similar line, \cite{Byrohl_2019} studied how the incorrect systemic redshift determination distorts the clustering of LAEs along the line of sight. In detail, an inaccurate \lya\ wavelength determination is translated into a imprecise redshift, thus into an uncertainty in its radial position in redshift space. This causes that LAEs look like they are more spread along the line of sight than what they actually are. In fact, this shuffling in the LAE radial position can be interpreted as a extra random radial velocity dispersion component, which translates into an additional `Finger-of-God' suppressing the apparent clustering along the line of sight. Furthermore,  \cite{Byrohl_2019}  found that the clustering distortion was mitigated after correcting the shift of the peak with different recipes, such as the those in \cite{Verhamme:2018aa}. However, the developed recipes in the literature \cite{Verhamme:2018aa,Byrohl_2019,Muzahid_2019} to estimate the \lya\ wavelength have limitations as the dispersion of the estimated \lya\ wavelength was around 1\AA{} (in rest frame). Such a large scatter introduces significant distortions to the apparent clustering of LAEs on scales $\sim 5 \; {\rm cMpc}/{h}$ in the monopole and up to $k\gtrsim 0.1 \; h/{\rm cMpc}$ in Fourier-space, as we will explicitly show in this paper.
    
We propose a novel approach to determine the systemic redshift from the \lya\ line profiles of LAEs using neural networks. This is motivated by the fact that, there must be information on the \lya\ wavelength in an entire spectral range of \lya\ line profiles. We explore whether or not a given survey that observes LAEs only through \lya\ emission (e.g., HETDEX \citep{Hill2008}) could benefit from acquiring a subsample with a systemic redshift (without using \lya), for example, by $\rm H_{\alpha}$ observations. Then, this subset could be used to train a neural network to predict the \lya\ wavelength in the rest of the main LAE population. In this work, we train different neural networks using subsets of the \lya\ line profiles computed by our model to predict their \lya\ wavelength.  

This work is part of a series of papers studying the impact of the \lya\ radiative transfer on the observed properties of LAEs. In our first work \citep{GurungLopez_2019a}, we focused on the \lya\ RT taking place in the ISM. There, we found that LAEs are a very peculiar population that exhibits a tight balance between star formation rate and metallicity. Then, in our second work \citep{GurungLopez_2020}, we implemented the \lya\ in the IGM and we focused on the different selection effects introduced by it. There, we studied the impact in the clustering on large scales due to the IGM-LAE coupling. In this third work we analyze the properties of the \lya\ stacked line profiles and study the impact on small scales of the miss-identification of the \lya\ wavelength. Nonetheless, we emphasize that, we adopt this simulation set for a proof of concept, and that our approach can be generally applied to any LAE spectroscopic observation, in principle. 

We briefly describe our model in section \S \ref{sec:model}, while for a more extensive description we refer the reader to \cite{GurungLopez_2019a,GurungLopez_2020}. Then, in \S \ref{sec:stacks} we study the properties of the \lya\ stacked line profiles. In section \S \ref{sec:algorithms} we describe the different methodologies used in this work to identify the systemic redshift of LAEs from their observed \lya\ line profiles. Then, in \S \ref{sec:ideal} we describe the effects of the \lya\ miss-identification in ideal line profiles. Meanwhile, in \S \ref{sec:real} we artificially reduce the quality of our \lya\ line profiles analyze the effects of the \lya\ miss-identification in realistic \lya\ line profiles. Then, we discuss our result in \S \ref{sec:Discussion}. Finally, we make our conclusions in \S \ref{sec:conclusions}. 

Throughout this paper, all the properties related to \lya\ line profiles are given in length units in the  rest frame of the LAEs.

\section{LAE theoretical model}\label{sec:model}

In this work, we adopt the LAEs simulated with a semi-analytic model in \citet{GurungLopez_2019a,GurungLopez_2020}. In this section, let us briefly describe the LAE model but emphasize on modeling the spectrum around the \lya\ emission. Our LAE model is based on four main ingredients: 
\begin{itemize}
    \item 1) The dark matter {\it N}-Body simulation, \pmill\  \citep{Baugh_2019}, that imprints the hierarchical growth of structures in the $\rm \Lambda CDM$ scenario. 
    This state-of-the-art cosmological simulation consists in $5040^3$ dark matter particles with mass of $1.061\times10^8\;{\rm M_{\odot}}\;h^{-1}$ distributed in a volume of $L_{\rm box}^{3}=(542.16\,{\rm cMpc}\,h^{-1})^3$. \pmill\ uses cosmological parameters: $H_{0}=67.77 \;{\rm km\;s^{-1}Mpc^{-1}}$, $\Omega_{\Lambda}=0.693$, $\Omega_{\rm M} = 0.307$ , $\sigma_{8}=0.8288$, consistent with \citet{Planck_2016}.
    
    \item 2) The model of galaxy formation and evolution, \galform\ \citep{cole00,lacey16,Baugh_2019}. 
    In short, \galform\ populates galaxies and gases within the dark matter halos and tracks their evolution through the cosmic history. \galform\  follows recipes to estimate a whole bunch of galaxy properties such as metallicity or star formation rate (SFR). 
    These recipes are calibrated to fit several observables, such as, the optical and near infrared luminosity functions at $z=0$ and its evolution up to $z=3$, the HI mass function at $z=0$, the sub-mm galaxy number counts and their redshift distributions among others. 
    Galaxies in \galform\ exhibit two components: the disks, where the quiescent star formation takes place and the bulges, where the strong star formation bursts take place. Each of these pieces exhibit different properties (such as metallicity, etc).
    
%%%%%%%%%%%%%%%%%%%%%%%%%%%%%%%%%%%%%%%%%%%%%%%%%%%%%%%%%%%%%%%%%%%%%%%%%%%%%%%%%%%%%%%%%%%%%%%%
\begin{figure*} 
\includegraphics[width=6.9in]{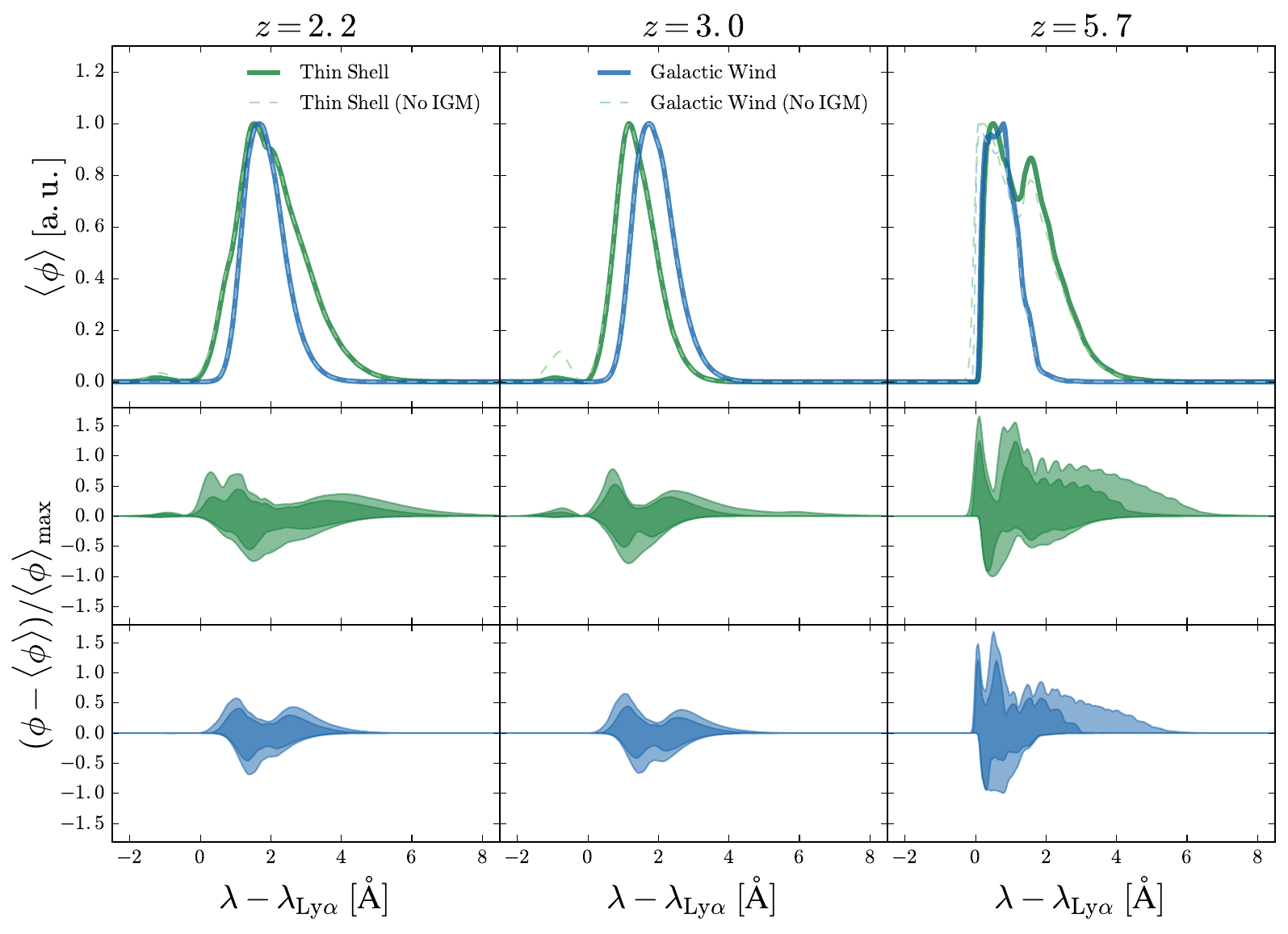}%
\caption{ Top: Stacked \lya\ line profile at redshift 2.2, 3.0, 5.7 from left to right. The \ThinShell\ geometry is displayed in thick green lines and the \GalacticWind\ in blue. Our default models, including the IGM \lya\ absorption, are shown in thick lines while, models without the IGM implementation are shown in thin dashed lines. The stacked line profiles are normalized so that their global maximums match unity. { Middle: Difference between individual and stack line profiles relative to the maximum of the stack line profile for the \ThinShell\ geometry including the IGM. The inner dark (light) regions show the area between the percentiles 16th and 84th (5th and 95th). Bottom: Same as middle but for the \GalacticWind\ outflow geometry including the IGM. } }
\label{fig:IGM_no_IGM_stacks}
\end{figure*}
%%%%%%%%%%%%%%%%%%%%%%%%%%%%%%%%%%%%%%%%%%%%%%%%%%%%%%%%%%%%%%%%%%%%%%%%%%%%%%%%%%%%%%%%%%%%%%%%    

    \item 3) The \lya\ radiative transfer in the ISM is implemented through the \texttt{Python} open source code, \flareon\ \citep{GurungLopez_2019b}. \flareon\ is based on a pre-computed grid of outflow models using \lyart\ \citep{orsi12}, spawning a wide range in neutral hydrogen column density ($\rm N_{H}$), outflow expansion velocity ($\rm V_{exp}$) and dust optical depth ($\rm \tau_a$). By using different machine learning and multidimensional interpolation algorithms, \flareon\ predicts the \lya\ escape fraction $f_{\rm esc}$ and line profile $\rm \phi(\lambda)$  with high accuracy for different outflow geometries. Here, we will focus on the `\ThinShell' and `\GalacticWind' outflow geometries.
    
    Through out this work, we define the escape fraction of a given medium as the ratio between the flux injected into the medium and the that emerges from it. For example, for the ISM escape fraction, $f^{\rm ISM}_{\rm esc}=L_{\rm  Ly\alpha,\,ISM}/L_{\rm Ly\alpha,\,0}$ where $L_{\rm Ly\alpha,\,0}$ and $L_{\rm Ly\alpha,\,ISM}$ are the intrinsic \lya\ luminosity and the luminosity after passing through the ISM region, respectively. Also, in our convention the line profile, $\phi(\lambda)$, is normalized as $\int\phi(\lambda)d\lambda=1$. 
    
    In practice, our model links the galaxy properties predicted by \galform\ to outflow features through simple recipes \citep{GurungLopez_2019a}. In this way, each component (disk and bulge) in each galaxy has a different parameter set of \{$N_{H}$ ,  $\rm V_{\rm exp}$, $\tau_a$\} through  their SFR, metallicity, cold gas mass and stellar mass. Then we use  \flareon\  to compute a \lya\ line profile and an escape fraction for each galaxy and component.
    
    \item 4) The radiative transfer in the IGM is implemented by computing the optical depth of \lya\ photons in the line of sight (fixed along $Z$ axis) between the observer and each galaxy. Therefore, the IGM transmission depends on the particular properties of the environment of each galaxy, and in particular, on the IGM density $\rho$, its density gradient a long the line of sight $\rm \partial_{Z}\rho$, the IGM line of sight velocity $\rm V_{Z}$ and its gradient $\rm \partial_{Z}V_{Z}$.

%%%%%%%%%%%%%%%%%%%%%%%%%%%%%%%%%%%%%%%%%%%%%%%%%%%%%%%%%%%%%%%%%%%%%%%%%%%%%%%%%%%%%%%%%%%%%%%%
\begin{figure*} 
\includegraphics[width=6.9in]{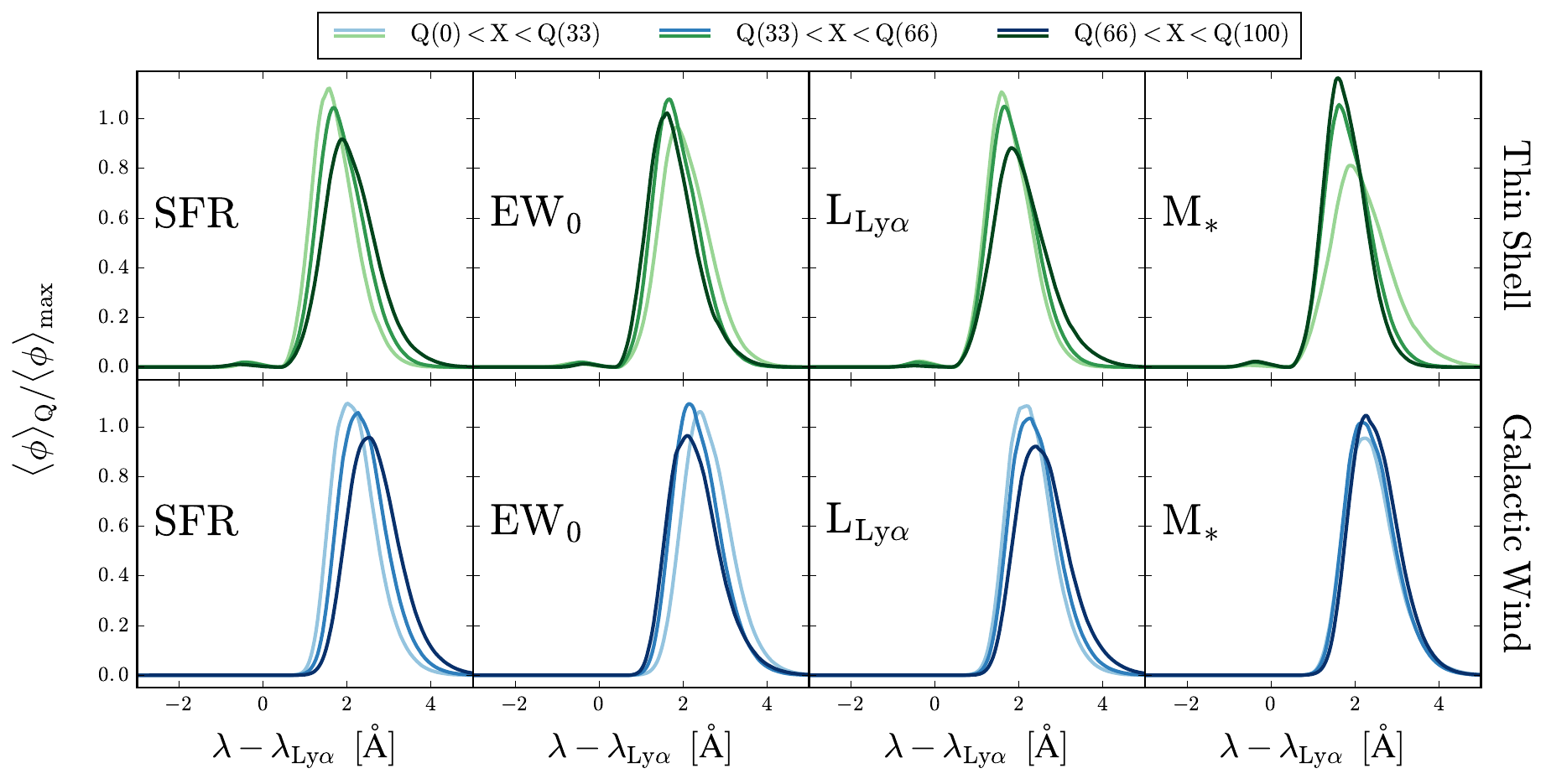}%
\caption{Break down of the \lya\ stacked line profiles at $z=3.0$ as a function of total star formation rate, \lya\ rest frame equivalent width, \lya\ luminosity and stellar mass from left to right. Top panels show the model with the \ThinShell\ outflow geometry, while bottom panels show the model using the \GalacticWind\ geometry. We rank our main LAE population by each of these galaxy properties and split it by the percentiles 33 ($Q(33)$) and 66 ($Q(66)$). The stacked line profiles are color coded according to the samples they are displaying. The lighter lines show the samples with the lowest values of the galaxy property X ($Q(0)<X<Q(33)$). The darkest lines show the samples with the highest values of each galaxy property ($Q(66)<X<Q(100)$). Meanwhile, the intermediate values ($Q(33)<X<Q(66)$) are shown in intermediate colors. The \lya\ stack profile of each subpopulation, $\langle \phi \rangle _{Q}$ is normalized to the maximum of the stacked line profile of the complete LAE population, $\langle \phi\rangle_{\rm max}$.}
\label{fig:split_galaxy}
\end{figure*}
%%%%%%%%%%%%%%%%%%%%%%%%%%%%%%%%%%%%%%%%%%%%%%%%%%%%%%%%%%%%%%%%%%%%%%%%%%%%%%%%%%%%%%%%%%%%%%%%

    Our model provides the \lya\ line profile and luminosity are computed by convolving the \lya\ line profile emerging from the ISM with the IGM transmission curve for each component of a galaxy.
    In practice, the  line profile and luminosity are given by 
%%%%%%  
    \begin{equation}\label{eq:phi_lya}
    \phi(\lambda) = {{{L_{\rm Ly\alpha }^{\rm Disk}}  \phi^{\rm Disk}_{\rm ISM}  + {L_{\rm Ly\alpha }^{\rm Bulge}}  \phi^{\rm Bulge}_{\rm ISM}}\over{{L_{\rm Ly\alpha }}}} ,
    \end{equation}
%%%%%% 
and 
%%%%%% 
\begin{equation}\label{eq:L_lya}
    L_{\rm Ly\alpha } = L_{\rm Ly\alpha }^{\rm Disk}+L_{\rm Ly\alpha }^{\rm Bulge}, 
\end{equation}
%%%%%% 
where the luminosity for each component is evaluated as
%%%%%%
\begin{equation}\label{eq:L_lya_component}
    L_{\rm Ly\alpha}^{X} = L_{\rm Ly\alpha,0}^{X} f_{\rm esc}^{\rm ISM, X} f_{\rm esc}^{\rm IGM, X}, 
\end{equation}
%%%%%%
with $X=\{{\rm Disk,\,Bulge}\}$. $f_{\rm esc}^{\rm IGM}$ is the escape fraction from the IGM. 

Furthermore, LAEs are defined as galaxies with a \lya\ emission line exhibiting a high contrast to the galaxy continuum. It is usually found in the literature that for a galaxy to be considered an LAE, it must exhibit a rest frame equivalent width $\rm EW_{0}>20$\AA{} \cite[e.g.][]{gronwall07,Konno_2018}. 
In this work we follow this criteria. In particular, we compute the $\rm EW_{0}$ for each galaxy as a function of its continuum luminosity per unit of wavelength around \lya\ wavelength $L_{\rm c}$ as
    
    \begin{equation}
    \label{eq:EW}
    {\rm EW}_{0} = L_{\rm Ly\alpha} / L_{\rm c} ,
    \end{equation} 
    
where $L_{\rm c}$ is directly provided by \galform\ and it is based on the full evolution of the stellar population given a galaxy.
    
The free parameters in the model, which depend on the outflow geometry model, are adjusted so that the simulated LAEs reproduce the observed \lya\ luminosity function at their corresponding redshift, as described in \cite{GurungLopez_2020}.

Finally, the samples studied here are selected by making a number density cut of $4\times10^{-3}(h/{\rm cMpc})^{3}$ in \lya\ luminosity. Given our simulation volume, each LAE population (combination of redshift and outflow geometry) is compound by  637444 galaxies.

\end{itemize}

\section{Stacked Lyman-\mbox{\boldmath$\alpha$} line profiles
}\label{sec:stacks}

Since the main goal of this paper is to study the determination of the systemic redshift of LAEs from their \lya\ line profile, we present here the detailed properties of the stacked \lya\ line profile. First we focus on how \lya\ radiative transfer impacts the observed \lya\ stacked line profile in our model. Then we discuss the \lya\ line profiles as a function of different galaxy and IGM properties to understand how they influence the stacked line profile.

Throughout this work we compute the stacked \lya\ line profiles in a consistent manner. First, we normalize all the \lya\ line profiles so that they have an area of unity, i.e., compute $\phi(\lambda)$. Then we evaluate the stacked line profile, $\langle \phi(\lambda)\rangle$, as the median of the line profile collection.

\subsection{The impact of radiative transfer}

In Fig.~\ref{fig:IGM_no_IGM_stacks} we compare the stacked \lya\ profile $\langle \phi(\lambda) \rangle$ before (dashed thin) and after (solid thick) being processed by the IGM. In general, at low redshift ($z=2.2$, 3.0), the IGM tends to absorb blue photons, i.e., $\lambda<\lambda_{\rm Ly\alpha}$, while it does not affect less for redder photons. Meanwhile, at $z=5.7$, the IGM optical depth to \lya\ is much greater and it affects to the stacked line profile up to wavelength 2\AA{} redder than \lya .

Furthermore, the impact of the IGM in the stacked line profiles is different for each geometry. We find that at $z=2.2$ and $z=3.0$ the IGM affects more the \ThinShell\ geometry than the \GalacticWind. In particular, the blue peak which is present before the line profiles are processed by the IGM (especially at $z=3.0$), is mostly vanished after the RT in the IGM. Meanwhile, at $z=5.7$ the IGM affects similarly both cases. This is consistent with the values of the \lya\ IGM escape fraction in our model (see Fig.10 and 11 in \cite{GurungLopez_2020}). These findings originate from the differences in the family of line profiles generated by each outflow geometry, as the IGM absorption depends on the wavelength of the photons, as discussed in detail in \citet{GurungLopez_2020}.

Our model predicts different shape of $\langle \phi(\lambda) \rangle$ for the \ThinShell\  and \GalacticWind\ models at all the redshift bins studied in this work. In general, we find that the stacked line profile for the \ThinShell\ model is bluer than that for \GalacticWind.  Meanwhile, at each different epoch,  the stacked line profiles for two outflow geometry models differ in unique fashions. For example, at $z=2.2$, the peaks of $\langle \phi \rangle$ of both geometries match, but the \ThinShell\ exhibits a broader $\langle \phi \rangle$ ($\sim 3$\AA{}) than the \GalacticWind ($\sim 1.5$\AA{}). At $z=3.0$, the width of the stacked profiles are comparable between both geometries, but the \GalacticWind\ profile is more redshifted than the \ThinShell one.Another difference between our two outflow models is that at redshift of 2.2 and 3.0 the \ThinShell\ exhibits a weak peak bluer than \lya, while the \GalacticWind\ lacks this blue peak. Finally, at $z=5.7$, the \ThinShell\ profile is broader than the \GalacticWind one. 

The differences between the `\ThinShell'\ and `\GalacticWind'\ outflow geometries arise due to two facts. First, the distribution of the ISM parameters, $\{V_{\rm exp}, \tau_{a}, N_{H}\}$ of the LAEs are different for both geometries \citep[see Fig.A1 and A2 of ][]{GurungLopez_2020}. Second, the radiative transfer in each geometry leads to a different line profile, even for the same parameter set of $\{V_{\rm exp}, \tau_{a}, N_{H}\}$.For example, the `\ThinShell'\ geometry model is more prone to exhibit a blue peak than the `\GalacticWind' model \citep{GurungLopez_2019b}, as we see in the stacked profile at redshifts  2.2 and, in  particular, at $z=3.0.$

In summary, the radiative transfer impacts the shape of the stacked profile including the peak position and the width from the ISM to IGM scales in a non-trivial manner. 

{ In the lower panels of Fig.\ref{fig:IGM_no_IGM_stacks} we show the diversity of line profiles for our models including the IGM. We display the relative difference to the maximum of the stack line profile of the line profile population. In the middle row we display the \ThinShell\ while the \GalacticWind\ is shown in the bottom. The dark (light) shaded region shows the area between the percentile 16th and 84th (5th and 95th) of the line profile distribution.  We find that the variety of lines is similar at redshift 2.2 and 3.0. At these redshifts, the \ThinShell\ geometry has some dispersion at wavelengths lower then \lya . This indicates that in these models, some lines exhibit a blue peak even after the IGM absorption. Meanwhile, the models using the  \GalacticWind\ lack  emission at bluer frequencies than \lya. At redshift 5.7 the diversity of lines increases, specially for the \ThinShell . The morphology of the relative difference (shaded areas)  changes with respect to z=2.2 and 3.0, getting wider in wavelength and exhibiting fluctuations with larger amplitudes. }

\subsection{Dependence of the stacked profile on galaxy and environmental properties}

So far we have considered the stacked profile for all LAEs in our simulation. 
In this section we split the \lya\ stacked line profile according to different galaxy and IGM properties. To this end, for a given galaxy or IGM property $X$, we compute the percentiles 33.33 ($Q(33)$) and 66.66 ($Q(66)$). Then, we split the LAEs into three subsamples of the same size containing the galaxies with the lowest ($Q(0)<X<Q(33)$), intermediate ($Q(33)<X<Q(66)$) end the greatest ($Q(66)<X<Q(100)$) values of $X$.

%%%%%%%%%%%%%%%%%%%%%%%%%%%%%%%%%%%%%%%%%%%%%%%%%%%%%%%%%%%%%%%%%%%%%%%%%%%%%%%%%%%%%%%%%%%%%%%%
\begin{figure*} 
\includegraphics[width=6.9in]{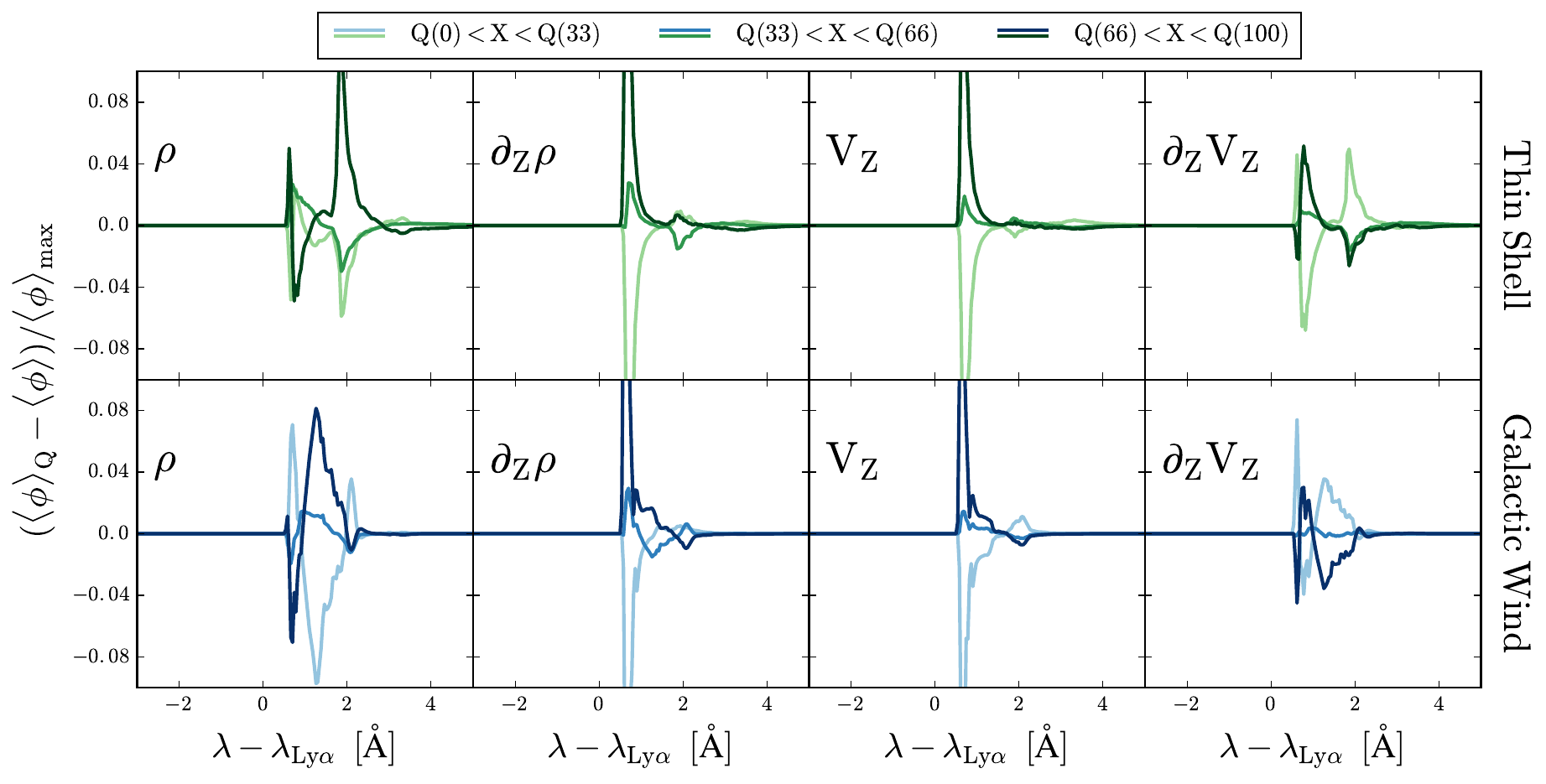}
\caption{ Break down of the \lya\ stacked line profiles at redshift $z=5.7$ as a function of the IGM large scale properties (density, density gradient along the line of sight, velocity along the line of sight and its gradient along the line of sight from left to right). For each of these IGM properties we rank and divide the LAE population in three samples of the same size. Here we show the difference between the stack of these subsamples and the stacked line profile of the full LAE sample. In the top (bottom) panels we show the \ThinShell\ (\GalacticWind). The color code is the same as in Fig.\ref{fig:split_galaxy}. }
\label{fig:split_IGM}
\end{figure*}
%%%%%%%%%%%%%%%%%%%%%%%%%%%%%%%%%%%%%%%%%%%%%%%%%%%%%%%%%%%%%%%%%%%%%%%%%%%%%%%%%%%%%%%%%%%%%%%%

\subsubsection{Imprints of the galaxy properties}

In Fig.~\ref{fig:split_galaxy} we split the \lya\ stacked line profile as a function of the star formation rate, rest frame \lya\ equivalent width, \lya\ luminosity and stellar mass (from left to right). The \lya\ stack line profile of each subpopulation is normalized to the maximum of the stacked line profile of the complete LAE sample, $\langle \phi\rangle_{\rm max}$. Here we show the snapshot at redshift 3.0 only, but have confirmed that the other two redshift bins ($z=2.2$ and 5.7) exhibit similar trends.

Overall our model predicts that the stacked \lya\ line profile properties depend on the galaxy properties. For instance, the peak of the \lya\ line profile correlates positively with the SFR and $L_{\rm Ly\alpha}$ for both outflow geometries. The peak of the line profile anti-correlates with $\rm EW_{0}$ for both outflow geometries. Interestingly, the dependence of the stacked \lya\ line profile on the stellar mass behaves differently for two outflow geometries;  In \ThinShell\ the peak anti-correlates with $M_{*}$, while there is no apparent trend in \GalacticWind.

We also find that, not only the peak position, but also the shape of the stacked line profile changes through the dynamical range of these galaxy properties. In particular, we find that for both outflow geometries, the stacked line profile becomes broader at higher SFR and $L_{\rm Ly\alpha}$ values, while it shrinks for high values of $\rm EW_0$. Meanwhile, when the `\ThinShell'\ is implemented, increasing the stellar mass leads to broader stacked profiles. In contrast, the width of the stacked line profile using the `\GalacticWind'\ remains constant through the stellar mass dynamical range.

These non-trivial differences between \ThinShell\ and \GalacticWind\ are consequences of the complicated interplay between the properties of galaxies and outflows (see equations XX and XX in \cite{GurungLopez_2020}). This highlights the importance of having different outflow models to model the RT in the ISM.

\subsubsection{Imprints of the IGM}

Here we study how the different large scale IGM properties  change the observed \lya\ line profile. In order to do so, we split the \lya\ stacked line profile $\langle \phi \rangle$ by IGM properties for both outflow geometries. In Fig.~\ref{fig:split_IGM} we show the difference between the stacked line profile of the full LAE population and the split samples.  The IGM properties used for dividing the LAE population are density $\rho$, density gradient along the line of light $\rm \partial_Z \rho$, velocity along the line of sight $\rm V_Z $ and its gradient along the line of sight $\rm \partial_Z V_Z$ from left to right. These properties were computed in a regular grid of cubic cells of $2\; {\rm cMpc}/h$ side, as in \cite{GurungLopez_2020}. Here we focus on the snapshot at $z=5.7$, where IGM is optically thicker to \lya\ photons than at $z=2.2$ and $z=3.0$. Note that we also find the same trends at $z=2.2$ and 3.0, but with a lower amplitude.

We find that, although differences are tiny ($\lesssim 10\%$ compared to $\langle \phi\rangle_{\rm max}$), both outflow geometries exhibit the same trends with a clear dependency of the IGM properties. This suggests that there is a smooth dependence between the IGM properties and the stacked line profile. For example, the higher the IGM density is, the more flux is absorbed at bluer wavelengths. This causes that the observed stacked line profile is slightly more redshifted in high IGM density regions. Also, our model predicts that LAEs located in regions with high $\partial_Z \rho$, $\rm V_Z$ and $\partial_Z V_Z$ exhibit a bluer stacked line profile, while the opposite is true for low values of these IGM properties. 

The trends in the stacked line profile are in agreement with the our previous work \citep[][]{GurungLopez_2020}, where the IGM transmission generally anti-correlates with $\rho$ and positively correlates with $\partial_Z \rho$, $\rm V_Z$ and $\partial_Z V_Z$ \citep[][]{GurungLopez_2020}. We also showed that, the bluer the wavelength around \lya\ , the more sensitive the line profile is to IGM absorption (see Fig.5 in \citet[][]{GurungLopez_2020}). Combining these two facts, the LAEs lying in regions with lower IGM transmission will exhibit a redder line profile than LAEs lying in regions with higher transmission, which is consistent with Fig.~\ref{fig:split_IGM}.

\section{Determining the redshift of LAEs}\label{sec:algorithms}

As we showed in the previous section, the wavelength of photons initially emitted at the \lya\ wavelength changes as they travel through the ISM and the IGM. In this way,  the \lya\ line profiles are modified in a non-trivial way \citep[e.g.]{zheng11} by the \lya\ RT. This complicates the determination of the \lya\ wavelength from an observed \lya\ line profile \citep[][]{Verhamme:2018aa,Byrohl_2019}. In general, in each \lya\ line profile, the true \lya\ wavelength (\wlya) and the wavelength set as \lya\ (\wlyaO ) can differ, as we show below. 

%Thus, each \lya\ line profile is shifted by $\rm \Delta \lambda = \lambda ^{Obs}_{Ly\alpha} - \lambda _{Ly\alpha} $ along wavelength.  In other words, the \lya\ line profiles transforms as  

%\begin{equation}
%\label{eq:observed_lya_profile}
%\rm 
%\phi ( \lambda ) \rightarrow \phi( \lambda + \Delta\lambda ),
%\end{equation}
%where $\phi( \lambda + \Delta\lambda )$ is the observed \lya\ line profile for given a \wlyaO . 
    
In the following we introduce the different methods that we use through this work to find the \lya\ wavelength (\wlyaO) directly from the \lya\ line profile. First, in \S \ref{subsec: Standard methodologies}, we describe two different methods to retrieve \wlyaO\ that have been already used in the literature. These algorithms depend on line profile characteristics such as the width and the position of the global maximum of the \lya\ line. Then, in \S \ref{subsec: NN}, we introduce a novel method that makes use of the full \lya\ line profile in order to predict \wlyaO\ through neural networks.

\subsection{ Standard methodologies }
\label{subsec: Standard methodologies}

\begin{itemize}
    \item  {\bf \GM } (Global Maximum): This is the simplest method to assign a \lya\ wavelength. Basically, the position of the global maximum ($\lambda_{\rm Ly\alpha , Max}$) is set as the \lya\ wavelength, i.e.,
    
    \begin{equation}\label{eq:algorithm_1}
    \lambda_{\rm Ly\alpha}^{\rm Obs} = \lambda_{\rm Ly\alpha , Max} .
    \end{equation}

    \item  {\bf \IC } (Intensity Center) : This method assigns the centroid of the line as the \lya\ wavelength the , i.e.,
    
    \begin{equation}\label{eq:algorithm_IC}
    \lambda_{\rm Ly\alpha}^{\rm Obs} = \frac{\sum \phi(\lambda)\lambda}{\sum \phi(\lambda)}
    \end{equation}
    
    Several works in the literature \citep[e.g.][]{steidel10,Rudie_2012} use a similar approach to \IC\ to estime the redshift of LAEs.

    \item  {\bf \VER } : This method takes into account that the \lya\ photons tend to be redshifted as they escape the galaxies through outflows. As a result, the position of the red peak is shifted from the \lya\ frequency. This shift depends on the outflow properties and can be related to the FWHM of the red peak \cite{Verhamme:2018aa,GurungLopez_2019b}. As we have shown  (see \S \ref{sec:stacks}) the \lya\ line profiles predicted by our model are clearly dominated by a prominent red peak and only a faint blue peak is found. Therefore,  $\lambda_{Ly\alpha , Max}$ and the maximum of the read peak matches. Thus, in this method, we compute the \lya\ wavelength as

   \begin{equation}\label{eq:algorithm_2}
    \rm 
    \lambda_{\rm Ly\alpha , obs} = \lambda_{\rm Ly\alpha , Max} - FWHM_{\rm Red},
    \end{equation}
    where $\rm FWHM_{ Red}$ is the FWHM of the red peak of the \lya\ line profile. This relation is compatible with the observational results found in \cite{Verhamme:2018aa}, whom first suggested this kind of correction. However, this trend depends strongly in the outflow geometry that is assumed \cite{GurungLopez_2019b}. In particular, this relation works well for the \ThinShell\ geometry, while the \GalacticWind\ deviates slightly from it. In this work we use Eq.\ref{eq:algorithm_2} for both, the \ThinShell\ and the \GalacticWind\ in order to qualitatively study the impact in the clustering of using a relation that is slight off. This mimics the observational framework in which an outflow geometry is assume for all the LAE population while the real escape channel has some differences with the assumed model. 
    
    \end{itemize}

    In general, the \GM , \IC\ and \VER\ algorithms, as presented here, are biased estimators of the redshift. For example, as the maximum of the \lya\ line is usually redder than \lya,  \GM\ and \IC\ usually provide $\lambda_{\rm Ly\alpha , obs}>\lambda_{\rm Ly\alpha}$. Therefore, the distribution of $\lambda_{\rm Ly\alpha , obs}$ is not centered at $\lambda_{\rm Ly\alpha}$. Also, in the case of the \VER\ algorithm, the relation between the FWHM and the offset of the global maximum might change depending on redshift and outflow geometry, leading to a biased estimation. In practice, these systematic biases can be corrected. For example, in \cite{steidel10} and \cite{Rudie_2012}, authors compared the redshift provided by the \IC\ method and the redshift provided by other spectral features, such as H$\alpha$. Then, they used the  systematic offset between these two distributions as a correction for the redshift inferred only from \lya . 
    
    In the following we leave the \GM , \IC\ and \VER\ methodologies uncorrected by systematic biases. In this way, it becomes more apparent the reason behind some of the trends that we find (see \S \ref{sec:ideal} ). This choice does not affect the recovered clustering, as it is insensitive to the mean of the $\lambda_{\rm Ly\alpha , obs}$ distribution \citep[e.g.][]{Byrohl_2019}.

\subsection{Neural networks}\label{subsec: NN}

    We propose a novel method to determine the systemic redshift of LAEs from their whole \lya\ line profile through a neural network. In this section we explain all the ingredients of the neural networks implemented in this work. First we describe the architecture and the different training sample. Then we study how the full line profile helps us determine the \lya\ wavelength.

    \subsubsection{Neural network architecture}

    The architecture of the neural network in this work consists in an input layer, a single hidden layer and an output layer. 
    We remark that this work does not focus on finding the best architecture to solve the \lya\ wavelength determination problem, as this would depend on the \lya\ observation characteristics (e.g. spectral resolution, signal to noise ratio, etc).  Instead, we adopt this simple architecture as a proof of concept. 
    
    In this work we seek for an algorithm that could be replicated in observational experiments. With this goal in mind, for a given \lya\ line profile $\phi(\lambda)$, we set as input $\phi(\Delta \lambda _{\rm Obs} )$, where we have mapped 
    
    \begin{equation}\label{eq:algorithm_N}
    \lambda \rightarrow \Delta \lambda _{\rm Obs}= \lambda - \lambda_{\rm Ly\alpha, Max}.
    \end{equation}
    In this way, the global maximum of the \lya\ line profile is always centered at $\Delta \lambda _{\rm Obs} = 0$, which can be easily replicated in observational experiments. Additionally, we rescale each individual $\phi(\Delta \lambda _{\rm Obs} )$ in a way that the minimum of the line profile is 0 and the maximum is 1. 
    
%%%%%%%%%%%%%%%%%%%%%%%%%%%%%%%%%%%%%%%%%%%%%%%%%%%%%%%%%%%%%%%%%%%%%%%%%%%%%%%%%%%%%%%%%%%%%%%%
\begin{figure} 
\includegraphics[width=3.3in]{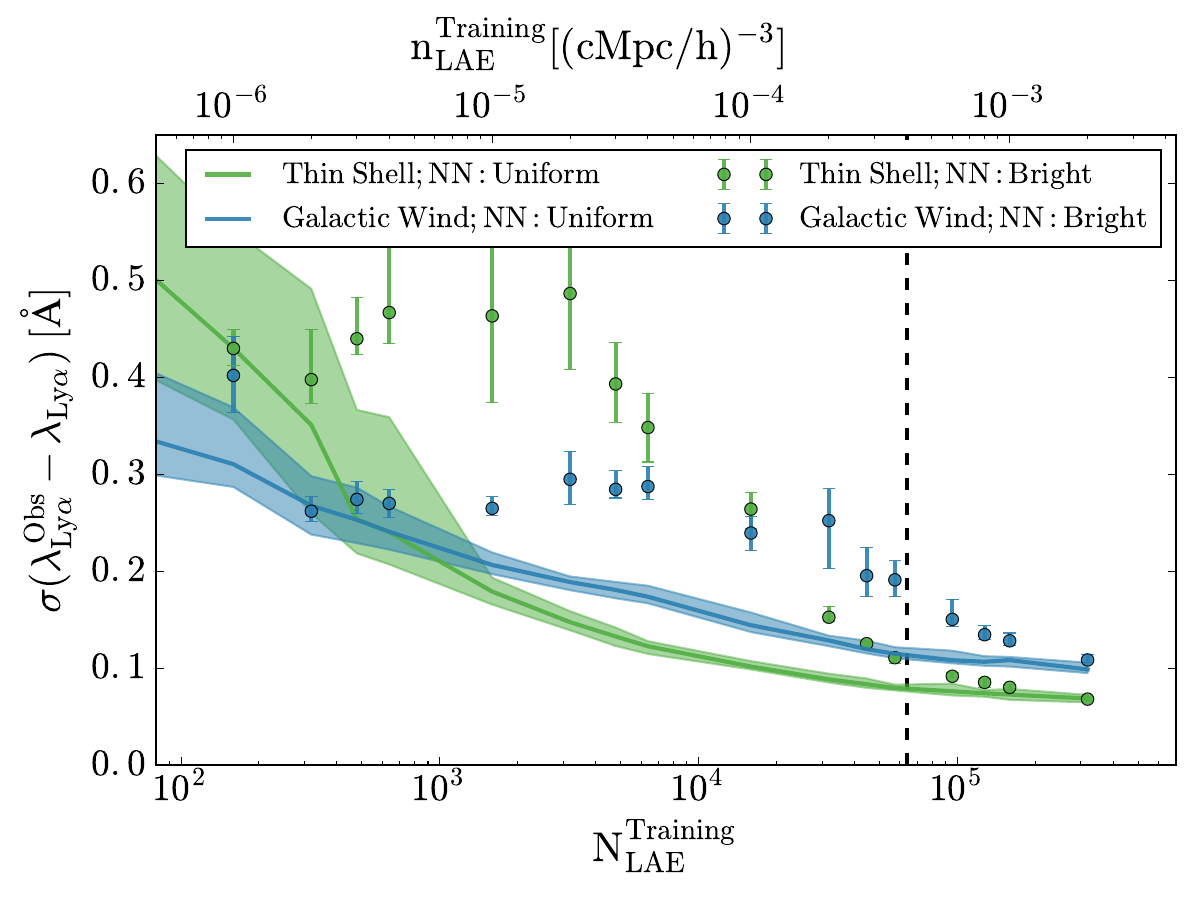}
\caption{Comparison of the accuracy of the neural network as a function of the training sample size at redshift $z=3.0$. The solid lines show the \NNU\ algorithm while the dots  show the \NNB . The shaded regions and error bars are the 1 sigma dispersion for the  \NNU\ and \NNB\ algorithms, respectively. The \ThinShell\ is represented in \colorThin\ while the \GalacticWind\ is plotted in \colorWind . The black dashed line shows the fiducial cut in number density for the training sample adopted through this work, which corresponds to a 10\% of the number density of the full LAE sample. }
\label{fig:accuracy_NN}
\end{figure}
%%%%%%%%%%%%%%%%%%%%%%%%%%%%%%%%%%%%%%%%%%%%%%%%%%%%%%%%%%%%%%%%%%%%%%%%%%%%%%%%%%%%%%%%%%%%%%%%
%%%%%%%%%%%%%%%%%%%%%%%%%%%%%%%%%%%%%%%%%%%%%%%%%%%%%%%%%%%%%%%%%%%%%
%%%%%%%%%%%%%%%%%%%%%%%%%%%%%%%%%%%%%%%%%%%%%%%%%%%%%%%%%%%%%%%%%%%%%%%%%%%%%%%%%%%%%%%%%%%%%%%%
\begin{table}
\caption{ Mean ($\mu(\Delta \lambda)$) and standard deviation ($\sigma(\Delta \lambda)$) of the difference between the \lya\ assigned as \lya\  and the true \lya\ frequency for the different \lya\ identification algorithms, redshifts and outflow geometries. }
\label{tab:parameters}
\begin{tabular}{ccccc}
Redshift   & Geometry      & Algorithm   & $\mu$                        & $\sigma$                    \\
           &               &             & [\AA{}]                      & [\AA{}]                     \\ \hline
2.2        & Thin Shell    & GM          & 1.822 & 0.66 \\
           &               & IC          & 2.1634 & 0.72 \\
           &               & GM-F        & -0.3059 & 0.45 \\
           &               & NN:Uniform  & 0.0032 & 0.07 \\
           &               & NN:Bright   & -0.0087 & 0.08 \\ \cline{2-5}
           & Galactic Wind & GM          & 1.8289 & 0.89 \\
           &               & IC          & 2.0189 & 0.94 \\
           &               & GM-F        & 0.4394 & 0.36 \\
           &               & NN:Uniform  & 0.0022 & 0.11 \\
           &               & NN:Bright   & -0.156 & 0.17 \\ \hline
3.0        & Thin Shell    & GM          & 1.4747 & 0.99 \\
           &               & IC          & 1.6347 & 1.07 \\
           &               & GM-F        & 0.0251 & 0.52 \\
           &               & NN:Uniform  & 0.0303 & 0.08 \\
           &               & NN:Bright   & 0.0202 & 0.1 \\ \cline{2-5}
           & Galactic Wind & GM          & 1.9134 & 0.99 \\
           &               & IC          & 2.1233 & 1.13 \\
           &               & GM-F        & 0.4757 & 0.34 \\
           &               & NN:Uniform  & 0.0007 & 0.11 \\
           &               & NN:Bright   & 0.0112 & 0.16 \\ \hline
5.7        & Thin Shell    & GM          & 1.5134 & 1.22 \\
           &               & IC          & 1.7079 & 0.8 \\
           &               & GM-F        & -0.7141 & 1.44 \\
           &               & NN:Uniform  & 0.0116 & 0.1 \\
           &               & NN:Bright   & 0.0094 & 0.12 \\ \cline{2-5}
           & Galactic Wind & GM          & 0.8751 & 0.97 \\
           &               & IC          & 1.131 & 0.85 \\
           &               & GM-F        & -0.5247 & 1.09 \\
           &               & NN:Uniform  & -0.0132 & 0.09 \\
           &               & NN:Bright   & -0.0185 & 0.17 \\ \hline
\end{tabular}
\end{table}
%%%%%%%%%%%%%%%%%%%%%%%%%%%%%%%%%%%%%%%%%%%%%%%%%%%%%%%%%%%%%%%%%%%%%%%%%%%%%%%%%%%%%%%%%%%%%%%%
%%%%%%%%%%%%%%%%%%%%%%%%%%%%%%%%%%%%%%%%%%%%%%%%%%%%%%%%%%%%%%%%%%%%%%%%%%%%%%%%%%%%%%%%%%%%%%%%

    \subsubsection{Training sets}
    
    Throughout this work we implement two neural networks. 
    Both of them have the same architecture, but different training sets:
    
    \begin{itemize}
    
        \item {\bf \NNU} : The training sample is randomly selected from the whole LAE population. In other words, there is no dependence on any LAE property. 
        
        \item {\bf \NNB} : The training sample is constructed only by the brightest LAEs. In practice, we rank our LAE population by their \lya\ luminosity. Then, we split the LAE population in two and use for training the brightest subset.
        
    \end{itemize}
    
    The motivation behind each training set is different. On one hand, \NNU\ represents an ideal scenario where, in a given survey, a subset of the whole LAE population is homogeneously selected and re-observed at a wavelength range that allows the measurement of spectral features (other than the \lya\ line) and the assignment of the true systemic rest frame. However, it is, in general, challenging to obtain systemic redshifts of such a homogeneous subsample, since the flux of other spectral features tend to be less prominent than \lya\ \citep{Trainor_2015} . On the other hand, \NNB\ is designed to study how well the LAE redshift determination works even only with the brightest LAEs re-observed at other wavelength, which is closer to a realistic situation than the \NNU\ case.

    We train both neural networks (\NNU\ and \NNB) for each combination of redshift, outflow geometry and spectral quality (see \S \ref{sec:ideal}). 
    
    The performance of the neural networks are linked to the size of the training set. In general, the larger the training set, the more accurate the neural network becomes. In Fig.~\ref{fig:accuracy_NN} we show the accuracy of our neural networks (\NNU\ and \NNB) for the different outflow geometries at redshift 3.0 as a function of the number density of LAEs used for the training, $\rm n^{Training}_{LAE}$, and the training set size $\rm N_{LAE}^{Training}$.  Here, we use the standard deviation of $\rm \lambda _{Ly\alpha}^{Obs} - \lambda_{Ly \alpha}$ (noted as $\sigma(\Delta \lambda)$) to quantify the quality of the neural network, as the clustering of LAEs is sensitive to the distribution of $\Delta \lambda$  \citep[][]{Byrohl_2019}. We remark that the mean and median of $\rm \lambda _{Ly\alpha}^{Obs} - \lambda_{Ly \alpha}$ are, in general, one order of magnitude smaller than $\sigma(\Delta \lambda)$, and they have little impact on the two-point statistics of the auto-correlation function. In order to estimate the variance of $\sigma( \Delta \lambda) $, for each value of $\rm N_{LAE}^{Training}$, we performed 100 iterations  if $\rm n^{Training}_{LAE}<4\times10^{-4}$ $({\rm cMpc}/h)^{-3}$ and 10 iteration if $\rm n^{Training}_{LAE}>4\times10^{-4}$ $({\rm cMpc}/h)^{-3}$. 
    
    Through out this work we use rest frame length units to quantify $\Delta \lambda$. However, It is common in the literature to provide this quantity in velocity units  \citep[e.g.][]{Byrohl_2019}. These two quantities are equivalent and can be transformed as 
    %\begin{equation}\label{eq:v_units}
    $\Delta v = c \Delta \lambda / \lambda_{\rm Ly\alpha} \simeq (247 km/s) \times  \Delta\lambda/1$\AA{} . 
    %\end{equation}
    
    As we show in Fig.~\ref{fig:accuracy_NN}, the neural networks are able to assign a \wlyaO\ close to \wlya . Overall, the accuracy increases as the training sample size is increased. However, the \NNU\ and \NNB\ algorithms behave slightly different. On one hand, in the \NNU\ algorithms, $\sigma(\Delta \lambda)$ decreases until $\rm n^{Training}_{LAE}\sim 4\times10^{-4}$ $({\rm cMpc}/h)^{-3}$ ($\sim 10\%$ of the total LAE sample), where it reaches a plateau around 0.1\AA{}.  This means that at this value of $\rm n^{Training}_{LAE}$, the training sample is big enough to cover the full variety of \lya\ line profiles. Thus, adding more galaxies to the training sample beyond $10^{-4}$ $({\rm cMpc}/h)^{-3}$ add little information, leaving the accuracy constant. On the other hand, the performance of the \NNB\ algorithm is worse at low $\rm n^{Training}_{LAE}$. Meanwhile, the \NNB\ converge to the \NNU\ accuracy when $\rm n^{Training}_{LAE}\sim 2\times10^{-3}$ $({\rm cMpc}/h)^{-3}$ ($\sim$50\% of the total sample).  This is because the \lya\ line profile depends on galaxy and IGM properties (see Figs.~\ref{fig:split_galaxy} and \ref{fig:split_IGM}). In particular, the \lya\ line profile of the bright LAEs is more redshifted than the \lya\ line profiles of the faint LAEs. Hence, the training sample in the \NNB\ is biased towards redshifted \lya\ line profiles and it does not contain typical line profiles from faint LAEs. This reduces the accuracy at low  $\rm n^{Training}_{LAE}$ in comparison to \NNU , which makes an uniform selection on $\llya$. Then, for larger values of $\rm n^{Training}_{LAE}$, the training samples of the \NNU\ and \NNB\ become more similar, which makes them converge to the same accuracy. Finally, the line profiles generated using the \GalacticWind\ outflow geometry seem slightly more complex than the \ThinShell\ counterparts, which results, typically, in a better accuracy for the \ThinShell\ geometry. 
    
    From now on, we fix the number density of the training sample to $\rm n^{Training}_{LAE}= 4\times10^{-4}$ $({\rm cMpc}/h)^{-3}$. We have chosen this value of $\rm n^{Training}_{LAE}$ for two main reasons: a) the information of the training sample of the \NNU\ saturates and increasing $\rm n^{Training}_{LAE}$ does not add new information to it, and b) the \NNB\ has not yet converged to the \NNU\ accuracy, so we can study the differences between these two methodologies.

\section{The effects of the \lya\ wavelength determination in ideal line profiles.}\label{sec:ideal}

Hereafter, we study how the miss-identification of the \lya\ wavelength modifies the clustering. In order to understand the physical consequences, in this section, we rather focus on an ideal case where we ignore binning artifacts and the instrumental noise in a LAE spectrum. We will consider more realistic situations in next section.

%%%%%%%%%%%%%%%%%%%%%%%%%%%%%%%%%%%%%%%%%%%%%%%%%%%%%%%%%%%%%%%%%%%%%%%%%%%%%%%%%%%%%%%%%%%%%%%%
%%%%%%%%%%%%%%%%%%%%%%%%%%%%%%%%%%%%%%%%%%%%%%%%%%%%%%%%%%%%%%%%%%%%%%%%%%%%%%%%%%%%%%%%%%%%%%%%
\begin{figure*}
\includegraphics[width=5.9in]{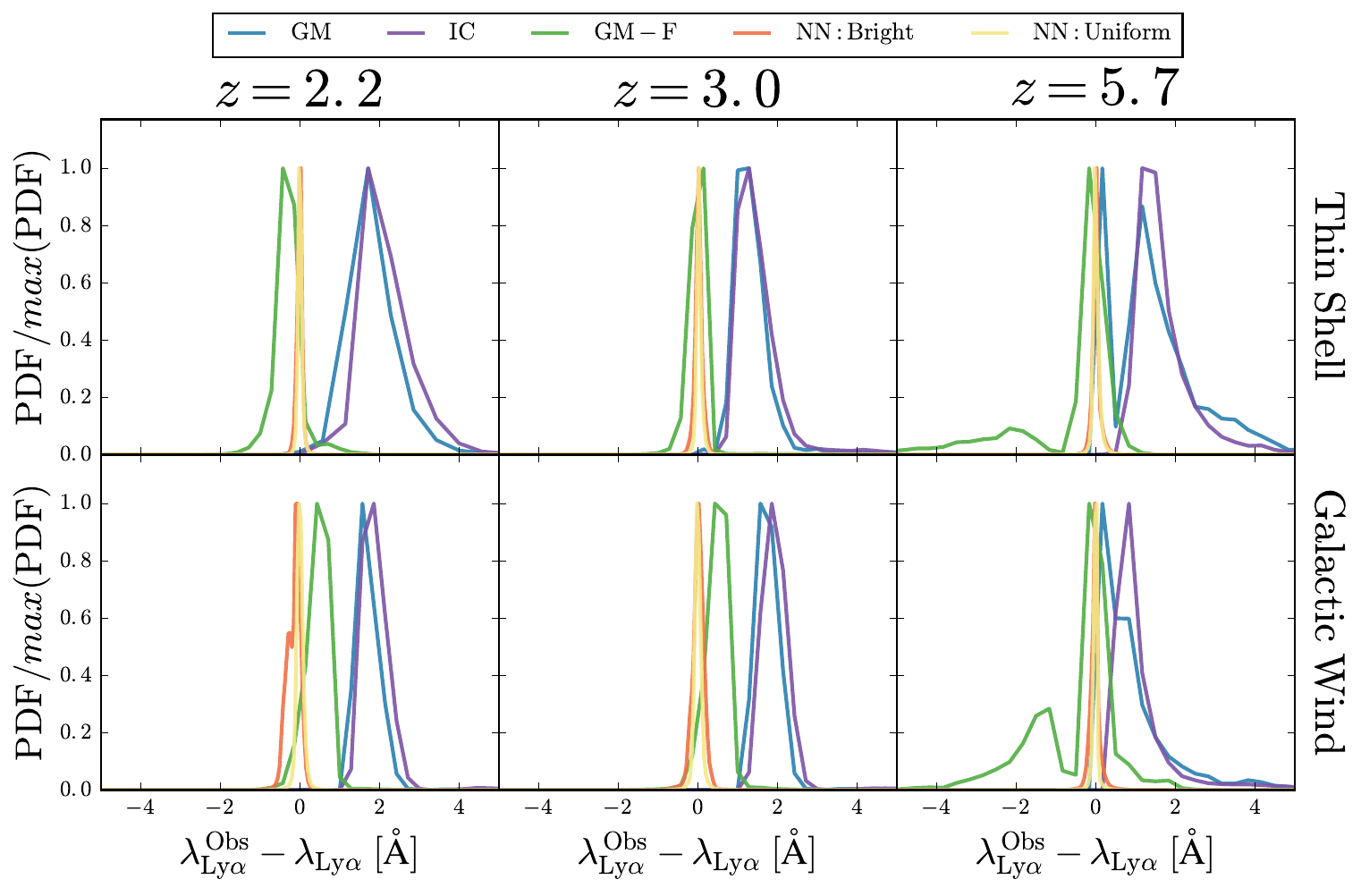}%
\caption{ Distribution of the difference between the assigned \lya\ wavelength and the intrinsic \wlya for each of the different \lya\ identification algorithms. The \GM\ algorithm is displayed in blue, \IC\ in purple, \VER\ in green,  \NNB\ in orange and \NNU\ in yellow. Each column shows a different redshift bin (2.2, 3.0 and 5.7 from left to right). The models using the \ThinShell\ (\GalacticWind) geometry are shown in the top (bottom) panels. }
\label{fig:performance_comparison}
\end{figure*}
%%%%%%%%%%%%%%%%%%%%%%%%%%%%%%%%%%%%%%%%%%%%%%%%%%%%%%%%%%%%%%%%%%%%%%%%%%%%%%%%%%%%%%%%%%%%%%%%
%%%%%%%%%%%%%%%%%%%%%%%%%%%

\subsection{Algorithm performances in ideal \lya\ line profiles}

In this section we compare the performance of the four methodologies to determine \wlya\ from a \lya\ line profile. In Fig.~\ref{fig:performance_comparison} we show the probability distribution function (PDF) of the deviation of \wlyaO\ from \wlya\ ($\Delta \lambda$) for the different redshifts, outflow geometries and algorithms to determine \wlyaO . Also, we list the mean ($\mu(\Delta \lambda)$) and standard deviation ($\sigma(\Delta \lambda)$) of these distribution in Tab.\ref{tab:parameters}. Overall, the algorithms using neural networks outperform the standard algorithms (\GM, \IC\ and \VER). We find that the best methodology to retrieve \wlya\ is \NNU\ (yellow), as the standard deviation of $\Delta \lambda$ is the smallest at all redshifts and outflow geometries. In detail, for all our models using \NNU , $\sigma(\Delta \lambda)$ is below 0.1\AA{} and $\mu(\Delta \lambda)$ is lower than 0.01\AA{}. The \NNU\ is followed closely by the \NNB (orange), which also exhibits a great performance, with $\sigma(\Delta \lambda) \sim 0.15$ and $\mu(\Delta \lambda) < 0.01$\AA{}. 

Regarding the standard methodologies, in general, \VER\ performs better than \GM . Meanwhile, \IC\ is the methodology with the worst performance at $z=2.2$ and $z=3.0$, while it performs better than \GM\ and \VER\ at $z=5.7$.  If we focus on \GM\ (blue) and \IC\ (purple), we find that \GM\ presents a smaller dispersion than \IC\ at redshifts 2.2 and 3.0, while the opposite is true at $z=5.7$. We also find that the performances of \GM\ and \IC\ are similar for both outflow geometries; the mean of the $\Delta \lambda$ distribution is shifted $\sim 1.7$\AA{} redwards \wlya . This is a direct consequence of the \lya\ RT, as the photons get redshifted when they travel through the neutral hydrogen of outflows. Then, as \GM\ takes the global maximum of the \lya\ line profile as \wlyaO, the whole $\Delta \lambda$ is systematically redshifted. In fact, we find that for the \GM\ algorithm, the peak of the $\Delta \lambda $ distribution is located at the same position than the peak of their corresponding stacked line profiles (see Fig.~\ref{fig:IGM_no_IGM_stacks}). Moreover, for the \ThinShell\ at $z=5.7$ the $\Delta \lambda$ distribution of the \GM\ algorithm exhibits a double peak shape, and the same position than the peaks present in the \lya\ stacked line profile of that model. 

The performance of \VER\ (green) depends strongly on the outflow geometry and redshift. On one hand, at low redshift ($z=2.2$ and $z=3.0$),  \VER\ performs better than \GM\ and \IC\  in both outflow geometries, as the \VER\ distributions of $\Delta\lambda$ are thinner than those of \GM\ and \IC\ . However, the \VER\ performance in the \ThinShell\ is better than in the \GalacticWind . In particular, $\mu (\Delta\lambda)$ is close to zero for the \ThinShell , while it is $\sim 0.5$\AA{} in the \GalacticWind . This is a consequence of the different relation between $\lambda_{\rm Ly\alpha , Max}$ and $\rm FWHM_{\rm Red}$ in the \ThinShell\ and \GalacticWind\ outflow geometries (see Fig.4 in \cite{GurungLopez_2019b}). On the other hand, at $z=5.7$, the $\Delta \lambda$ distribution becomes bimodal, with a second less prominent peak centered around $\Delta\lambda=-2$\AA{} (see Fig.~\ref{fig:IGM_no_IGM_stacks}). This is caused by the IGM modifying the \lya\ line profile in such a way that the $\rm FWHM_{\rm Red}$ of the observed \lya\ line profile is larger than the initial\footnote{For an example of this, see Fig.7 of \citep{GurungLopez_2020}}. As a consequence, \VER\ overcorrects the shift of the \lya\ peak, causing the peak in the $\Delta \lambda$ distribution at $\Delta \lambda<0$ . 

%%%%%%%%%%%%%%%%%%%%%%%%%%%%%%%%%%%%%%%%%%%%%%%%%%%%%%%%%%%%%%%%%%%%%%%%%%%%%%%%%%%%%%%%%%%%%%%%%

\subsection{Impact on the redshift-space clustering of LAEs }\label{ssec:clustering_ideal}

The misidentification of the \lya\ wavelength has a non-negligible impact on the three-dimensional clustering of LAE samples that rely only on their \lya\ line profile to determine their redshift, and hence, their radial position \citep{Byrohl_2019}. Indeed, the measured redshift of LAEs have three main contributions: a) the geometric redshift given by the Hubble flow, b) the redshift or blueshift given by the peculiar velocity of the galaxy along the line of sight \citep{kaiser87}, normally dabbed RSD, and c) a redshift or blueshift rising from the \lya\ wavelength misidentification, i.e.,  $\Delta \lambda \neq 0$.

%%%%%%%%%%%%%%%%%%%%%%%%%%%%%%%%%%%%%%%%%%%%%%%%%%%%%%%%%%%%%%%%%%%%%%%%%%%%%%%%%%%%%%%%%%%%%%%%
\begin{figure*} 
\includegraphics[width=6.9in]{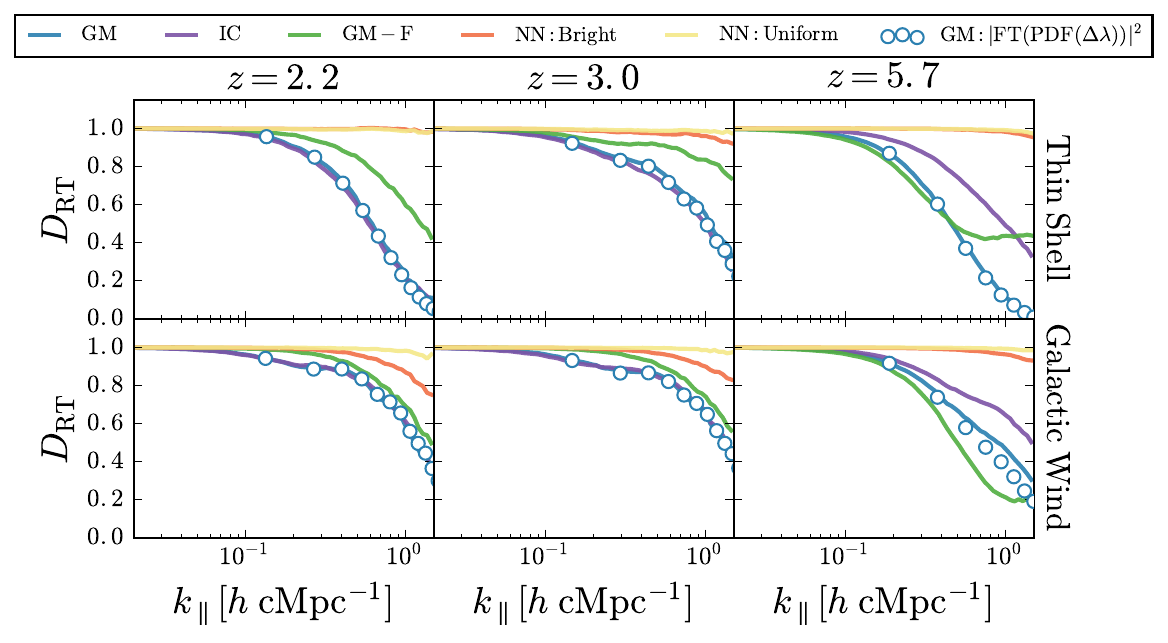}%
\caption{Damping of the power spectrum along the line of sight at redshift 2.2, 3.0 and 5.7 from left to right. In the top panels we show the models with the \ThinShell\ and \GalacticWind\ outflow geometry in the top and lower panels respectively. The colored solid lines show the different algorithms to identify the \lya\ wavelength (the color code is the same as in Fig.\ref{fig:performance_comparison}). The blue empty dots display the damping of the power spectrum computed from the PDF of $\Delta\lambda$ through Eq.\ref{eq:power_spectrum}.}
\label{fig:damping_comparison}
\end{figure*}
%%%%%%%%%%%%%%%%%%%%%%%%%%%%%%%%%%%%%%%%%%%%%%%%%%%%%%%%%%%%%%%%%%%%%%%%%%%%%%%%%%%%%%%%%%%%%%%%

In order to characterize the clustering of a galaxy population, it is useful to define the galaxy overdensity field
%%%%%%%
\begin{equation}
\label{eq:delta}
\delta_{\rm g} = \displaystyle{ \frac{ n_{\rm g}(\vec{x}) }{ \langle n_{\rm g} \rangle} - 1  },
\end{equation} 
%%%%%%%
where $n_{g}(\vec{x})$ is the number density galaxies at the position $\vec{x}$, and $\langle n_{\rm g} \rangle$ is its average value.  
Then the two-point correlation function (2PCF), $\xi_{\rm g}$, is defined as 
%%%%%%%
\begin{equation}
\label{eq:2PCF}
1 + \xi(\vec{r}) = \displaystyle{  \langle  [1+\delta_{\rm g}(\vec{x})] [1+\delta_{\rm g}(\vec{x}+\vec{r})]\rangle   } 
\end{equation} 
%%%%%%%
where $\vec{r}$ is the pair vector between two points separated a distance $r$. 
We also consider the power spectrum $P_{\rm g}(\vec{k})$, which is the Fourier transform of the 2PCF, i.e., 
%%%%%%%
\begin{equation}
\label{eq:Pk}
P_{\rm g}(\vec{k}) \; (2\pi)^{3}\; \delta_{D}(\vec{k}+\vec{k'})  = \displaystyle{ \langle \delta_{\rm g}(\vec{k}) \, \delta_{\rm g}(\vec{k'}) \rangle }
\end{equation} 
%%%%%%%
where $\vec{k}$ is the wavenumber, and $\delta_{D}(\vec{k})$ is the Dirac delta function.

Throughout this work we will focus on the clustering in \textit{redshift space}.
To incorporate the three redshift contributions to our clustering analysis, we recompute the position of our LAEs in redshift space as
%%%%%%%
\begin{equation}
\label{eq:position_LoS}
\vec{s} = \vec{r} + \displaystyle{ { {\rm V_{LoS} + \Delta V_{Ly\alpha}}\over{a(z)H(z)} }\hat{Z}}, 
\end{equation} 
%%%%%%%
where we take $Z$ as the direction of the line of sight, assuming the global plain-parallel approximation \citep{Beutler:2013aa}, $\rm V_{\rm LoS}$ is the velocity along the line of sight of the galaxy.
$a(z)$ and $H(z)$ are the scale factor and the Hubble parameter at redshift $z$, respectively. Additionally, 
%%%%%%%
\begin{equation}
\label{eq:position_Lya}
{\rm \Delta V_{Ly\alpha} } = c \; {\rm \left( 1 - {\frac{\lambda_{Ly\alpha}}{ \lambda^{Obs}_{Ly\alpha}} } \right) }\,,
\end{equation} 
%%%%%%%
where $c$ is the speed of light. Finally, there is a fraction of the LAE population that are shifted outside of the box after transforming their line position to redshift space, i.e., including the contributions of $\rm V_{\rm LoS}$ and $\rm \Delta V_{Ly\alpha}$. For these galaxies we assume that our simulation box is periodic along the line of sight. Therefore, galaxies with $\pi<0$, they are assigned $\pi={L_{\rm Box}}+\pi$ and galaxies with $\pi>{L_{\rm Box}}$, they are assigned $\pi=\pi-{L_{\rm Box}}$.
%, where $L_{\rm Box}$ is the side length of the simulation box. \\

\subsubsection{Clustering damping in Fourier space}

First, we focus on the impact of the \lya\ misidentification in the power spectrum. 
In \cite{Byrohl_2019}, the authors analytically showed that, due to the \lya\ misidentification, the amplitude of the power spectrum is reduced along the line of sight at large wavenumber ($k_{\parallel}$) values, i.e., at small scales. We define the damping of the power spectrum due to the \lya\ miss-identification as in \cite{Byrohl_2019}, i.e., 

\begin{equation}
\label{eq:power_spectrum}
D_{\rm RT} (k_{\parallel}) = {P^{\rm RT}(k)\over{P(k)}} = |FT(PDF(\Delta\lambda))|^2 ,
\end{equation}
where $P(k_{\parallel})$ is the intrinsic redshift-space power spectrum of the LAE population, i.e., setting ${\rm \Delta V_{Ly\alpha} }=0$, in Eq.~(\ref{eq:position_LoS}). Also, $P^{\rm RT}(k_{\parallel})$ is the power spectrum of the LAE sample after including the displacement along the line of sight due to the \lya\ misidentification. Finally, the Fourier transformation is indicated as $FT$. We estimate the power spectrum from the simulated LAEs by making use of the Fast Fourier Transformation (FFT). We set the number of grids as $512^{3}$ with which the Nyquist wavenumber is $k_{\rm Nyq}\sim 3\,h{\rm cMpc^{-1}}$. The last equality holds only if i) the moments of $PDF(\Delta\lambda)$ are scale independent and ii) $\Delta\lambda$ is uncorrelated with the large scale density and velocity fields \citep{Byrohl_2019}. 

%%%%%%%%%%%%%%%%%%%%%%%%%%%%%%%%%%%%%%%%%%%%%%%%%%%%%%%%%%%%%%%%%%%%%%%%%%%%%%%%%%%%%%%%%%%%%%%%
\begin{figure*} 
\includegraphics[width=5.9in]{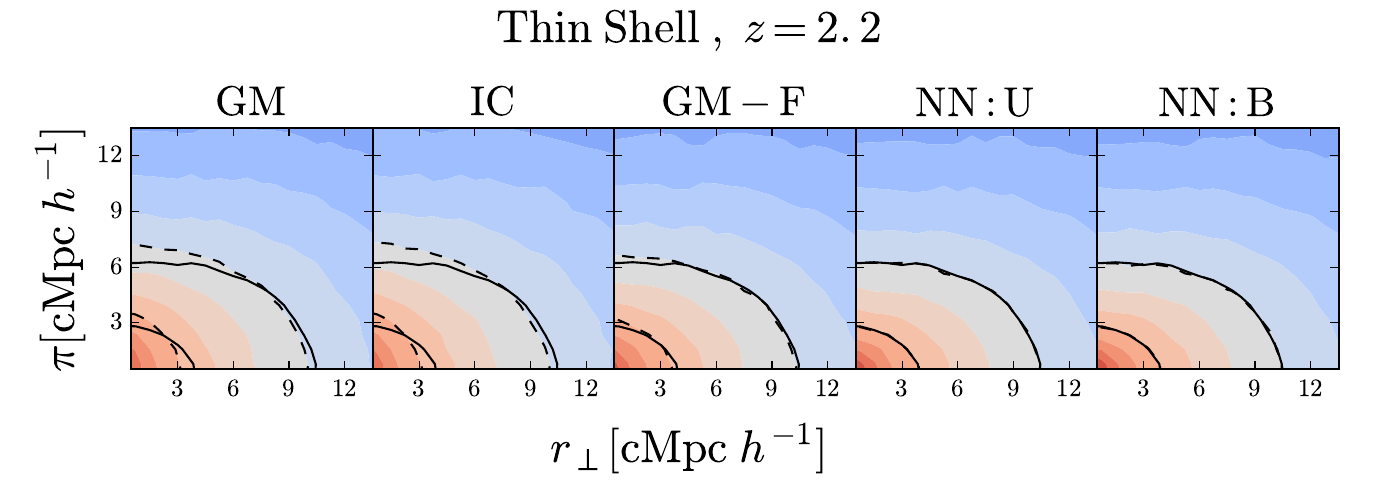}%
\caption{Redshift space clustering divided the   parallel ($\pi$) and perpendicular ($r_{\perp}$) directions to the line of sight at $z=2.2$ for the \ThinShell\ outflow geometry. From left to right, in each panel, we display the \GM , \IC , \VER , \NNU\ and \NNB\ algorithms, respectively. For each algorithm, we show the clustering levels $\xi(r_{\perp},\pi)=10^{-0.6}$ and $10^{0.0}$  in dashed black lines in their corresponding panels. The solid black lines is the same in every panel and corresponds with the same clustering levels in the case in which there no \lya\ wavelength miss-identification. }
\label{fig:2d_clustering}
\end{figure*}
%%%%%%%%%%%%%%%%%%%%%%%%%%%%%%%%%%%%%%%%%%%%%%%%%%%%%%%%%%%%%%%%%%%%%%%%%%%%%%%%%%%%%%%%%%%%%%%%

In Fig.~\ref{fig:damping_comparison} we show the damping in the power spectrum for our different \lya\ wavelength recovering algorithms and multiple models. In general, we find that the amplitude of the power spectrum including the \lya\ misidentification is lower than the intrinsic power spectrum at the scales relevant to the BAO and RSD measurements, as $D_{\rm RT} < 1 $ at $0.1<k_{\parallel}\,[h\,{\rm cMpc^{-1}}]<1$.In particular, the impact is greater on smaller scales (larger $k_{\parallel}$), while at large enough scales it disappears. This damping can be interpreted as the Finger-of-God effect and matches the results from \citet[][]{Byrohl_2019}, although the detailed suppressions behave differently as the PDFs are different.

$D_{\rm RT}$ has been computed in two different ways: i) computing the power spectra directly from the LAE positions in our simulation box (solid lines) and ii) by computing the Fourier transform of the one point PDF of $\Delta\lambda$ (open circle points). We compare the two methods only for the GM in Fig.~\ref{fig:damping_comparison}, and confirm that they are in a good agreement. This ensures that the misidentification is uncorrelated with the large scale density or velocity field, while there is a small hint of the IGM interaction at $k_{\parallel}\sim 1$ for the Galactic Wind at $z=5.7$. We have checked similar results for the other algorithms, and hence omitted them in the figure.

We find that, the algorithms with a higher accuracy for recovering the \lya\ wavelength from the line profile show a shallower damping of the power spectrum. In particular, the \NNU\ is the algorithm that is the least affected by the \lya\ wavelength misidentification. In fact, the recovered power spectrum agrees at the 1\% level up to $k_{\parallel}=1$ for both outflow geometries and at all redshifts. The second best performance is achieved by the \NNB , which exhibits up to $\sim 0.2$ decreases in the power spectrum amplitude in the \GalacticWind, while in the \ThinShell , the damping is slight stronger than in the \NNU . Then, the \GM , \IC\ and \VER\ algorithms are heavily affected by the \lya\ misidentification, as the amplitude of the power spectrum decreases dramatically on small scales at all redshifts and outflow geometries. In particular, \VER\ is less affected than \GM\ and \IC , at redshift 2.2 and 3.0. Meanwhile at $z=5.7$,  \IC\ behaves better than the other two  and the performance of \VER\ is comparable (in the \ThinShell) or slight worst (in the \GalacticWind) than \GM .

%%%%%%%%%%%%%%%%%%%%%%%%%%%%%%%%%%%%%%%%%%%%%%%%%%%%%%%%%%%%%%%%%%%%%%%%%%%%%%%%%%%%%%%%%%%%%%%%
\begin{figure*} 
\includegraphics[width=5.4in]{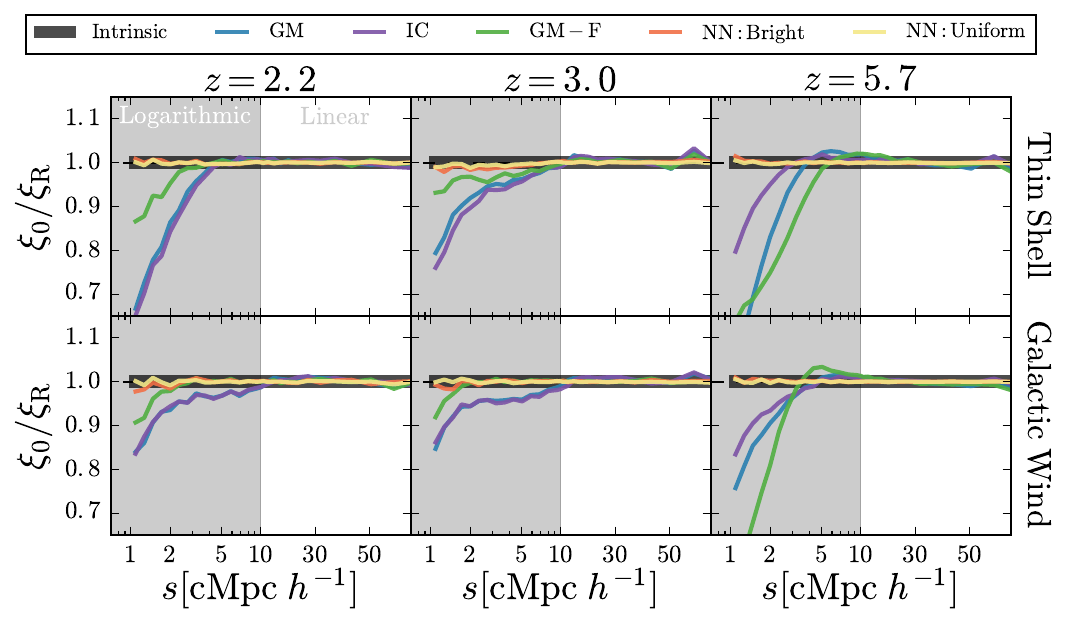}%
\caption{ Ratio between the monopole in redshift space of the LAEs samples using different \lya\ frequency identification algorithms (thin colored lines) and the one assuming a perfect accuracy in the \lya\ wavelength identification (thick black line). Each column displays a redshift bin (2.2, 3.0 and 5.7 from left to right). Top (bottom) panels panels display the \ThinShell\ (\GalacticWind) outflow geometry models. The scale in the grey shaded region is logarithmic, while in the white region is linear. }
\label{fig:monopole_comparison}
\end{figure*}
%%%%%%%%%%%%%%%%%%%%%%%%%%%%%%%%%%%%%%%%%%%%%%%%%%%%%%%%%%%%%%%%%%%%%%%%%%%%%%%%%%%%%%%%%%%%%%%%

\subsubsection{Impact on the 2D clustering in configuration space}

In this section we explore the 2PCF to illustrate qualitatively the clustering distortion produced by the misidentification of the \lya\ wavelength. Later,  we will further quantify the anisotropic distortions using the Legendre multipole moments.

We estimate 2PCF with the standard Landy-Szalay estimator \citep{landy93}. Fig.~\ref{fig:2d_clustering} shows the clustering divided into parallel and perpendicular to the line of sight components for our different algorithms. In particular, we display the model at redshift 2.2 using the \ThinShell\ geometry, but similar results are found for the other models too. To compare the performance of the algorithms, we show the contours with the clustering amplitude of $\xi(r_{\perp},\pi)=10^{-0.6}$ and $10^{0.0}$  for no \lya\ wavelength misidentification (solid) and for each algorithm (dashed).  Overall, the misidentification of the \lya\ wavelength causes an elongation of the LAE clustering along the line of sight. This elongation is more prominent for the algorithms with worst performance recovering the \lya\ wavelength. In concordance with our previous findings, the \GM , \IC\ and \VER\ algorithm fail to recover the intrinsic redshift-space clustering of LAEs. Meanwhile, \NNU\ and \NNB\ achieve almost a perfect recovery of the 2PCF .

\subsubsection{Impact on the monopole}

The multipole 2PCF is given by
%%%%%%
\begin{equation}
\label{eq:quadrupole}
\xi_{\ell}(s) = {{2\ell+1}\over{2}} \int _{-1}^{1} d\mu \; \xi(s,\mu) \; \mathbcal{L} _{\ell} (\mu),
\end{equation}
%%%%%%
where $\ell$ is the multipole degree and $\mathbcal{L} _{\ell}$ is the Legendre's polynomial of degree $\ell$ and $\mu$ is the cosine between the line of sight and the separation vector of a galaxy pair.

In Fig.~\ref{fig:monopole_comparison} we show the ratio of $\xi_{\rm 0}$, the monopole ($\ell=0$) of our LAE populations by using our different algorithms to determine the \lya\ wavelength, and $\xi_{\rm R}$, the monopole when the \lya\ wavelength is identified perfectly, i.e.,  $\Delta\lambda=0$ for every galaxy.  Overall we find that the method used to determine the \lya\ wavelength change the retrieved monopole of LAEs at scales below $10\,{\rm cMpc}/h$ for both outflow geometries at all redshifts. Meanwhile, the large scale clustering ($>10\,{\rm cMpc}/h$) remain unchanged.  In particular, at large scales, the monopole of the different algorithms converged to the intrinsic one with nearly no difference among them.

The clustering for the traditional algorithms (\GM , \IC\ and \VER) is suppressed at small scales. In contrast, the measured monopole using both our neural networks (\NNU and \NNB) match extraordinary well the intrinsic clustering of LAEs at all scales. These differences in the monopole between the standard approaches and the neural networks are driven by the much better performance of \NNU\ and \NNB\ when determining the \lya\ frequency (see Fig.~\ref{fig:performance_comparison}). In particular, the large dispersion of $\Delta \lambda$ given by \GM , \IC\ and \VER\ translates in to a large scatter of $\rm \Delta V_{Ly\alpha}$. Which means that, the position of the galaxies in redshift space are quite spread along the line of sight with respect to their original position in redshift space. This dilutes the clustering on small scales along the line of sight, which causes a decrease of power in the monopole on  scales $\gtrsim 1\,{\rm cMpc}/h$.

%%%%%%%%%%%%%%%%%%%%%%%%%%%%%%%%%%%%%%%%%%%%%%%%%%%%%%%%%%%%%%%%%%%%%%%%%%%%%%%%%%%%%%%%%%%%%%%%
\begin{figure*} 
\includegraphics[width=5.4in]{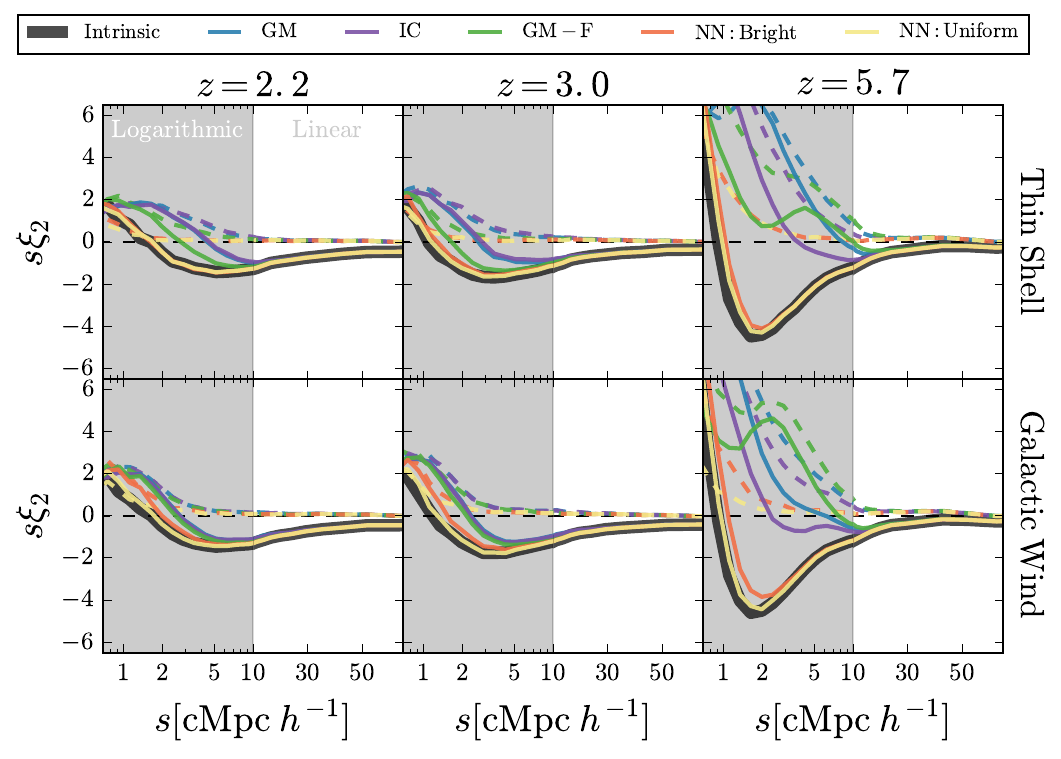}%
\caption{  Quadrupole of the LAE samples using \lya\ identification algorithms (\GM\ in blue, \IC\ in purple, \VER\ in green, \NNB\ in orange and \NNU\ in yellow), at different redshifts (2.2, 3.0 and 5.7 from left to right) and implementing different outflow geometries (\ThinShell\ in the top and \GalacticWind\ in the bottom). The colored solid lines are computed with both, the contribution of the \lya\ miss-identification and the peculiar motion of the galaxies. Meanwhile, the colored dashed line only include the shift due to the \lya\ line profile. The scale in the grey shaded region is logarithmic, while in the white region is linear.}
\label{fig:quadrupole_comparison}
\end{figure*}
%%%%%%%%%%%%%%%%%%%%%%%%%%%%%%%%%%%%%%%%%%%%%%%%%%%%%%%%%%%%%%%%%%%%%%%%%%%%%%%%%%%%%%%%%%%%%%%%

Moreover, we find that, on the scales studied here, both neural networks produce very similar monopoles at all redshifts and for both outflow geometries. Meanwhile, we see that the power suppression in the \VER\ monopole is generally smaller than the suppression in the \GM\ monopole. Again, this comes from the different performance among these algorithms. \VER\ performs better than \GM\, since it provides a tighter distribution of $\Delta \lambda$ in the PDF (Fig.~\ref{fig:damping_comparison}).

\subsubsection{Impact on the quadrupole}

The quadrupole ($\ell=2$) is more sensitive to the anisotropic clustering than the monopole. In Fig.\ref{fig:quadrupole_comparison} we show the quadrupole at different redshifts for both outflow geometries. The solid lines indicate the quadrupole when the peculiar motion of galaxies and the \lya\ wavelength misidentification are implemented (using Eq.~\ref{eq:position_LoS}).  In contrast, in order to isolate the contribution of the \lya\ shift, we show with dashed lines the quadrupole using Eq.~\ref{eq:position_LoS} but assuming that galaxies have no peculiar motion along the line of sight ($\rm V_{Los}=0$).

Overall, we find that the uncertainty in the \lya\ wavelength determination changes the ratio between the clustering parallel and perpendicular to the line of sight. In this way, the quadrupole amplitude is enhanced at scales $\lesssim 10\,{\rm cMpc}/h$. This result is consistent with the Finger-of-God effect as we have already confirmed in Fourier space. The \lya\ misidentification can be regarded as an additional random shuffling along the line of sight. This interpretation is further assured by the fact that the dashed lines have negligible quadrupole amplitudes on large scales, $\gtrsim 10\,{\rm cMpc}/h$. Additionally, all methodologies converge on large scales, $\gtrsim 10\,{\rm cMpc}/h$, suggesting that the quadrupole 2PCF on such scales can be safely used to infer the peculiar velocity contribution in LAE surveys. Finally, The negative amplitude of the quadrupole 2PCF at large scales is qualitatively consistent with the Kaiser effect.

Focusing on the \lya\ wavelength misidentification contribution, we find the same trend as the results in Fourier space. The different algorithms produce different quadrupole predictions, reflecting its algorithms' performance. 

On one hand, the quadrupole recovered by the standard algorithms (\GM , \IC\ and \VER)  is heavily distorted with respect the intrinsic one (black). We find that the amplitude of the suppression evolves with redshift, being lighter at low redshift. In fact, the typical scale at which the intrinsic and the observed LAE quadrupole converge is $\sim 5 {\rm cMpc}/h$, $\sim 8 {\rm cMpc}/h$ and $\sim 15 {\rm cMpc}/h$ at redshifts 2.2, 3.0 and 5.7 respectively. Additionally, at small scales, the quadrupole amplitude is enhanced, and its sign is flipped, specially at redshift 5.7.  

Our neural network approaches work pretty well recovering the intrinsic quadrupole at all redshift and for both outflow geometries. In detail, we find very small differences between \NNU\ and \NNB , as \NNU\ performs slightly better. In other words, the contribution to the quadrupole given by the \lya\ wavelength misidentification becomes negligible when the \lya\ wavelength is computed using \NNU\ and \NNB .

\section{The effects of the \lya\ wavelength determination in realistic line profiles.}\label{sec:real}

In the previous sections we have studied the properties the \lya\ line profiles directly predicted by our model. 
These \lya\ line profiles are ideal in terms of i) signal to noise, which is effectively infinite and constant across all the \lya\ luminosity range of our models (down to $10^{41.5}{erg\;s^{-1}}$), and ii) the size of the independent wavelength bins (0.1\AA{}) \footnote{This value comes from the bin size used in \flareon\ to store the \lya\ line profiles.}. However, in observational data sets, reaching these conditions is challenging and/or impossible nowadays. 

In this section we study how the quality of the \lya\ line profiles affects the clustering measurements in \lya\ focused spectroscopic galaxy surveys. We focus on the snapshot at redshift 3.0 using the \ThinShell\ outflow geometry. We have checked that, the same analysis of the other redshift bins and outflow geometries gives, qualitatively, the same trends.

\subsection{Mocking measured Ly$\alpha$ line profiles}

In the following we explain how the \lya\ line profiles produced by our model are deteriorated with three distinct steps, and we illustrate them in Fig.~\ref{fig:noisy_line}.

\begin{description}
    \item[Step 1) Spectral resolution:] {We degrade the wavelength resolution of the line profile. For this end, we convolve the \lya\ line profile produced by our model with a Gaussian kernel of full width half maximum $\rm W_{g}$. In this way, the line profile gets broader and the features are diluted. For example, the blue peak that is clearly present in the original line profile is hardly seen after a Gaussian filter of 1.0\AA{} (dark blue solid line in Fig.~\ref{fig:noisy_line}). We implement three different values of $\rm W_{g}$: 0.5\AA{} , 1.0\AA{} and 2.0\AA{} in the LAE's rest frame. }

    \item[Step 2) Pixelization:] { We pixelize the \lya\ line profile into wavelength bins of width $\Delta \lambda _{\rm pix}$. In practice, the pixelized \lya\ line profile $\phi_{\rm pix}$ is computed as 
    
    \begin{equation}
    \label{eq:pixelization}
    \rm
    \displaystyle
    \phi_{pix}(\lambda_{pix})  =  
    {
    \displaystyle
    {\int ^{\lambda_{pix}+\Delta \lambda_{pix}/2} 
          _{\lambda_{pix}-\Delta \lambda_{pix}/2}
          {\rm \phi (\lambda) \;  d\lambda }} 
    \over 
    {\Delta \lambda_{pix} } 
    }
    ,
    \end{equation} 
    where $\lambda_{\rm pix}$ is the wavelength of each wavelength bin. As the width of the wavelength bins increases, it becomes progressively more difficult to resolve features in the \lya\ line profile (see the red solid line in Fig.~\ref{fig:noisy_line}). 
    In this work we implement three different values of $\Delta \lambda _{\rm pix}$: 0.25\AA{}, 0.5\AA{} and 1.0\AA{} in the LAE's rest frame. Fore each of these, the full LAE population is convolved with the same value.}
    
    \item[Step 3) Noise:]{ Finally, we include noise in the \lya\ line profile. 
    In detail, we assign the signal to noise ratio of the faintest (in \lya) object of our catalog to $\rm S/N_{F}$. Then, for each galaxy, its signal to noise ratio is scaled as 
    
%%%%%%%%%%%%%%%%%%%%%%%%%%%%%%%%%%%%%%%%%%%%%%%%%%%%%%%%%%%%%%%%%%%%%%%%%%%%%%%%%%%%%%%%%%%%%%%%
\begin{figure} 
\includegraphics[width=3.2in]{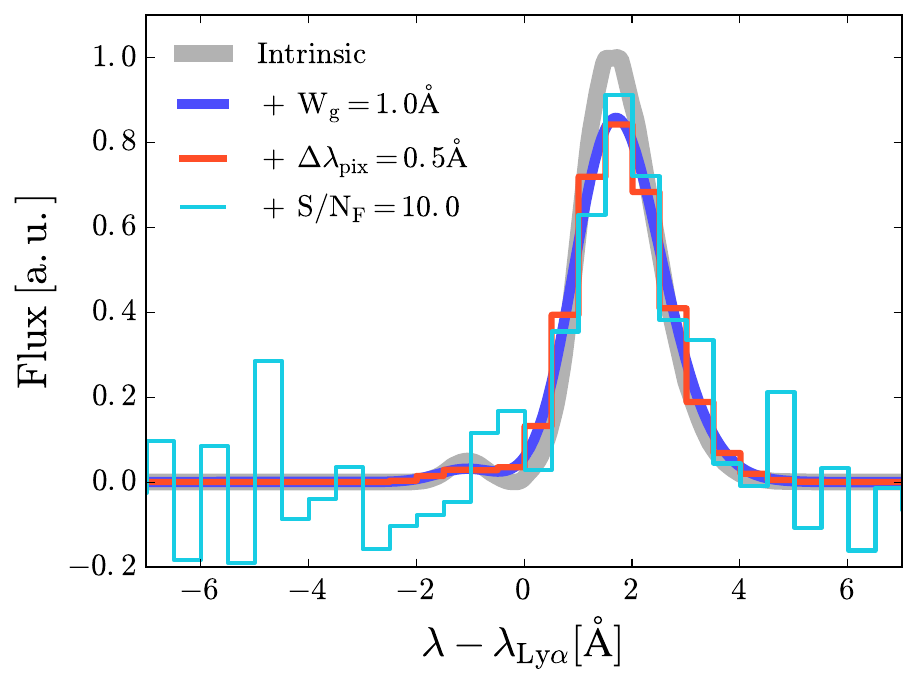}%
\caption{Illustration of the line profile quality down grade. In grey we show a particular line profile predicted by our model. The other line displays progressively (and cumulative) lower quality. In blue we include a gaussian kernel of $\rm FWHM=1$\AA{}, then we pixelize this line in wavelength bins of size $0.5$\AA{}. Finally we add gaussian noise to each pixel with an amplitude so that $\rm S/N=7$. }
\label{fig:noisy_line}
\end{figure}
%%%%%%%%%%%%%%%%%%%%%%%%%%%%%%%%%%%%%%%%%%%%%%%%%%%%%%%%%%%%%%%%%%%%%%%%%%%%%%%%%%%%%%%%%%%%%%%% 
    
    \begin{equation}
    \label{eq:noise}
    \rm
    \displaystyle
    S/N = S/N_{F} \times {L_{Ly\alpha}\over{L_{Ly\alpha,F}}} \; ,
    \end{equation} 
    where $L_{\rm Ly\alpha}$ is the \lya\ luminosity of each galaxy and $L_{\rm Ly\alpha,F}$ is the \lya\ luminosity of the faintest galaxy. Note that our models, by construction, provides good estimates of the LAE luminosity function \citep{GurungLopez_2020}. Therefore, the $\rm S/N$ distribution should also mimic observations. Next, to each pixel we add Gaussian noise with an amplitude that corresponds to the $\rm S/N$ of that LAE. In this way, LAEs with lower $\rm S/N$ have a noisier \lya\ line profile and vice versa. Moreover, the larger $\rm S/N_{F}$ is, the better signal to noise ratio has the faintest LAE and the whole LAE population. On the other side, for low $\rm S/N_{F}$ values, some information from the \lya\ line profile (e.g. light blue line) is vanished. In practice, the larger are $\rm W_{g}$ and $\Delta \lambda_{\rm pix}$ the more information is destroyed for a fixed value of $\rm S/N_{F}$. We use four values for $\rm S/N_{F}$ : 6.0, 7.0, 10.0 and 15.0. Also, for the LAE sample studied here,  $\rm L_{Ly\alpha,F} \sim 1.63\times10^{42}\;erg \; s^{-1} $.}
    
\end{description}

For each combination of \{$\rm W_{g}$, $\Delta \lambda _{\rm pix}$, $\rm S/N_{F}$\} , we produce a catalog of \lya\ line profiles. In practice, we could determine directly the \lya\ central wavelength from these line profiles, as we did in the ideal case in Sec.~\ref{sec:ideal}.However, as the line profiles are progressively downgraded it becomes more difficult to measure properly the $\rm FWHM_{red}$ (necessary for the \VER ). Also, the global maximum gets more and more discretized as $\rm \Delta \lambda_{pix}$ increases (needed for both, \GM , \VER ). In order to alleviate these problems, we fit a Gaussian curve to the most prominent peak of the \lya\ line profile and then, measure $\rm FWHM_{red}$ and $\lambda_{\rm Ly\alpha , Max}$ from the Gaussian. Note that, as the line profiles are convolved with a Gaussian kernel of width $\rm W_g$, $\rm FWHM_{red}$ will be, in general, overestimated, which would translate into a systematic bias for the \VER\ method. In order to correct this, we compute the FWHM as

\begin{equation}
\label{eq:new_FWHM}
\rm
FWHM_{red,c} = \sqrt{  FWHM_{red}^2 - W_g^2 }.
\end{equation} 

There are a few cases
\footnote{ The fraction of cases depends on the quality of the line profiles. For example, if $\rm W_g \leq 1.0$\AA{} and $\rm \Delta\lambda_{pix} \leq 0.5$\AA{} less than 2\% of the sample experience this issue for any $\rm S/N_{F}$ value. However, when $\rm W_g=2$\AA{} and $\rm S/N_{F}=6.0$ the percentage increases up to a 25\% for  $\rm \Delta\lambda_{pix}=1.0\AA$}
where, due to noise in the line profile, $\rm FWHM_{red} < W_g$. In these cases we assume that  $\rm FWHM_{red,c}=FWHM_{red}$. Then, for the standard \VER\ algorithm we use $\rm FWHM_{red,c}$ to correct the global maximum offset through Eq.\ref{eq:algorithm_2}. For completeness we also show the results for this algorithm without this correction, i.e., using $\rm FWHM_{red}$ instead of $\rm FWHM_{red,c}$. We refer to this last method as  \uVER .

This proceeding can be applied to observational data too. Note that, we only use the Gaussian fitting to compute $\rm FWHM_{red,c}$ and $\lambda_{\rm Ly\alpha , Max}$. In this way, the neural networks use directly the modeled line profiles and not the  Gaussian resulting from the fitting.
 
Another consideration is that, the performance of the \IC\ algorithm in realistic line profiles depends on more variables than just $\rm W_{g}$, $\Delta \lambda _{\rm pix}$ and $\rm S/N_{F}$. For example, its accuracy depends on the spectral range considered for computing the centroid of the line profile. If the spectral range is too narrow, then the centroid might be heavily affected by noise. Meanwhile, for a large spectral range the noise might be average out and the \IC\ accuracy would increase. Also, as it was illustrated in \S \ref{sec:ideal}, the behaviour of \IC\ is relatively similar to that of \GM , specially at redshift 2.2 and 3.0. Due to these reasons we drop the \IC\ algorithm in this section.

%%%%%%%%%%%%%%%%%%%%%%%%%%%%%%%%%%%%%%%%%%%%%%%%%%%%%%%%%%%%%%%%%%%%%%%%%%%%%%%%%%%%%%%%%%%%%%%%
\begin{figure*} 
\includegraphics[width=6.95in]{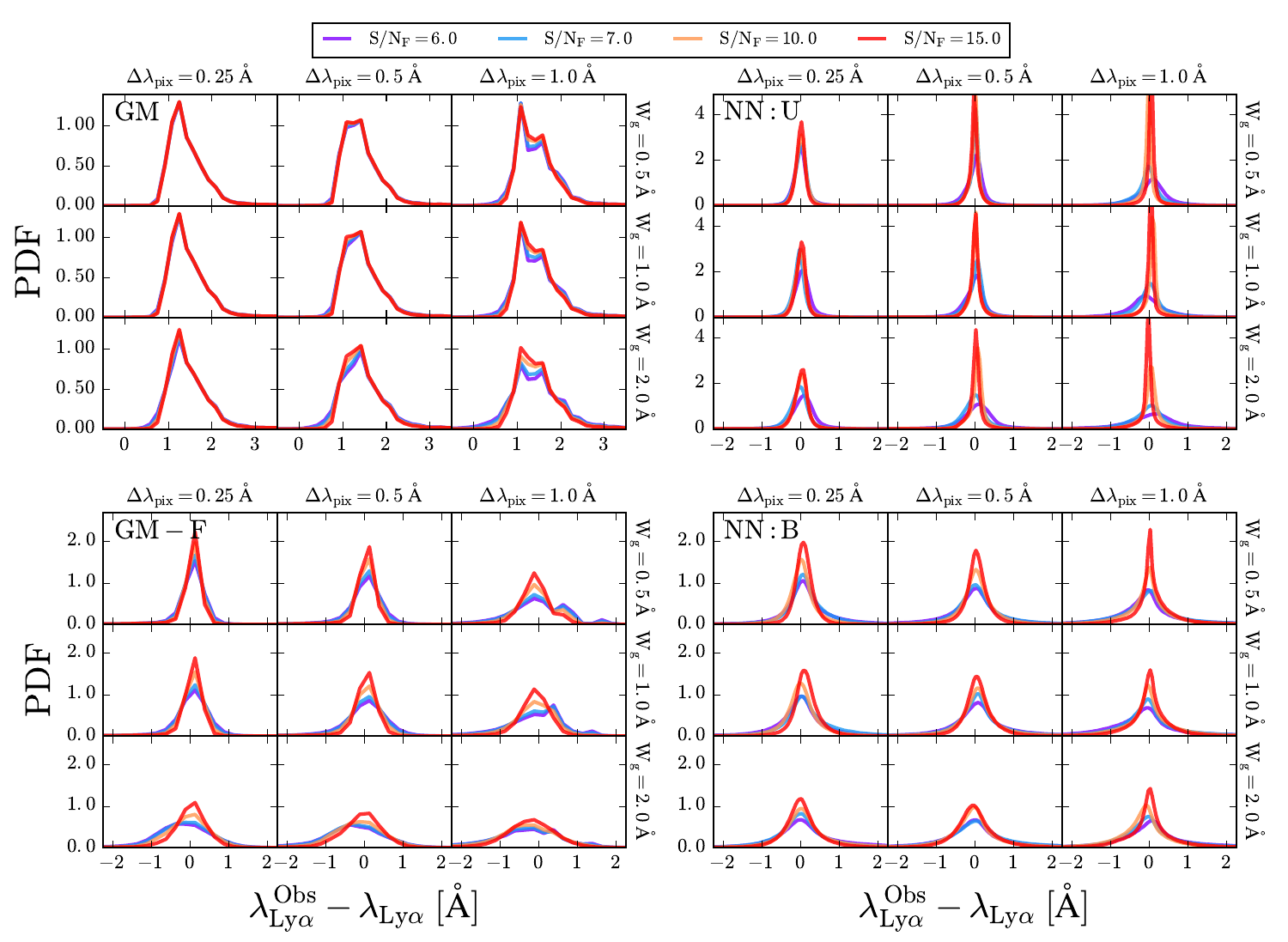}
\caption{ Distribution of the wavelength shift due to the miss-identification of the \lya\ wavelength from \lya\ line profiles for our different spectral qualities and the \GM\ algorithm. The columns show $\Delta\lambda_{\rm pix}=0.25$\AA{}, 0.5\AA{} and 1.0\AA{} from left to right. Meanwhile, rows show $\rm W_g=0.5$\AA{}, 1.0\AA{} and 2.0\AA{} from top to bottom. Additionally, in each panel, $\rm S/N_{F}=6.0$ is displays in purple, 7.0 in blue, 10.0 in orange and 15.0 in red. This corresponds to the analysis of the snapshot at  redshift 3.0 using the \ThinShell . Top left: \GM , bottom left: \VER , top right: \NNU , bottom right \NNB .  } 
\label{fig:D_lambda_1}
\end{figure*}

%%%%%%%%%%%%%%%%%%%%%%%%%%%%%%%%%%%%%%%%%%%%%%%%%%%%%%%%%%%%%%%%%%%%%%%%%%%%%%%%%%%%%%%%%%%%%%%%

\begin{table}
\caption{Standard deviation ($\sigma$) of the difference between the wavelength assigned as \lya\  and the true \lya\ frequency for different quality configurations and \lya\ identification algorithms. This corresponds to the analysis of the snapshot at  redshift 3.0 using the \ThinShell .} 
\label{tab:real_3_thin_Shell}
\begin{tabular}{ccccccccc}
\multicolumn{3}{c}{ $z=3.0$ , Thin Shell}                                                     & GM             & GM-F            & uGM-F     & NN:U           & NN:B   \\ \hline
$\rm W_g ^{Rest}$                              & $\rm \Delta \lambda _{pix}^{Rest}$ & $\rm S/N_{F}$ & $\sigma$       & $\sigma$        & $\sigma$  & $\sigma$       & $\sigma$    \\
$\;$[\AA{}]                                    & [\AA{}]                            &               & [\AA{}]        & [\AA{}]         & [\AA{}]   & [\AA{}]        & [\AA{}]        \\ \hline
0.5                                            & 0.25                               & 6.0         & 1.33 & 0.7  & 0.69 & 0.52 & 0.94 \\
                                               &                                    & 7.0         & 1.28 & 0.67  & 0.66 & 0.42 & 0.74 \\
                                               &                                    & 10.0        & 1.23 & 0.61  & 0.6 & 0.29 & 0.49 \\
                                               &                                    & 15.0        & 1.21 & 0.58  & 0.57 & 0.2 & 0.3 \\ \cline{2-8}
                                               & 0.5                                & 6.0         & 1.37 & 0.83  & 0.81 & 0.7 & 1.24 \\
                                               &                                    & 7.0         & 1.34 & 0.76  & 0.75 & 0.53 & 1.04 \\
                                               &                                    & 10.0        & 1.27 & 0.66  & 0.65 & 0.36 & 0.6 \\
                                               &                                    & 15.0        & 1.24 & 0.61  & 0.6 & 0.23 & 0.41 \\ \cline{2-8}
                                               & 1.0                                & 6.0         & 1.55 & 1.24  & 1.2 & 1.07 & 1.91 \\
                                               &                                    & 7.0         & 1.46 & 1.07  & 1.04 & 0.82 & 1.5 \\
                                               &                                    & 10.0        & 1.32 & 0.83  & 0.8 & 0.46 & 0.87 \\
                                               &                                    & 15.0        & 1.26 & 0.69  & 0.67 & 0.29 & 0.66 \\ \hline
1.0                                            & 0.25                               & 6.0         & 1.31 & 0.77  & 0.72 & 0.57 & 0.99 \\
                                               &                                    & 7.0         & 1.26 & 0.72  & 0.68 & 0.44 & 0.88 \\
                                               &                                    & 10.0        & 1.23 & 0.66  & 0.62 & 0.31 & 0.56 \\
                                               &                                    & 15.0        & 1.18 & 0.61  & 0.58 & 0.21 & 0.39 \\ \cline{2-8}
                                               & 0.5                                & 6.0         & 1.38 & 0.93  & 0.88 & 0.81 & 1.43 \\
                                               &                                    & 7.0         & 1.33 & 0.84  & 0.8 & 0.61 & 1.05 \\
                                               &                                    & 10.0        & 1.27 & 0.71  & 0.67 & 0.37 & 0.74 \\
                                               &                                    & 15.0        & 1.2 & 0.64  & 0.61 & 0.25 & 0.46 \\ \cline{2-8}
                                               & 1.0                                & 6.0         & 1.56 & 1.34  & 1.32 & 1.18 & 2.44 \\
                                               &                                    & 7.0         & 1.47 & 1.14  & 1.12 & 0.92 & 1.57 \\
                                               &                                    & 10.0        & 1.34 & 0.86  & 0.84 & 0.51 & 0.9 \\
                                               &                                    & 15.0        & 1.27 & 0.71  & 0.69 & 0.3 & 0.59 \\ \hline
2.0                                            & 0.25                               & 6.0         & 1.29 & 0.97  & 0.85 & 0.79 & 1.85 \\
                                               &                                    & 7.0         & 1.25 & 0.89  & 0.77 & 0.62 & 1.22 \\
                                               &                                    & 10.0        & 1.19 & 0.78  & 0.64 & 0.37 & 0.88 \\
                                               &                                    & 15.0        & 1.12 & 0.69  & 0.56 & 0.26 & 0.54 \\ \cline{2-8}
                                               & 0.5                                & 6.0         & 1.41 & 1.22  & 1.16 & 1.05 & 2.21 \\
                                               &                                    & 7.0         & 1.36 & 1.07  & 0.99 & 0.81 & 1.86 \\
                                               &                                    & 10.0        & 1.25 & 0.88  & 0.76 & 0.43 & 0.89 \\
                                               &                                    & 15.0        & 1.17 & 0.76  & 0.63 & 0.27 & 0.62 \\ \cline{2-8}
                                               & 1.0                                & 6.0         & 1.67 & 1.79  & 1.76 & 1.46 & 2.98 \\
                                               &                                    & 7.0         & 1.52 & 1.53  & 1.48 & 1.13 & 2.22 \\
                                               &                                    & 10.0        & 1.35 & 1.1  & 1.01 & 0.61 & 1.03 \\
                                               &                                    & 15.0        & 1.22 & 0.86  & 0.73 & 0.35 & 0.78 \\ \hline
\end{tabular}
\end{table}

\subsection{\lya\ wavelength displacement}\label{ssec:w_lya_real}

In this section we compare how the performance of the different algorithms to determine the \lya\ wavelength from a \lya\ line profile vary with the quality of the spectrum. In Figs.~\ref{fig:D_lambda_1} we show the distributions of the displacement between the assigned \lya\ wavelength and the true one for the \GM , \VER , \NNU\ and \NNB\ algorithms at redshift 3.0 and using the \ThinShell . Then, the we sum up the performance of all the redshifts and outflow geometries combinations in Tabs.\ref{tab:real_3_thin_Shell}, \ref{tab:real_2_thin_Shell}, \ref{tab:real_2_galactic_Wind},  \ref{tab:real_3_galactic_Wind}, \ref{tab:real_5_thin_Shell} and \ref{tab:real_5_galactic_Wind}.  Overall, for all our four algorithms decreasing the quality of the \lya\ line profiles, i.e., decreasing $\rm S/N_F$ and increasing $\Delta \lambda_{\rm pix}$ and $\rm W_g$, causes a larger uncertainty in the identification of the \lya\ wavelength. 

%%%%%%%%%%%%%%%%%%%%%%%%%%%%%%%%%%%%%%%%%%%%%%%%%%%%%%%%%%%%%%%%%%%%%%%%%%%%%%%%%%%%%%%%%%%%%%%%
\begin{figure*} 
\includegraphics[width=3.5in]{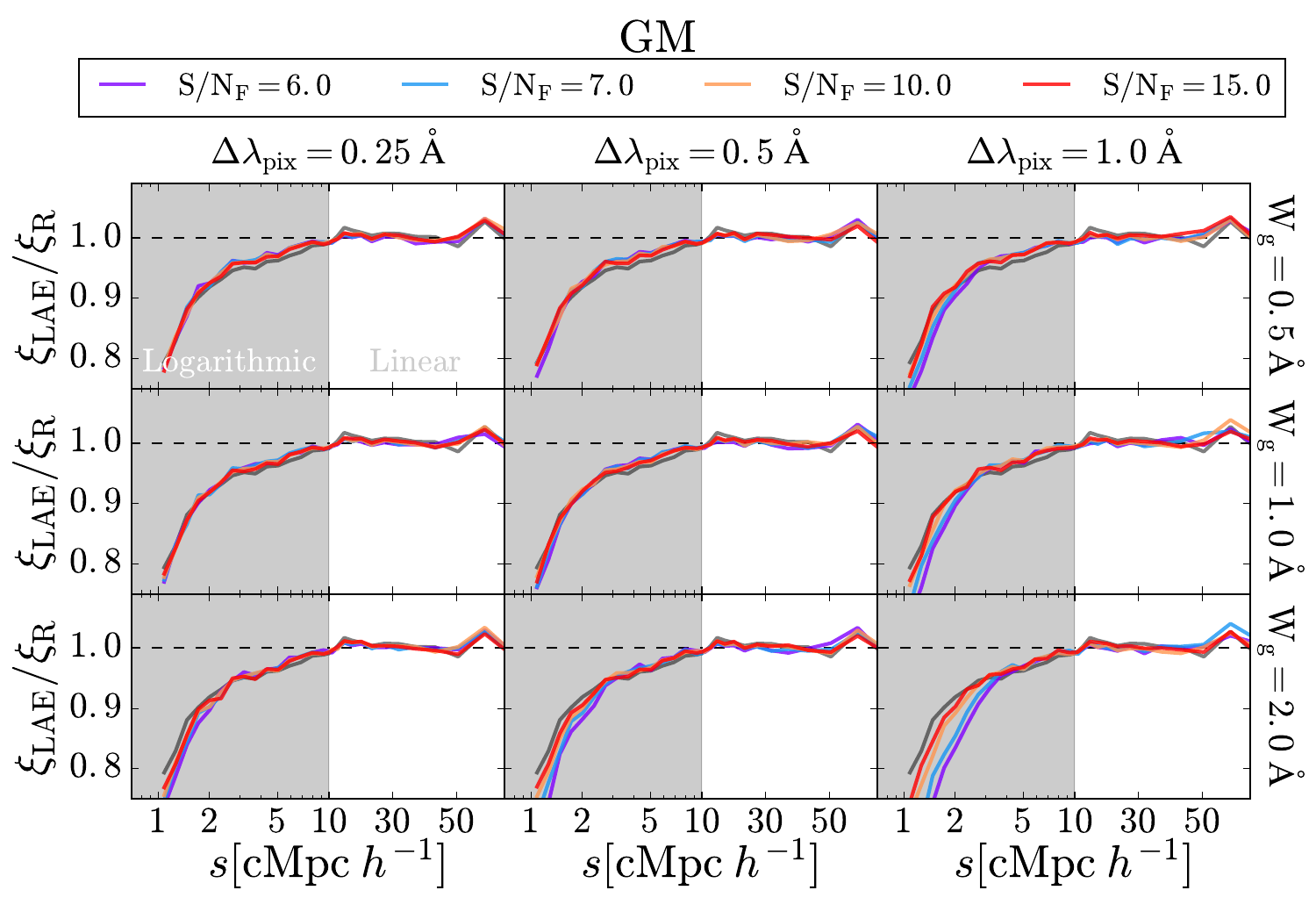}%
\includegraphics[width=3.5in]{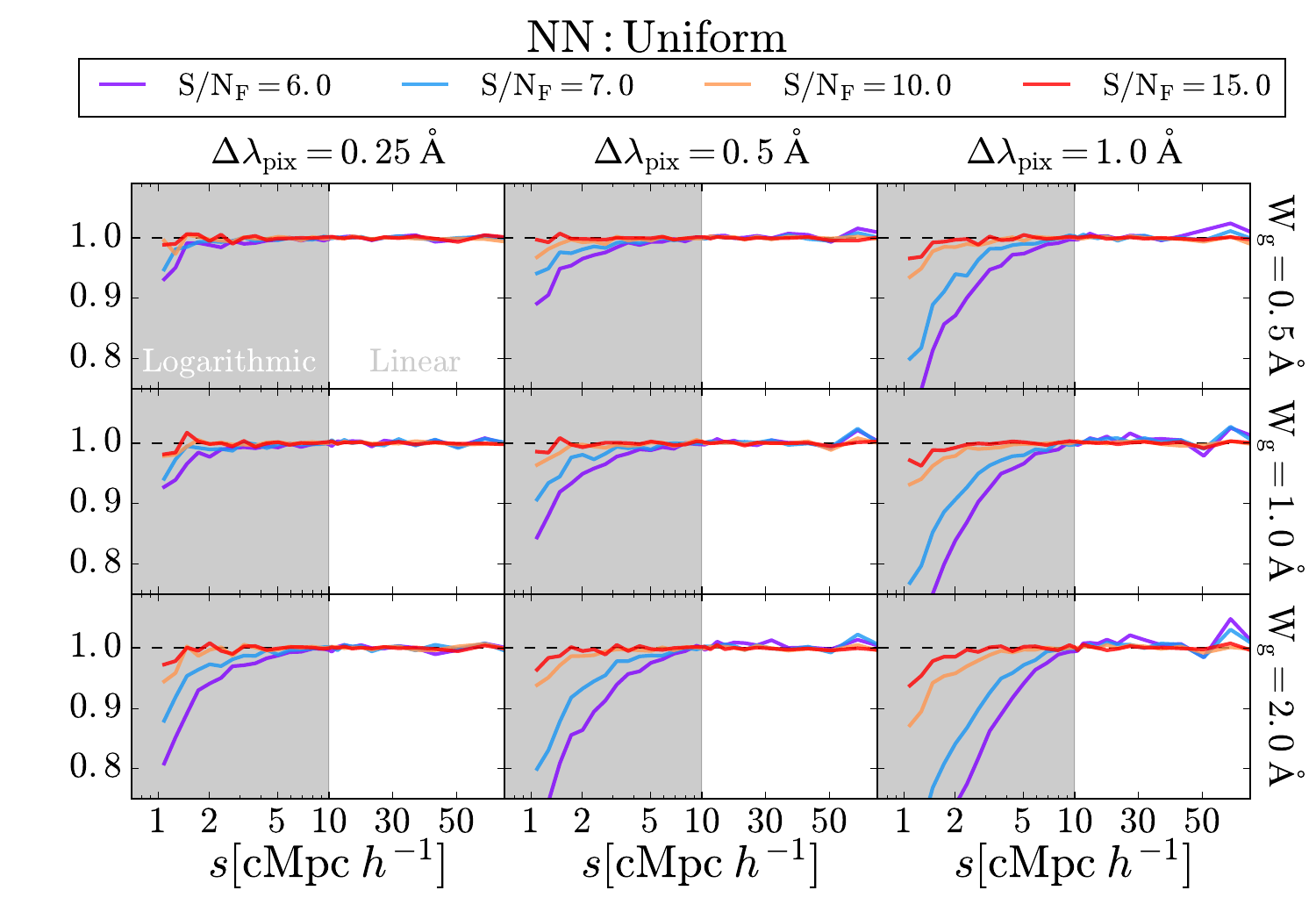}
\caption{   Ratio between the observed monopole of LAE samples with ($\rm \xi_{LAE}$) and without ($\rm \xi_{R}$) the miss-identification of the \lya\ frequency using the \GM\ (left) and \NNU\ (right) algorithms. This corresponds to the analysis of the snapshot at  redshift 3.0 using the \ThinShell . The columns show $\Delta\lambda_{\rm pix}=0.25$\AA{}, 0.5\AA{} and 1.0\AA{} from left to right. Meanwhile, rows show $\rm W_g=0.5$\AA{}, 1.0\AA{} and 2.0\AA{} from top to bottom. Additionally, in each panel, $\rm S/N_{F}=6.0$ is displays in purple, 7.0 in blue, 10.0 in orange and 15.0 in red. The solid grey line is the monopole computed when the ideal profiles are used (same as Fig.~\ref{fig:monopole_comparison}). The dashed black line signalizes unity.  }
\label{fig:monopole_1}
\end{figure*}
%%%%%%%%%%%%%%%%%%%%%%%%%%%%%%%%%%%%%%%%%%%%%%%%%%%%%%%%%%%%%%%%%%%%%%%%%%%%%%%%%%%%%%%%%%%%%%%%
%%%%%%%%%%%%%%%%%%%%%%%%%%%%%%%%%%%%%%%%%%%%%%%%%%%%%%%%%%%%%%%%%%%%%%%%%%%%%%%%%%%%%%%%%%%%%%%%

Focusing on the \GM\ (Fig.~\ref{fig:D_lambda_1} top left), we find that this algorithms is insensitive to lowering the quality of the \lya\ line profile. In fact, the mean of the distribution is always entered around 1.5\AA{}. This is due to the fact that \lya\ photos are redshifted as they escape through the galaxy outflows. Moreover, the width of the $\Delta\lambda$ remains also constant ($\sim 1$\AA{}) until $\Delta\lambda_{\rm pix} = 0.5$\AA{} and $\rm W_g=1.0$\AA{}. From that point, it steadily grows. Additionally, the shape of the distribution changes with the quality of the \lya\ line profile. On one hand, low values of $\Delta\lambda_{\rm pix}$ and $\rm W_g$ the distribution is skewed, exhibiting a tail towards large $\Delta\lambda$ values.  On the other hand, as $\Delta\lambda_{\rm pix}$ and $\rm W_g$ increase, the distribution becomes more symmetric. 

Next, we find that the \VER\ (Fig.\ref{fig:D_lambda_1} bottom left) algorithm performance is heavily affected by the quality of the \lya\ line profile. In the first place, when  $\Delta\lambda_{\rm pix}$ and $\rm W_g$ are low, the $\Delta\lambda$ distribution is centered around 0. This suggests that correction of the Gaussian kernel works in this range. However, as the quality  of the line profile decreases, the $\Delta\lambda$ distribution moves progressively towards negative  $\Delta\lambda$ values and gets broader. This is a consequence of the way in which \VER\ derived the \lya\ wavelength. \VER\ corrects the displacement in the \lya\ frequency found in \GM\ by  using a relation between the wavelength shift and the width of the red peak of the line. As the quality of the line decreases, the determination of  $\rm FWHM_{red}$ becomes more noisy and the fraction of cases with $\rm FWHM_{red} < W_g$ rises. This causes that the $\Delta\lambda$ distributions is shifted to negative values and that it becomes broader.
%$\rm FWHM_{red} < W_g$
%$\rm W_{g}$ in creases, the width of the red peak also increases, resulting in \lya\ wavelength that are over corrected by the shift due to the RT.  In the second place, the spread of the $\Delta\lambda$ distribution increases as the quality of the \lya\ line profiles becomes lower. 
In fact, we find that the standard deviation of the distribution reaches 2\AA{} when $\Delta\lambda_{\rm pix} = 1.0$\AA{}. $\rm W_g=2.0$\AA{} and $\rm S/N_F = 6.0$ .

The standard deviations of the \uVER\ algorithm corresponding to redshift 3.0 and the \ThinShell\ geometry are listed in \ref{tab:real_3_thin_Shell}. \uVER\ exhibits an almost identical scatter than \VER\ for  $\rm W_g \leq 1.0$\AA{} and $\rm \Delta\lambda_{pix} \leq 0.5$\AA{}. In this regime the cases in which $\rm FWHM_{red} < W_g$ is negligible. However, for lower qualities  the $\rm FWHM_{red} < W_g$ cases rise and increases the dispersion of \VER .  This shows that for clustering purposes, when the quality of the line profile is low, it is better to leave $\rm FWHM_{red}$ uncorrected by the instrument point spread function, although this creates a systematic bias in the mean of the $\Delta\lambda$ towards negative values.

Moreover, among the two neural network algorithms studied in this work, \NNB\ (Fig.\ref{fig:D_lambda_1} bottom right) is the one which is the most affected by the decrease of quality in the line profile. In general the shape of the  $\Delta\lambda$ distribution is composed by a prominent peak located at $\Delta\lambda = 0 $ and extended wings bluewards and redwards of \lya . Increasing  $\Delta\lambda_{\rm pix}$ and $\rm W_g$ causes that the wings become more elongated, and hence the accuracy of the algorithm is reduced. Meanwhile, for a set of fixed $\Delta\lambda_{\rm pix}$ and $\rm W_g$, decreasing $\rm S/N_F$ lowers the peak contribution while the wings remain constant, which increases significantly the width of the distributions. The large dependence on $\rm S/N_{F}$ comes from the fact that \NNB\ uses as a training set the 10\% brightest LAEs. As a consequence, this algorithm is trained to reproduce lines with a much higher signal to noise ratio than average in the LAE population.  This has little effect when the faintest LAE in the sample has $\rm S/N_{F}=15$, since all the galaxy population is going to have a very good S/N and quality. In detail, when $\rm S/N_{F}=15$, the S/N of the brightest LAEs is $\sim 1000$, but the difference in quality between the faint and the bright ends are very small, as the noise for the faintest galaxy is already very tiny. In other words, the quality of the \lya\ line profile used for the training set is very similar to the one of the whole LAE sample, even though they exhibit a quite different S/N. However, when $\rm S/N_{F}$ decreases, the differences in quality in the \lya\ line profile become larger, as the faintest LAEs become more and more noisier, while the quality of bright LAEs remains almost unchanged. 

Also, we find that \NNU\ (Fig.\ref{fig:D_lambda_1} top right) is the algorithm with the best performance in most of the range of the \{$\Delta\lambda_{\rm pix}$, $\rm W_g$, $\rm S/N_{F}$\} volume studied here.  Increasing $\Delta\lambda_{\rm pix}$ and $\rm W_g$ and decreasing $\rm S/N_{F}=15$ reduce the performance of \NNU . However, it is remarkable how little the spread of the $\Delta\lambda$ distribution is increased through the varied $\Delta\lambda_{\rm pix}$ and $\rm W_g$ range when $\rm S/N_{F}=15$ is kept fixed. In fact $\sigma(\Delta\lambda)$ varies from 0.2\AA{} in the best case to only 0.35\AA{} in the worst scenario. This highlights that this level of pixelization and line diluting (due to a $\rm FWHM \neq 0$), the \lya\ line profiles still contain the necessary information to identify the true \lya\ wavelength. However, when $\rm S/N_{F}$ is reduced, the noise level increases and destroys part of this information. In particular, the higher $\Delta\lambda_{\rm pix}$ and $\rm W_g$ values, the more likely is to lose this information due to the noise. 

Finally, as shown in Tabs.\ref{tab:real_2_thin_Shell}, \ref{tab:real_2_galactic_Wind},  \ref{tab:real_3_galactic_Wind}, \ref{tab:real_5_thin_Shell} and \ref{tab:real_5_galactic_Wind}. , the \NNU\ is the best methodology at the other redshifts and geometries too.

\subsection{Monopole artifacts}

 %%%%%%%%%%%%%%%%%%%%%%%%%%%%%%%%%%%%%%%%%%%%%%%%%%%%%%%%%%%%%%%%%%%%%%%%%%%%%%%%%%%%%%%%%%%%%%%%
\begin{figure*} 
\includegraphics[width=3.5in]{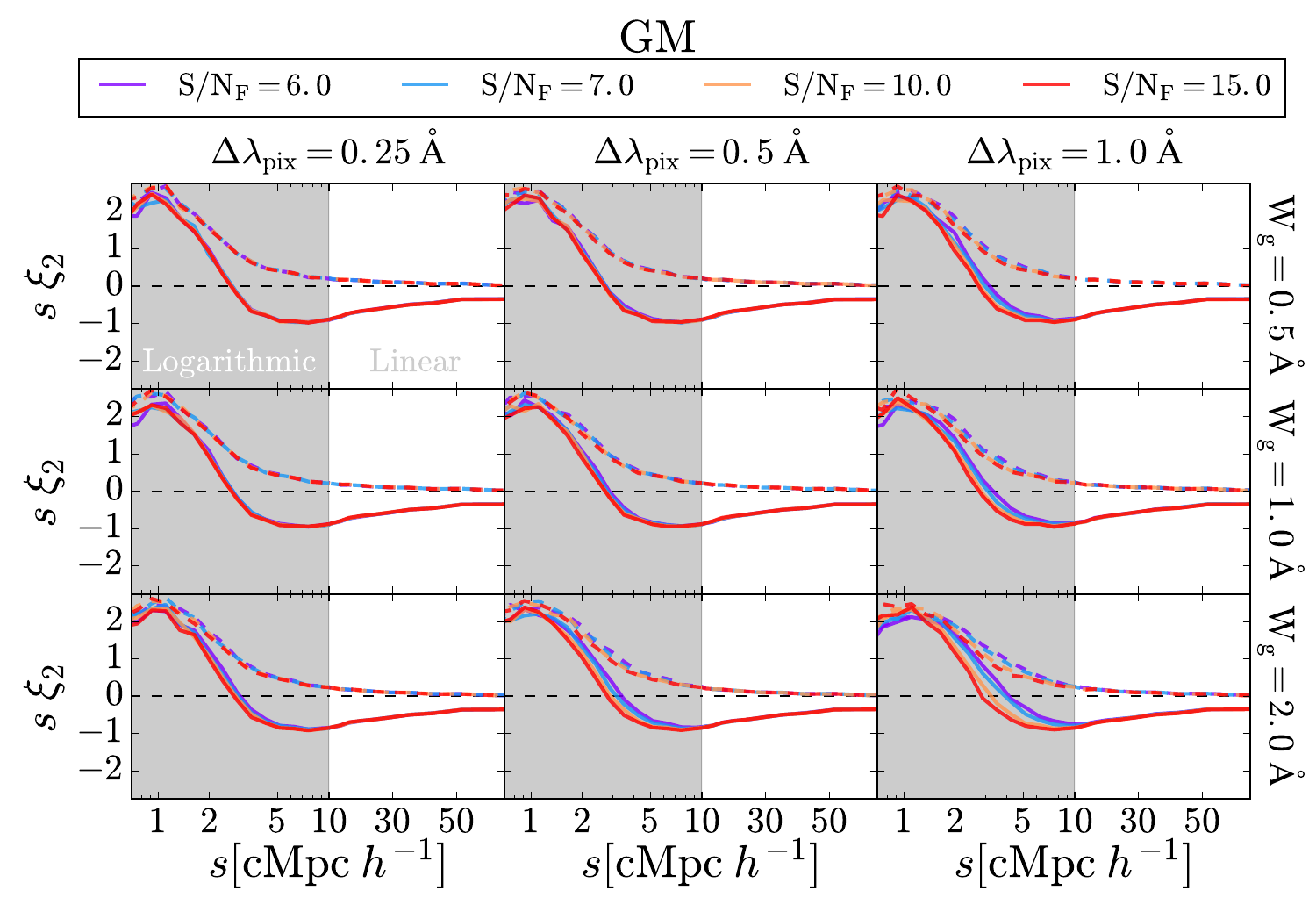}%
\includegraphics[width=3.5in]{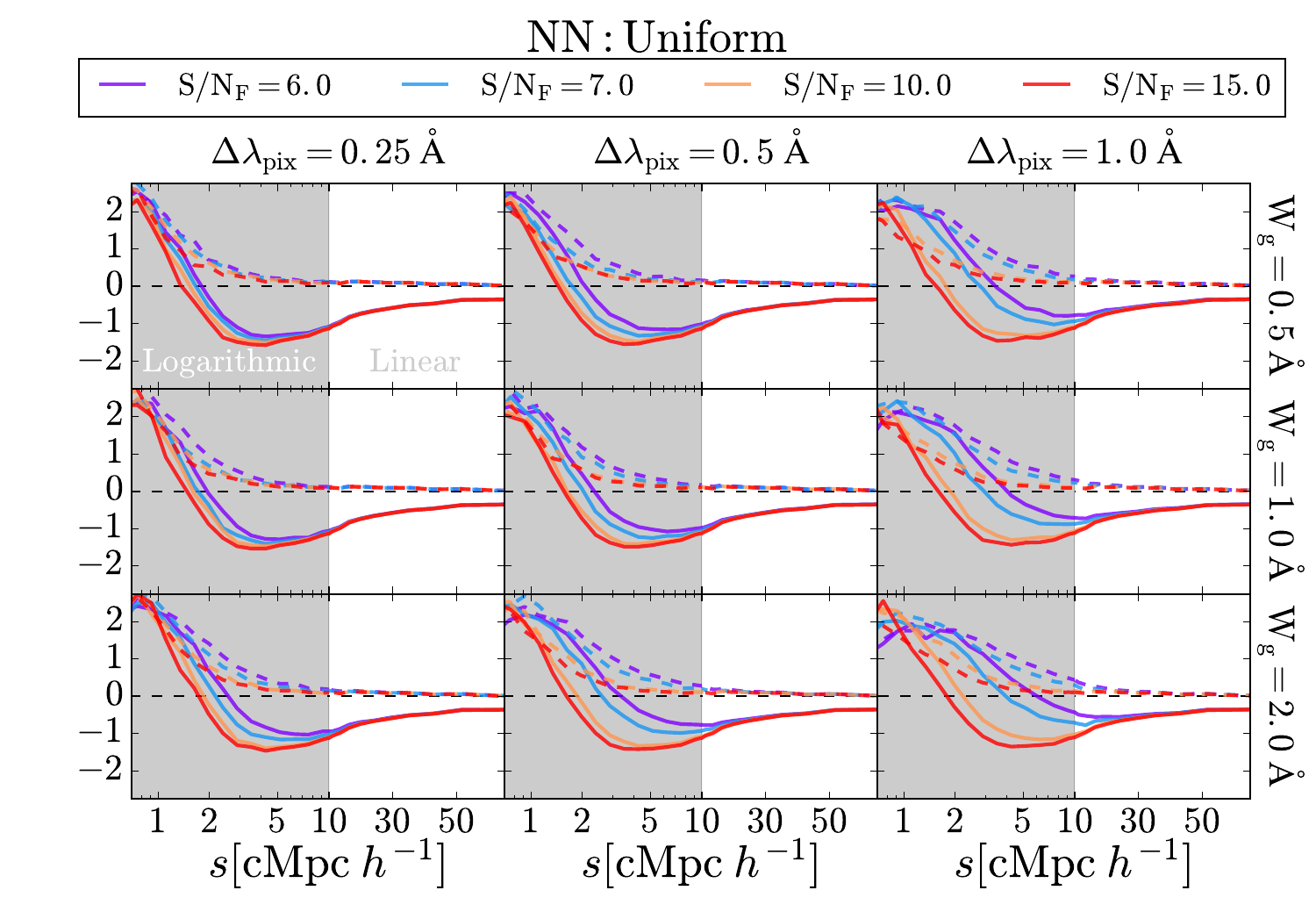}
\caption{  Quadrupole of LAE samples with (solid) and without (dashed) the peculiar motion of galaxies. The plot structure is the same as Fig.\ref{fig:monopole_1}. Here the black dashed line indicates zero. This corresponds to the analysis of the snapshot at  redshift 3.0 using the \ThinShell .} 
\label{fig:quadrupole_1}
\end{figure*}
%%%%%%%%%%%%%%%%%%%%%%%%%%%%%%%%%%%%%%%%%%%%%%%%%%%%%%%%%%%%%%%%%%%%%%%%%%%%%%%%%%%%%%%%%%%%%%%%

In general, the misidentification of the \lya\ wavelength translates into an incorrect redshift determination and, hence, into a shift in the position of the LAE along the line of sight. This has a direct impact in the measured clustering of LAEs on small scales, as we showed in \S\ref{ssec:clustering_ideal}. Also, we have just shown in \S\ref{ssec:w_lya_real} how the quality of a given set of observed \lya\ line profiles is mirrored into the identification of the \lya\ wavelength. In fact, the lower the quality, the more spread the $\Delta\lambda$ distribution becomes. Here we study how the quality of the \lya\ line profiles imprints the clustering on small scales, focusing on the 2PCF.In the following, we will show the results only for two extreme cases, i.e., the worst (\GM) and the best (\NNU) algorithms. 

In Fig.~\ref{fig:monopole_1} we show the monopole 2PCF for all the different combinations of \{$\Delta\lambda_{\rm pix}$, $\rm W_g$, $\rm S/N_{F}$\}  including the contribution of peculiar velocities and \lya\ misidentification (through Eq.\ref{eq:position_LoS}) given by the \NNU\ algorithm. Overall, we find a clustering suppression on small scales ($\xi_0$, colored lines). At larger scales, the observed LAE clustering converges to the clustering that would be observed if the \lya\ frequency was known at infinite precision ($\xi_{\rm 0,R}$, dashed black line). For the \GM\ algorithm we find that there is a strong clustering suppression across the quality range studied here ($\gtrsim 20\%$ at $s\sim 1\,{\rm cMpc}\,h^{-1}$). Additionally, the suppression depends slight on the quality of the \lya\ line profiles. In fact, the suppression remain quite constant through all the \{$\Delta\lambda_{\rm pix}$, $\rm W_g$, $\rm S/N_{F}$\} range, expect at $\Delta\lambda_{\rm pix}=1$\AA{} and $\rm W_g=2$\AA{}, where the lower $\rm S/N_F$, the stronger the suppression. In fact, we also show the monopole 2PCF for ideal line profiles (solid grey curve) to illustrate how little the clustering recovered by the GM algorithms is perturbed by the line quality. This shows that the Gaussian fitting that we are applying after downgrading the quality to recover the global maximum of the lines profiles works well.

For the \NNU\ (Fig.~\ref{fig:monopole_1} right), the convergence scale depends strongly on the quality of the observed \lya\ line profiles. In general, the lower the quality, (i.e., the greater $\Delta\lambda_{\rm pix}$ and $\rm W_g $, and the lower $\rm S/N_{F}$) the larger is the clustering suppression and hence the larger is the convergence scale. It is remarkable how well the \NNU\ algorithm performs when the signal to noise ratio of the \lya\ spectrum is good. In fact, the monopole is affected only on scales lower than $2{\rm cMpc}/h$ when $\rm S/N_F=15$. This highlights that, although diluted, the \lya\ line profiles still contain the information about the \lya\ wavelength. However, the addition of noise easily destroys progressively this information, causing a greater suppression. In general, the higher \{$\Delta\lambda_{\rm pix}$, $\rm W_g$\}, the most sensitive to noise becomes the clustering of LAEs.

\subsection{Quadrupole artifacts}

In Fig.~\ref{fig:quadrupole_1} we display the quadrupole for all the multiple quality configuration for the \GM\ and \NNU\ algorithms, respectively. At the same time we show the samples including only the \lya\ misidentification shift along the line of sight (dashed lines) and the samples including also the shift due to peculiar galaxy velocities (solid lines). In general, we find similar trends to the monopole. In particular, the lower the quality of a given set of \lya\ line profiles, the larger is the clustering suppression along the line of sight (and the more positive the quadrupole becomes).However, the quadrupole appears to be more sensitive to the quality of the observed line profiles. In fact, for a given algorithm and quality configuration, similarly to the ideal case, the suppression in the quadrupole extents to larger scales than in the monopole. 

In detail, the quadrupole of the LAE samples for \GM\  (Fig.~\ref{fig:quadrupole_1}) exhibits a significant suppression of the clustering along the line of sight, which increases its amplitude. As well as in the monopole, the quadrupole of LAE samples using the \GM\ algorithm are only slightly affected by the quality of the \lya\ line profile. In fact, for most of the \{$\Delta\lambda_{\rm pix}$, $\rm W_g$, $\rm S/N_{F}$\} combinations, the quadrupole remains almost unchanged. Although, for the worst combination of  $\Delta\lambda_{\rm pix}$ and $\rm W_g$ studied here, the amplitude of the quadrupole correlates with $\rm S/N_{F}$ on scales smaller than ${\rm \sim 20cMpc}/h$.

Moreover, \NNU\ (Fig.\ref{fig:quadrupole_1}) exhibits the best performance over most of the \{$\Delta\lambda_{\rm pix}$, $\rm W_g$, $\rm S/N_{F}$\} volume covered here. In contrast to the monopole, where the \NNU\ algorithm provided a measurement without a suppression above ${\rm \sim 2cMpc}/h$ for  $\rm S/N_{F}=15$, the quadrupole convergence scale for this $\rm S/N_{F}$ is extended to ${\rm \sim 5cMpc}/h$ in general. In agreement with the monopole, the clustering suppression along the line of sight in this algorithm depends strongly in the quality of the line. In this way, the suppression gets larger as the quality gets lower.

%%%%%%%%%%%%%%%%%%%%%%%%%%%%%%%%%%%%%%%%%%%%%%%%%%%%%%%%%%%%%%%%%%%%%%%%%%%%%%%%%%%%%%%%%%%%%%%%
\begin{figure*} 
\includegraphics[width=3.2in]{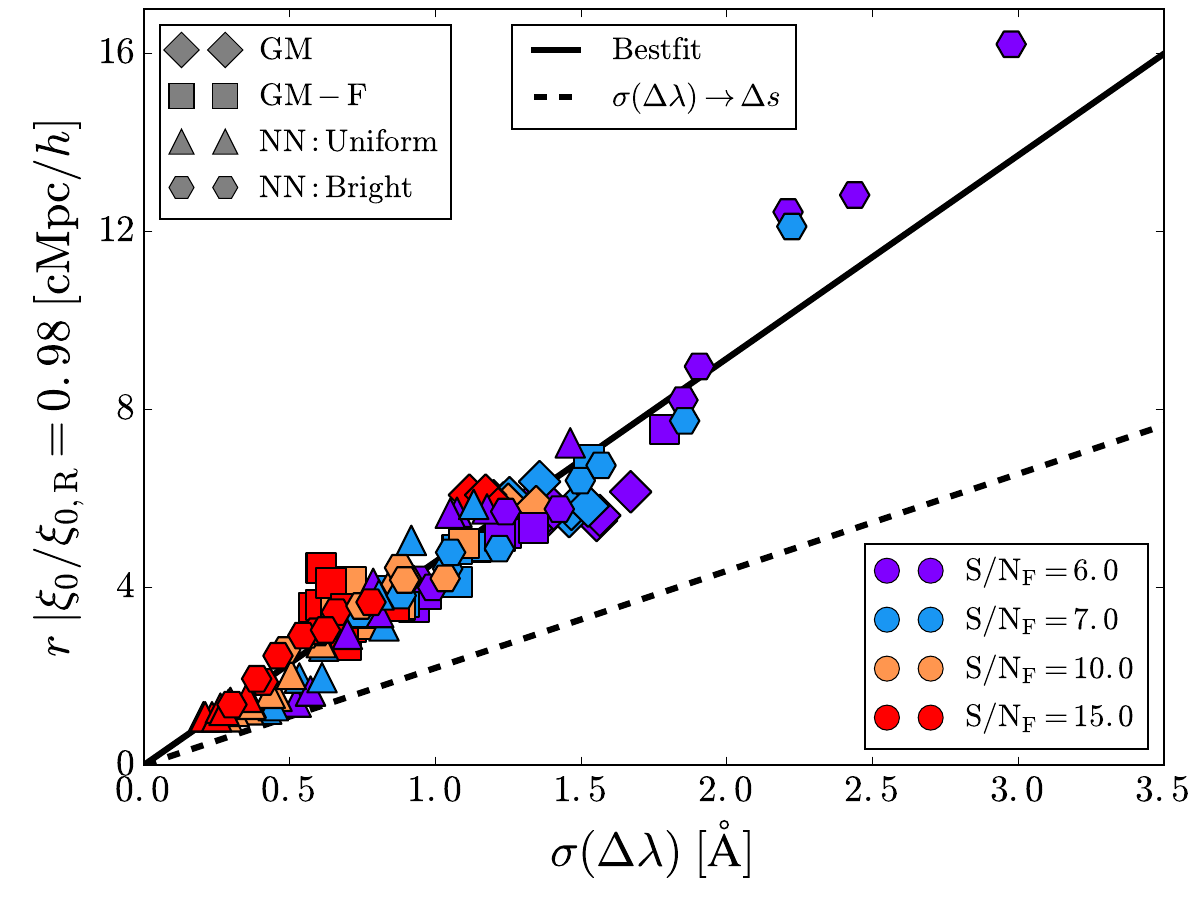}%
\hspace{0.5cm}
\includegraphics[width=3.2in]{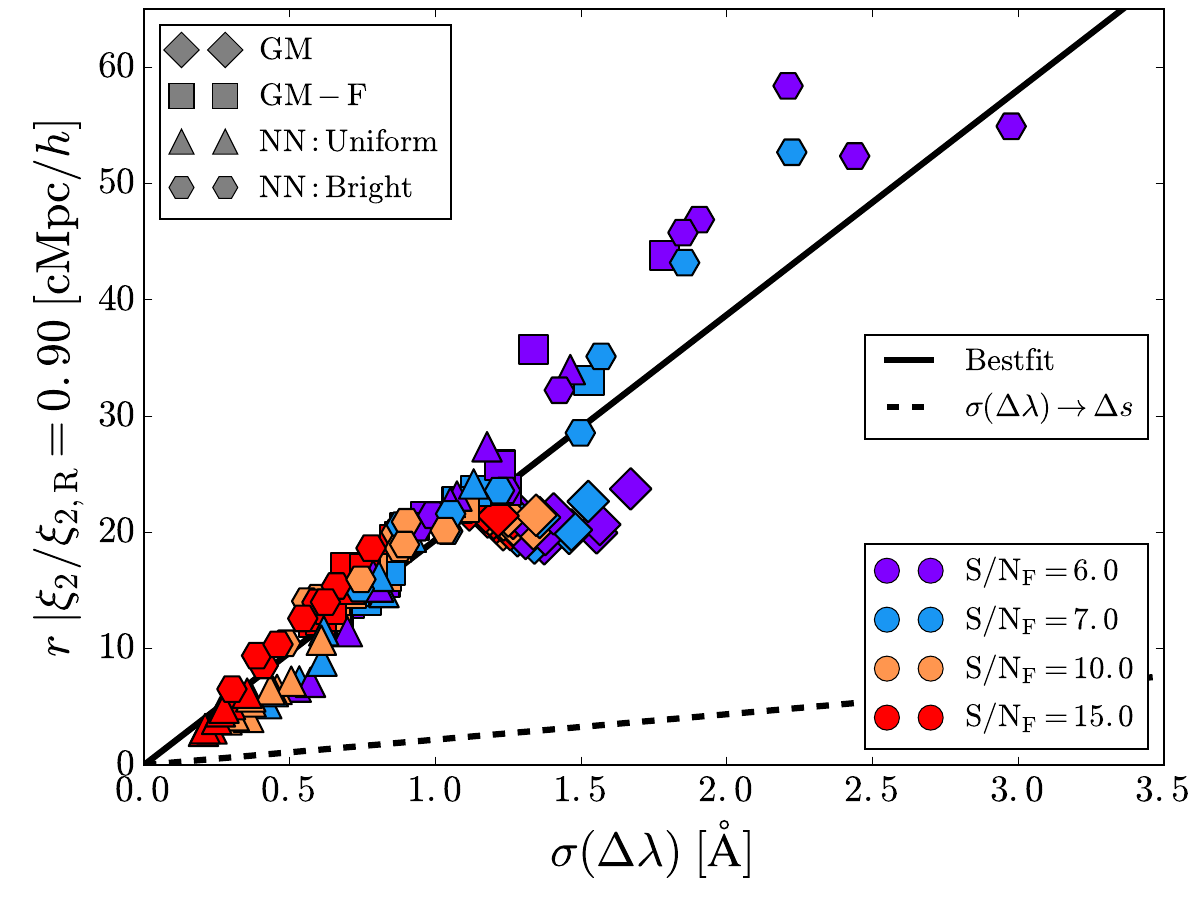}
\caption{ Left : Monopole convergence scale defined such as $\rm \xi_{0}/|xi_{0,R}=0.98$ as a function of the standard deviation of the $\Delta\lambda$ distribution for all our quality configurations and \lya\ identification algorithms. The \GM\ are displayed as diamonds, \VER\ as squares, \NNU\ as triangles and \NNB\ as hexagons. Then the $\rm S/N_F$ of the sample is colored coded. Purple means $\rm S/N_F=6.0$, blue 7.0, orange 10.0 and red 15.0. Additionally we show the linear best fit to these data points in black solid line. The black dashed line indicates the direct conversion from $\Delta\lambda$ to distance shift due to \lya\ miss-identification. Right: Same as left, but for the quadrupole. The convergence distance for the quadrupole is defined as $\rm \xi_{2}/|xi_{0,R}=0.90$.}
\label{fig:supression_monopole}
\end{figure*}

%\begin{figure} 
%\includegraphics[width=3.5in]{supression_scale_quadrupolepole.pdf}
%\caption{Same as Fig.\ref{fig:supression_monopole} but for the quadrupole. The converge distance has a slightly different definition: $r|\rm \xi_2/\xi_{2,R}=0.90$. }
%\label{fig:supression_quadrupole}
%\end{figure}
%%%%%%%%%%%%%%%%%%%%%%%%%%%%%%%%%%%%%%%%%%%%%%%%%%%%%%%%%%%%%%%%%%%%%%%%%%%%%%%%%%%%%%%%%%%%%%%%

\section{ Discussion }\label{sec:Discussion}

\subsection{ Relation between \lya\ stack line profile and galaxy properties}

As studied in \label{sec:stacks}, our model reproduces the trends observed between $\Delta \lambda$ and different galaxy properties. Observational studies  have split  the \lya\ stacked line profile as function of several galaxy properties. For example, \cite{Guaita_2017} studied the stacked spectrum of LAEs with spectroscopic observations from VIMOS ULTRA-DEEP SURVEY \citep[VUDS,][]{LeFevre_2015,Tasca_2017}. Their sample consisted on 76 galaxies between redshift $z=2$ and $z=4$, exhibiting, both, \lya\ and CIII] (1908\AA{}) as emission lines. The galaxy systemic redshifts were determined by the observed wavelength of the CIII] lines. Then, authors split their galaxy sample  by different observed properties. They found that $\Delta\lambda$ anti-correlated with the \lya\ rest frame equivalent width and stellar mass, while it correlated with the galaxy overdensity. These trends are in agreements with our model predictions.  \cite{Guaita_2017}  also found that the width of the \lya\ stacked line profile anti-correlated with the \lya\ rest frame equivalent width and stellar mass, as our model predicts too. Moreover, \cite{Muzahid_2019} studied the relation between $\Delta\lambda$ and the SFR in 96 LAE at $z\sim3$ with spectroscopic observations with MUSE \citep{bacon10}.  \cite{Muzahid_2019}, found the an anti-correlation between $\Delta\lambda$ and the \lya\ EW, as well as, a correlation between $\Delta\lambda$ and the SFR and \lya\ luminosity. 

\subsection{ Clustering convergence scale }

In Fig.~\ref{fig:supression_monopole} (left) we show the scale where $\xi_0$ and $\xi_{\rm 0,R}$ converge as a function of the standard deviation of the PDF of $\Delta\lambda$ ($\sigma(\Delta\lambda)$) for all the algorithms and \lya\ line profile quality configuration explored in this work. Here we define the monopole convergence scale as the scale at which $\xi_0 / \xi_{0,R}=0.98$. Additionally we show the best fitting linear relation between this convergence scale and $\sigma(\Delta\lambda)$ with a slope $4.56 {\rm cMpc}/(h$\AA{}) and null origin. We find that our samples follow quite well this linear relation. In detail, the samples with good $\rm S/N_F$ tend to cluster at lower $\sigma(\Delta\lambda)$ values and vice versa. As a reference, we also show the direct conversion from $\sigma(\Delta\lambda)$ to distance (dashed black line), computed using Eq.\ref{eq:position_Lya} and assuming $\rm X_{LoS}=0$ and $\rm \rm V_{LoS}=0$. The monopole convergence distance (as defined here) has a slope a factor of $\sim 2$ larger than the conversion from $\sigma(\Delta\lambda)$ to distance. Fig.~\ref{fig:supression_monopole} illustrates the strong parallelism between the behaviours of the PDF of $\Delta\lambda$ and the observed clustering on small scales \citep{Byrohl_2019}. 

Then, in the right panel of Fig.\ref{fig:supression_monopole} we show the relation between the convergence scale for the quadrupole (defined as $r|\xi_{2}/\xi_{2,R}=0.9$). We find the same trends than in the monopole case. Basically, the lower is the accuracy identifying the \lya\ frequency, the larger is the suppression on the observed LAE quadrupole. Additionally, the convergence scale for a fix set \{$\Delta\lambda_{\rm pix}$, $\rm W_g$, $\rm S/N_{F}$\} is larger in the quadrupole than in the monopole, even though, the definition of convergence scale is more relaxed in the quadrupole. In fact, following the same procedure as in the monopole we fitted a one degree polynomial with null origin ordinate to our samples. We find an slope of $19.2 {\rm cMpc}/h$\AA{}, which is a factor $\sim 10$ larger than the direct conversion between $\sigma(\Delta \lambda)$ and distance.

Overall, the performance in determining the \lya\ wavelength is fundamentally limited by the spectral quality. A close comparison between the different models implemented in this work shows that the methodology with the lowest monopole suppression on small scales, in general, is \NNU .  Then it is followed by \NNB\ for high signal to noise ratios. However, \NNB\ is the methodology the most affected by the reduction on $\rm S/N_F$. In fact, for low $\rm S/N_F$ \NNB\ gives the worst results (also depending on the $\Delta\lambda_{\rm pix}$, $\rm W_g$ values). Then, in terms of general performance \VER\ works better than \GM . However, for very low quality \lya\ line profiles ($\Delta\lambda_{\rm pix}=1.0$\AA{}, $\rm W_g=2.0$\AA{}, $\rm S/N_{F}=6.0$) the monopole suppression for the LAE samples identified using \GM\ is lowest among its counterparts using different algorithms. This change of trend shows that eventually, for very low quality spectrum, finding the global maximum and setting it as the \lya\ wavelength is the most robust proceeding. 

%%%%%%%%%%%%%%%%%%%%%%%%%%%%%%%%%%%%%%%%%%%%%%%%%%%%%%%%%%%%%%%%%%%%%%%%%%%%%%%%%%%%%%%%%%%%%%%%

\subsection{ Implications for HETDEX }

The Hobby-Eberly Telescope Dark Energy Experiment~\citep[HETDEX]{Hill2008,Adams2011} is a spectroscopic survey chasing LAEs between redshift $\sim 1.9$ and $\sim 3.5$. In principle, HETDEX rely only in the measured \lya\ line profile to assign a redshift to an LAE.Therefore, the LAE clustering measured by HETDEX would be sensitive to the clustering distortions studied in this work.

The typical spectral resolution and pixel size of HETDEX observations are, respectively, 5\AA{} and 2\AA{} in the observed frame. As the spectral quality is fixed at the observed frame, the spectral quality in the LAE's rest frame depends on the LAE redshift. Considering HETDEX's redshift range, the spectral resolution varies between $\sim1.7$\AA{} and $\sim1.1$\AA{} and in the LAE's rest frame. Meanwhile, the pixel size ranges from $\sim 0.66$\AA{} to $\sim 0.44$\AA{} rest frame. For these values of rest frame spectral resolution and pixel size we find that the \NNU\  algorithm exhibits a better performance than the \GM\ and \VER\ algorithms. Thus, in principle, HETDEX would benefit from following the machine learning approach presented in this work. 

However, it is challenging to compute the precise impact in the clustering in HETDEX. Since HETDEX LAEs populate smoothly over a given redshift window, the spectral quality is slightly different for every LAE and evolves with redshift. We plan to study this in a follow up paper in which, we will implement LAEs in a simulation lightcone. Meanwhile, here we just give a brief calculation of how much the recovered clustering can improve in HETDEX by using our methodology. 

If we consider that HETDEX observations will exhibit an average $W_g=1.5$\AA{} and $\Delta \lambda_{\rm pix}=0.5$\AA{} (rest frame) and $\rm S/N_{F}=6.0$. Then, the $\sigma(\Delta\lambda)$ values \footnote{These values are calculated taking the mean of $\sigma(\Delta\lambda)$ for $\rm W_g=1.0$\AA{} and $\rm W_g=2.0$\AA{} rest frame.} for the \GM , \VER , \NNU\ and \NNB\ algorithms are $1.39$\AA{}, $1.02$\AA{}, $0.93$\AA{} and $1.43$\AA{}. For this particular configuration the \NNB\ algorithm is outperformed by the other algorithms. 
This is caused by the fact that, \NNB\ is only trained with the brightest LAEs. Also, \VER\ performs better than \GM , as in general. Meanwhile, the \NNU\ is algorithm with the highest accuracy, exhibiting a 10\% better performance than \VER. Following our results in the previous subsection, these $\sigma(\Delta\lambda)$ values translate into a convergence scale ($r|\xi_{0}/\xi_{0,R}=0.98$) for the monopole of ${6.33\rm cMpc/}h$ , ${4.65\rm cMpc/}h$ , ${\rm 4.24 cMpc/}h$ and ${\rm 6.52 cMpc/}h$ for the \GM , \VER , \NNU\ and \NNB\ algorithms respectively. Meanwhile, the convergence scale of the quadrupole ($r|\xi_{0}/\xi_{0,R}=0.98$) are ${26.7\rm cMpc/}h$ , ${19.6\rm cMpc/}h$ , ${\rm 17.9 cMpc/}h$ and ${\rm 27.5 cMpc/}h$ . 

\subsection{Training set sizes}

\begin{figure} 
\includegraphics[width=3.37in]{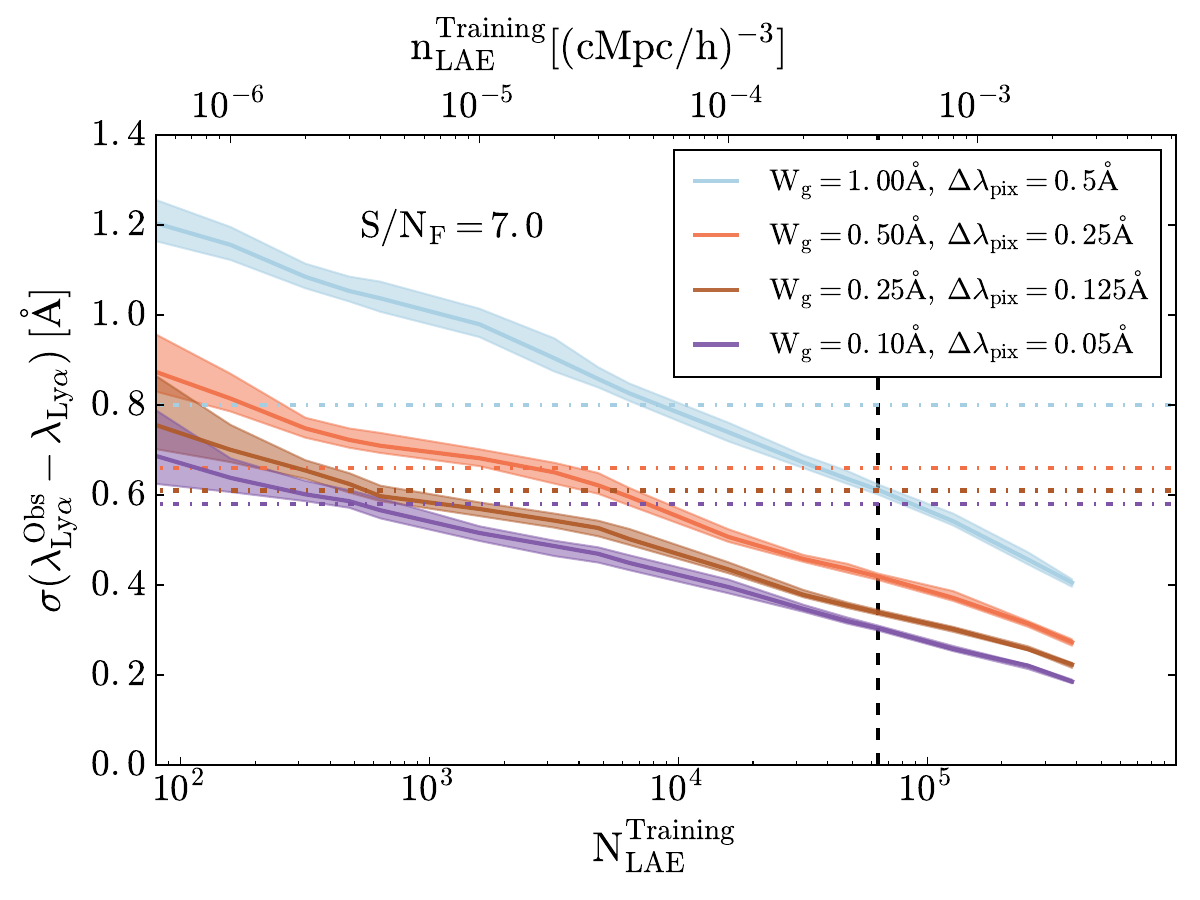}%
\caption{ {Accuracy of \NNU\ at redshift 3.0 using the \ThinShell\ geometry.  Each color indicates a line profile quality configuration $\{\rm W_g , \Delta\lambda_{pix}\}$ keeping $\rm S/N_F = 7.0$, being blue \{1.0\AA{},0.5\AA{}\}, red \{0.5\AA{},0.25\AA{}\}, brown \{0.25\AA{},0.125\AA{}\} and purple \{0.1\AA{},0.05\AA{}\}. For each value of $\rm n_{LAE}^{Training}$ we compute 100 iterations if ${\rm n_{LAE}^{Training} \leq 4\times10^{-4} } ({\rm cMpc}/h)^{-3}$ and 20 if ${\rm n_{LAE}^{Training} > 4\times10^{-4} } ({\rm cMpc}/h)^{-3}$.  The solid lines show the median and the shaded regions the 1 sigma scatter. The horizontal colored dashed lines show the accuracy of \VER\ for each quality. The vertical black line show the fiducial value of  $\rm n_{LAE}^{Training}$ .} }
\label{fig:accuracy_realistic}
\end{figure}

We acknowledge that obtaining the training sample used in \S \ref{sec:ideal} and \ref{sec:real} would be extremely challenging nowadays. However, we argue that this is mainly because the LAE population presented in this series of works is, by itself, extremely challenging to obtain. Given our number density cut, each model contains  637444 LAEs down to a Lyman-alpha luminosity limit of $\sim 10^{42} erg/s$, depending on redshift. 

It is important to note that the training set size necessary to reach convergence depends on the LAE population for which one wants to estimate the systemic redshift. In general, a given LAE population with a larger diversity of lines would need a larger training set. For example, if you would consider LAEs in a narrow Lyman-alpha luminosity range, most probably, the diversity of lines would be smaller than in our sample. For this kind of population, the  number of galaxies with a well constrained systemic redshift necessary to reach convergence would be smaller. In principle, this would be in general the case as  \cite{Guaita_2017} and \cite{Muzahid_2019}  found that the line profile properties depend on different galaxy properties such as the Lyman-alpha luminosity, or the  equivalent width. 

Moreover, neither the \NNU\ or the \NNB\ training samples are  selected optimally. In \S\ref{sec:algorithms}, \ref{sec:ideal} and \ref{sec:real} it is made clear that the  \NNU\ performs much better than \NNB\ due to the difference in training sets. However there is still plenty of space for improvement in the way of selecting the training sample. For example, a more efficient selection criteria would be to perform a latin hypercube sampling of different properties, such as the Lyman-alpha luminosity, the  FWHM of the line or its  global maximum. In principle this could lead to an faster convergence, in terms of  $\rm N_{LAE}^{Training}$, in comparison with NN:U. 

Also, in general, it is not necessary to reach convergence to improve the results from previous methodologies. For example, focusing in the ideal case using the \ThinShell\ at redshift 3.0 (Fig.\ref{fig:accuracy_NN}), the performance of the best traditional methodology (in this case \VER) gives $\Delta \lambda = 0.52$\AA{} (see Tab.\ref{tab:parameters}). The performance of NN:U in the range of training set sizes considered (down to 100 galaxies) is, { in almost every case}, better than $\Delta \lambda = 0.52$\AA{}. This is also the case for the \GalacticWind. 

{
Finally, we explore the \lya\ wavelength determination accuracy as a function of the training set size and line profile quality in Fig.\ref{fig:accuracy_realistic} for the \ThinShell\ at redshift 3.0 and using the \NNU\ algorithm. For this, we picked two of the $\{\rm W_g , \Delta\lambda_{pix} , S/N_F\}$ combinations studied in \S\ref{sec:real}:  \{1.0\AA{},0.5\AA{},7.0\} (blue) and \{0.5\AA{},0.25\AA{},7.0\} (red). We also added \{0.25\AA{},0.125\AA{},7.0\} (brown) and  \{0.1\AA{},0.05\AA{},7.0\} (purple) in order to have a smooth transition from realistic to ideal line profiles. We kept $\rm S/N_F = 7.0$ fixed because it is a low intermediate value. We have check that we obtain qualitatively the same trends with the other $\rm S/N_F$ values. Overall, we find that as we increase the line profile quality, $\sigma(\Delta\lambda)$ decreases for fix $\rm N_{LAE}^{Training}$ values. This means that the better line profile quality, the smaller the training sample has to be in order to obtain the same accuracy. 

For comparison, we show $\sigma(\Delta\lambda)$ of the best standard methodology for these configurations, \VER , in horizontal dashed lines. For the training set size used in sections \S\ref{sec:ideal} and \S\ref{sec:real} (vertical black dashed line), \NNU\ always exhibits a better accuracy than \VER , as shown previously. However,  the accuracy of \NNU\ depends on the training set sample. If  $\rm N_{LAE}^{Training}$ is too small, the accuracy of \VER\ becomes better than the one of \NNU .  The $\rm N_{LAE}^{Training}$ range in which \NNU\ performs better than \VER\ depends on the line profile quality. In fact, the better the line profile quality, the lower $\rm N_{LAE}^{Training}$ is necessary to improve \VER . For example, for the case with $\{\rm W_g , \Delta\lambda_{pix} , S/N_F\}=$\{0.1\AA{},0.05\AA{},7.0\}, about 500 LAEs with a good redshift determination are required to improve \VER . However, for a sample with $\{\rm W_g , \Delta\lambda_{pix} , S/N_F\}=$\{1.0\AA{},0.5\AA{},7.0\}, around $10^4$ LAEs would be necessary. 

}

\section{ Conclusions }\label{sec:conclusions}

    In this work we have addressed the clustering distortions in LAE samples due to the misidentification of the \lya\ wavelength in the \lya\ line profile and how to mitigate them using neural networks. With this goal, we have analyzed the \lya\ line profiles from our previous LAE theoretical model \citep{GurungLopez_2020}, that includes the \lya\ RT in the ISM and in the IGM. Our LAE model reproduces by construction the observed LAE luminosity function and a bunch of their observables, such us the \lya\ escape fraction, the metalicity distribution and clustering amplitude of LAEs \citep{GurungLopez_2019a}.
    
    After analysing the stacked \lya\ line profiles of our LAE model we find that:
\begin{itemize}    
    \item{ The stacked \lya\ line profile is affected by the IGM. In particular, as the IGM \lya\ optical depth increases with redshift, the higher the redshift, the more the stacked \lya\ line profile is modified by the IGM. At low redshift ($z=2.2$ and $z=3.0$), we find that the stacked \lya\ line profile changes slightly and only bluewards the \lya\ wavelength. Meanwhile, at redshift 5.7, the radiative transfer in the IGM modifies the stacked \lya\ line profile up to 2\AA{} redder than the \lya\ wavelength.}
    
    \item{ The  stacked  \lya\ line profile depends on both, galaxy and IGM properties. On one hand, our model predicts that, for the \ThinShell\ and the \GalacticWind\ outflow geometries, the \lya\ stacked line profile is centered at large wavelength for higher values of SFR and \llya .  Additionally, LAEs with higher values of \lya\ equivalent width exhibit a bluer \lya\ stacked line profile. Meanwhile, in the \GalacticWind\ we find only a small dependence on the stellar mass, while in the \ThinShell , the lower $M_{*}$, the more redshifted is the peak of the stacked \lya\ line profile. Over all, these trends are in good agreement with observational works \citep{Guaita_2017,Muzahid_2019}. On the other hand, the stacked \lya\ line profiles are more redshifted in high density environments and low IGM large scale line of sight velocity, its gradient and density gradient. }
\end{itemize}

    Then, we have introduced a novel approach to measure the systemic redshift of LAEs from their \lya\ line using neural networks. In this frame, given a survey that only observed the \lya\ line, a fraction of the LAE population could be re-observed to find other features to determine their systemic redshifts. 
    Then, this sub-population would be used to train a neural network that predicts the systemic redshift. 
    In particular, we have explored two different ways of building the training set: i) the re-observed sources are chosen uniformly across their properties (\NNU), and ii) only the brightest LAEs are re-observed (\NNB).In order to asses the performance of these methodologies we compare them with others found in the literature. In particular, we use i) \GM , that uses the global maximum of the \lya\ line profile to assign a systemic redshift \citep{Verhamme:2018aa,Byrohl_2019,Muzahid_2019} and ii) \VER , suggested by \cite{Verhamme:2018aa}, that uses the width of the \lya\ line correct for the redshift due to the \lya\ RT in the ISM. 
    
    First, we focus on the \lya\ line profiles produced by our model, which are ideal in terms of signal to noise and pixelization. We find that the \NNU\ and \NNB\ algorithms performs better than \GM\ and \VER . In fact the distributions of the displacement of \lya\ ($\Delta \lambda$)   if the broadest for the \GM , followed by the \VER\ a and \NNB , while for the \NNU , it is the thinnest. 
    Then we study how each of these methods impacts the observed clustering of LAEs with ideal line profiles. We find that:
    
\begin{itemize}
    \item{ In general, the power spectrum exhibits a damping at small scales that disappears at large enough distances, as found in  \cite{Byrohl_2019}.  This damping is directly linked to the performance recovering the systemic redshift of the LAEs. In fact, the clustering of samples using \GM\ and \VER\ exhibits a decrease of 80\% of power at $k_{\parallel}=1.0$. In contrast, the samples using \NNU\ and \NNB\ exhibit a much shallower damping. Typically, samples using \NNB\ have a damping of 10\% or less at $k_{\parallel}=1.0$. Meanwhile, the samples using \NNU\  are mostly unaffected at $k_{\parallel}=1.0$, as the power spectrum damping is of the order of the 1\% . }
    
    \item{ The monopole also exhibits a damping at small scales parallel to the one observed in the power spectrum. In particular, the power suppression in the monopole is of the order of 1\% at 1cMpc/$h$. Meanwhile, for \GM\ and \VER\ the monopole damping can go up to the 60\% at 1cMpc/$h$. Also, for the \GM\ and \VER , the suppression extends up to 10cMpc/$h$ , where, in general, the intrinsic and the observed monopole converge. }
    
    \item{ The quadrupole is also sensitive to the systemic redshift determination. In fact, we find that, the lower the accuracy recovering the \lya\ wavelength, the more power exhibits the quadrupole between 1cMpc/$h$ and 10cMpc/$h$. The quadrupole of the samples using \NNB\ and, specially, \NNU\ are mostly identical to the quadrupole of the underling LAE population.   }
\end{itemize}

    Next, we explore the benefits of using \NNU\ and \NNB\ in comparison with \GM\ and \VER\ in realistic line profiles. With this goal, we lower the quality of the \lya\ line profile mocking several artifacts in observations. In practice, i) we dilute the line assuming different instrumental FWHM, ii) we reduce the wavelength resolution by pixelizing the line and iii) we include noise in the \lya\ line profile. Then, we study the properties of these samples, finding:
    
\begin{itemize}
    \item{ The performance, i.e., the distributions $\Delta \lambda$ is tightly connected to the spectral quality. Overall, we find that in the spectral quality range covered in this work, the \NNU\ is the best methodology. Additionally, the \NNB\ is very affected by the noise in the spectrum. This is a result of using only the brightest LAEs as training sample, as this one lacks faint LAEs in which their \lya\ line profile is noisy. In this way,  \NNB\ is the second best algorithm for high signal to noise lines. Meanwhile, \VER\ progressively gets worst as the spectral quality decreases. \VER\ is mostly affected by the instrumental FWHM, as this modifies the width of the \lya\ line and over-corrects the RT in the ISM. Finally, \GM\ is the algorithm with the lowest performance in most of the spectral quality regime studied here. However, \GM\ is quite insensitive to lowering the spectral quality, specially, increasing the noise. This, in the worst spectral quality considered here, translates into a better accuracy of \GM\ in comparison with \NNB\ and \VER , while it is similar to \NNU .  }
    
    \item{ Consequently, the monopole and the quadrupole are affected by the reduction of the spectral quality. We find that \NNU\ is algorithm that recovers better the clustering of the underlying LAE population. However, as the spectral quality is reduce, the damping of power at small scales increases, as in \VER\ and \NNB . Meanwhile, the clustering damping at small scales is quite constant for \GM\ through the dynamical range of the spectral quality studied here.  }
    
    \item{ There is a linear relation between the algorithm performance and the typical scale up to which the clustering power is decreased. This will be usful for the desing of future surveys based on the \lya\ line.}    
 
\end{itemize}

Therefore, we conclude that spectroscopic \lya\ based surveys such, as HETDEX, might benefit from measuring the systemic redshift of a relative small subsample of LAEs, using other spectral features. And then, using this subsample to train machine learning algorithms to predict the systemic redshift of the rest of the observed LAE population.

%%%%%%%%%%%%%%%%%%%%%%%%%%%%%%%%%%%%%%%%%%%%%%%%%%%%%%%%%%%%%%%%%%%%%%%%%%%%%%%%%%%%%%%%%%%%%%%%
%\begin{figure} 
%\includegraphics[width=3.3in]{fig_damping_median_Thin_Shell_z_2_2_All_space_1_FWHM_0_0_PIX_0_0_SN_0_0.pdf}%
%\caption{ }
%\label{fig:PDF_FT}
%\end{figure}
%%%%%%%%%%%%%%%%%%%%%%%%%%%%%%%%%%%%%%%%%%%%%%%%%%%%%%%%%%%%%%%%%%%%%%%%%%%%%%%%%%%%%%%%%%%%%%%%

%%%%%%%%%%%%%%%%%%%%%%%%%%%%%%%%%%%%%%%%%%%%%%%%%%%%%%%%%%%%%%%%%%%%%%%%%%%%%%%%%%%%%%%%%%%%%%%%

\begin{table}
\caption{ Same as Tab.\ref{tab:real_3_thin_Shell} but displaying the details about the \ThinShell\ geometry at redshift 2.2 . }
\label{tab:real_2_thin_Shell}
\begin{tabular}{ccccccccc}
\multicolumn{3}{c}{ $z=2.2$ , Thin Shell}                                                     & GM             & GM-F            & uGM-F     & NN:U           & NN:B   \\ \hline
$\rm W_g ^{Rest}$                              & $\rm \Delta \lambda _{pix}^{Rest}$ & $\rm S/N_{F}$ & $\sigma$       & $\sigma$        & $\sigma$  & $\sigma$       & $\sigma$    \\
$\;$[\AA{}]                                    & [\AA{}]                            &               & [\AA{}]        & [\AA{}]         & [\AA{}]   & [\AA{}]        & [\AA{}]        \\ \hline
0.5                                            & 0.25                               & 6.0         & 1.57 & 0.79  & 0.78 & 0.67 & 1.63 \\
                                               &                                    & 7.0         & 1.46 & 0.68  & 0.67 & 0.52 & 1.03 \\
                                               &                                    & 10.0        & 1.33 & 0.58  & 0.57 & 0.28 & 0.49 \\
                                               &                                    & 15.0        & 1.25 & 0.53  & 0.52 & 0.22 & 0.36 \\ \cline{2-8}
                                               & 0.5                                & 6.0         & 1.84 & 1.17  & 1.16 & 0.99 & 1.81 \\
                                               &                                    & 7.0         & 1.65 & 0.94  & 0.93 & 0.73 & 1.5 \\
                                               &                                    & 10.0        & 1.39 & 0.67  & 0.66 & 0.34 & 0.61 \\
                                               &                                    & 15.0        & 1.24 & 0.56  & 0.55 & 0.22 & 0.38 \\ \cline{2-8}
                                               & 1.0                                & 6.0         & 2.43 & 1.91  & 1.88 & 1.31 & 2.45 \\
                                               &                                    & 7.0         & 2.15 & 1.55  & 1.53 & 1.06 & 1.89 \\
                                               &                                    & 10.0        & 1.63 & 0.95  & 0.93 & 0.52 & 1.04 \\
                                               &                                    & 15.0        & 1.35 & 0.65  & 0.64 & 0.27 & 0.41 \\ \hline
1.0                                            & 0.25                               & 6.0         & 1.46 & 0.83  & 0.79 & 0.72 & 1.65 \\
                                               &                                    & 7.0         & 1.33 & 0.72  & 0.68 & 0.55 & 1.27 \\
                                               &                                    & 10.0        & 1.21 & 0.59  & 0.56 & 0.31 & 0.56 \\
                                               &                                    & 15.0        & 1.09 & 0.52  & 0.5 & 0.23 & 0.44 \\ \cline{2-8}
                                               & 0.5                                & 6.0         & 1.78 & 1.26  & 1.21 & 1.03 & 2.01 \\
                                               &                                    & 7.0         & 1.57 & 1.02  & 0.98 & 0.75 & 1.46 \\
                                               &                                    & 10.0        & 1.29 & 0.7  & 0.67 & 0.39 & 0.68 \\
                                               &                                    & 15.0        & 1.14 & 0.57  & 0.54 & 0.24 & 0.43 \\ \cline{2-8}
                                               & 1.0                                & 6.0         & 2.44 & 2.03  & 1.99 & 1.37 & 2.62 \\
                                               &                                    & 7.0         & 2.1 & 1.66  & 1.62 & 1.1 & 2.15 \\
                                               &                                    & 10.0        & 1.58 & 1.01  & 0.97 & 0.54 & 0.96 \\
                                               &                                    & 15.0        & 1.25 & 0.68  & 0.65 & 0.25 & 0.45 \\ \hline
2.0                                            & 0.25                               & 6.0         & 1.24 & 1.04  & 0.92 & 0.87 & 1.97 \\
                                               &                                    & 7.0         & 1.12 & 0.9  & 0.77 & 0.65 & 1.73 \\
                                               &                                    & 10.0        & 0.93 & 0.69  & 0.57 & 0.39 & 0.73 \\
                                               &                                    & 15.0        & 0.82 & 0.56  & 0.48 & 0.27 & 0.52 \\ \cline{2-8}
                                               & 0.5                                & 6.0         & 1.65 & 1.56  & 1.48 & 1.23 & 2.47 \\
                                               &                                    & 7.0         & 1.42 & 1.28  & 1.18 & 0.94 & 2.26 \\
                                               &                                    & 10.0        & 1.08 & 0.87  & 0.75 & 0.45 & 0.98 \\
                                               &                                    & 15.0        & 0.9 & 0.66  & 0.56 & 0.27 & 0.43 \\ \cline{2-8}
                                               & 1.0                                & 6.0         & 2.5 & 2.42  & 2.37 & 1.6 & 2.81 \\
                                               &                                    & 7.0         & 2.05 & 2.01  & 1.94 & 1.29 & 2.55 \\
                                               &                                    & 10.0        & 1.41 & 1.28  & 1.18 & 0.62 & 1.35 \\
                                               &                                    & 15.0        & 1.03 & 0.85  & 0.72 & 0.31 & 0.52 \\ \hline
\end{tabular}
\end{table}

\begin{table}
\caption{ Same as Tab.\ref{tab:real_3_thin_Shell} but displaying the details about the \GalacticWind\ geometry at redshift 2.2 . }
\label{tab:real_2_galactic_Wind}
\begin{tabular}{ccccccccc}
\multicolumn{3}{c}{ $z=2.2$ , Galactic Wind}                                                     & GM             & GM-F            & uGM-F     & NN:U           & NN:B   \\ \hline
$\rm W_g ^{Rest}$                              & $\rm \Delta \lambda _{pix}^{Rest}$ & $\rm S/N_{F}$ & $\sigma$       & $\sigma$        & $\sigma$  & $\sigma$       & $\sigma$    \\
$\;$[\AA{}]                                    & [\AA{}]                            &               & [\AA{}]        & [\AA{}]         & [\AA{}]   & [\AA{}]        & [\AA{}]        \\ \hline
0.5                                            & 0.25                               & 6.0         & 1.71 & 0.73  & 0.72 & 0.46 & 0.84 \\
                                               &                                    & 7.0         & 1.69 & 0.7  & 0.69 & 0.39 & 0.73 \\
                                               &                                    & 10.0        & 1.65 & 0.65  & 0.65 & 0.3 & 0.53 \\
                                               &                                    & 15.0        & 1.62 & 0.62  & 0.61 & 0.24 & 0.38 \\ \cline{2-8}
                                               & 0.5                                & 6.0         & 1.76 & 0.85  & 0.84 & 0.61 & 1.24 \\
                                               &                                    & 7.0         & 1.73 & 0.78  & 0.77 & 0.47 & 0.78 \\
                                               &                                    & 10.0        & 1.63 & 0.68  & 0.68 & 0.35 & 0.57 \\
                                               &                                    & 15.0        & 1.61 & 0.63  & 0.63 & 0.26 & 0.43 \\ \cline{2-8}
                                               & 1.0                                & 6.0         & 1.92 & 1.28  & 1.24 & 0.92 & 1.78 \\
                                               &                                    & 7.0         & 1.8 & 1.09  & 1.05 & 0.69 & 1.25 \\
                                               &                                    & 10.0        & 1.67 & 0.85  & 0.82 & 0.4 & 0.65 \\
                                               &                                    & 15.0        & 1.6 & 0.71  & 0.7 & 0.26 & 0.51 \\ \hline
1.0                                            & 0.25                               & 6.0         & 1.63 & 0.76  & 0.74 & 0.53 & 0.99 \\
                                               &                                    & 7.0         & 1.61 & 0.72  & 0.7 & 0.41 & 0.76 \\
                                               &                                    & 10.0        & 1.56 & 0.65  & 0.64 & 0.31 & 0.65 \\
                                               &                                    & 15.0        & 1.53 & 0.61  & 0.6 & 0.23 & 0.41 \\ \cline{2-8}
                                               & 0.5                                & 6.0         & 1.71 & 0.92  & 0.88 & 0.71 & 1.49 \\
                                               &                                    & 7.0         & 1.67 & 0.82  & 0.79 & 0.51 & 1.11 \\
                                               &                                    & 10.0        & 1.59 & 0.71  & 0.69 & 0.36 & 0.81 \\
                                               &                                    & 15.0        & 1.53 & 0.64  & 0.63 & 0.27 & 0.47 \\ \cline{2-8}
                                               & 1.0                                & 6.0         & 1.9 & 1.33  & 1.32 & 1.04 & 2.02 \\
                                               &                                    & 7.0         & 1.79 & 1.12  & 1.11 & 0.78 & 1.42 \\
                                               &                                    & 10.0        & 1.64 & 0.83  & 0.81 & 0.4 & 0.84 \\
                                               &                                    & 15.0        & 1.55 & 0.7  & 0.68 & 0.3 & 0.52 \\ \hline
2.0                                            & 0.25                               & 6.0         & 1.52 & 0.92  & 0.83 & 0.71 & 1.91 \\
                                               &                                    & 7.0         & 1.47 & 0.85  & 0.76 & 0.54 & 1.24 \\
                                               &                                    & 10.0        & 1.41 & 0.75  & 0.66 & 0.35 & 0.86 \\
                                               &                                    & 15.0        & 1.36 & 0.66  & 0.6 & 0.26 & 0.61 \\ \cline{2-8}
                                               & 0.5                                & 6.0         & 1.63 & 1.17  & 1.13 & 1.01 & 2.3 \\
                                               &                                    & 7.0         & 1.56 & 1.02  & 0.96 & 0.73 & 1.56 \\
                                               &                                    & 10.0        & 1.45 & 0.83  & 0.74 & 0.43 & 1.08 \\
                                               &                                    & 15.0        & 1.39 & 0.73  & 0.64 & 0.3 & 0.74 \\ \cline{2-8}
                                               & 1.0                                & 6.0         & 1.92 & 1.79  & 1.76 & 1.36 & 3.12 \\
                                               &                                    & 7.0         & 1.75 & 1.48  & 1.45 & 1.06 & 2.13 \\
                                               &                                    & 10.0        & 1.55 & 1.04  & 0.97 & 0.54 & 1.14 \\
                                               &                                    & 15.0        & 1.44 & 0.82  & 0.73 & 0.32 & 0.65 \\ \hline
\end{tabular}
\end{table}

\begin{table}
\caption{ Same as Tab.\ref{tab:real_3_thin_Shell} but displaying the details about the \GalacticWind\ geometry at redshift 3.0 . }
\label{tab:real_3_galactic_Wind}
\begin{tabular}{ccccccccc}
\multicolumn{3}{c}{ $z=3.0$ , Galactic Wind}                                                     & GM             & GM-F            & uGM-F     & NN:U           & NN:B   \\ \hline
$\rm W_g ^{Rest}$                              & $\rm \Delta \lambda _{pix}^{Rest}$ & $\rm S/N_{F}$ & $\sigma$       & $\sigma$        & $\sigma$  & $\sigma$       & $\sigma$    \\
$\;$[\AA{}]                                    & [\AA{}]                            &               & [\AA{}]        & [\AA{}]         & [\AA{}]   & [\AA{}]        & [\AA{}]        \\ \hline
0.5                                            & 0.25                               & 6.0         & 2.37 & 0.92  & 0.92 & 0.54 & 0.89 \\
                                               &                                    & 7.0         & 2.32 & 0.86  & 0.87 & 0.47 & 0.74 \\
                                               &                                    & 10.0        & 2.31 & 0.8  & 0.81 & 0.33 & 0.57 \\
                                               &                                    & 15.0        & 2.28 & 0.76  & 0.77 & 0.25 & 0.43 \\ \cline{2-8}
                                               & 0.5                                & 6.0         & 2.39 & 1.02  & 1.02 & 0.69 & 1.42 \\
                                               &                                    & 7.0         & 2.34 & 0.94  & 0.94 & 0.53 & 0.99 \\
                                               &                                    & 10.0        & 2.29 & 0.84  & 0.84 & 0.39 & 0.61 \\
                                               &                                    & 15.0        & 2.26 & 0.78  & 0.79 & 0.28 & 0.48 \\ \cline{2-8}
                                               & 1.0                                & 6.0         & 2.5 & 1.44  & 1.4 & 0.99 & 2.2 \\
                                               &                                    & 7.0         & 2.43 & 1.24  & 1.21 & 0.75 & 1.49 \\
                                               &                                    & 10.0        & 2.32 & 1.0  & 0.98 & 0.46 & 0.85 \\
                                               &                                    & 15.0        & 2.25 & 0.87  & 0.86 & 0.33 & 0.54 \\ \hline
1.0                                            & 0.25                               & 6.0         & 2.27 & 0.93  & 0.92 & 0.56 & 0.98 \\
                                               &                                    & 7.0         & 2.27 & 0.89  & 0.89 & 0.46 & 0.89 \\
                                               &                                    & 10.0        & 2.22 & 0.8  & 0.81 & 0.36 & 0.66 \\
                                               &                                    & 15.0        & 2.22 & 0.76  & 0.78 & 0.28 & 0.5 \\ \cline{2-8}
                                               & 0.5                                & 6.0         & 2.33 & 1.09  & 1.07 & 0.78 & 1.68 \\
                                               &                                    & 7.0         & 2.28 & 0.98  & 0.97 & 0.63 & 1.12 \\
                                               &                                    & 10.0        & 2.22 & 0.85  & 0.85 & 0.38 & 0.7 \\
                                               &                                    & 15.0        & 2.19 & 0.79  & 0.8 & 0.29 & 0.52 \\ \cline{2-8}
                                               & 1.0                                & 6.0         & 2.46 & 1.5  & 1.49 & 1.12 & 2.76 \\
                                               &                                    & 7.0         & 2.35 & 1.28  & 1.28 & 0.86 & 1.77 \\
                                               &                                    & 10.0        & 2.26 & 0.99  & 0.99 & 0.5 & 0.7 \\
                                               &                                    & 15.0        & 2.2 & 0.85  & 0.85 & 0.33 & 0.55 \\ \hline
2.0                                            & 0.25                               & 6.0         & 2.13 & 1.06  & 1.02 & 0.8 & 1.74 \\
                                               &                                    & 7.0         & 2.1 & 0.99  & 0.95 & 0.62 & 1.39 \\
                                               &                                    & 10.0        & 2.04 & 0.88  & 0.85 & 0.41 & 0.94 \\
                                               &                                    & 15.0        & 2.0 & 0.79  & 0.8 & 0.31 & 0.6 \\ \cline{2-8}
                                               & 0.5                                & 6.0         & 2.23 & 1.32  & 1.3 & 1.1 & 2.38 \\
                                               &                                    & 7.0         & 2.15 & 1.16  & 1.13 & 0.81 & 1.81 \\
                                               &                                    & 10.0        & 2.08 & 0.98  & 0.94 & 0.49 & 1.01 \\
                                               &                                    & 15.0        & 2.0 & 0.85  & 0.83 & 0.32 & 0.73 \\ \cline{2-8}
                                               & 1.0                                & 6.0         & 2.48 & 1.93  & 1.92 & 1.43 & 2.78 \\
                                               &                                    & 7.0         & 2.33 & 1.64  & 1.62 & 1.1 & 2.79 \\
                                               &                                    & 10.0        & 2.14 & 1.17  & 1.14 & 0.61 & 1.35 \\
                                               &                                    & 15.0        & 2.05 & 0.96  & 0.93 & 0.37 & 0.71 \\ \hline
\end{tabular}
\end{table}

\begin{table}
\caption{ Same as Tab.\ref{tab:real_3_thin_Shell} but displaying the details about the \ThinShell\ geometry at redshift 5.7 . }
\label{tab:real_5_thin_Shell}
\begin{tabular}{ccccccccc}
\multicolumn{3}{c}{ $z=5.7$ , Thin Shell}                                                     & GM             & GM-F            & uGM-F     & NN:U           & NN:B   \\ \hline
$\rm W_g ^{Rest}$                              & $\rm \Delta \lambda _{pix}^{Rest}$ & $\rm S/N_{F}$ & $\sigma$       & $\sigma$        & $\sigma$  & $\sigma$       & $\sigma$    \\
$\;$[\AA{}]                                    & [\AA{}]                            &               & [\AA{}]        & [\AA{}]         & [\AA{}]   & [\AA{}]        & [\AA{}]        \\ \hline
0.5                                            & 0.25                               & 6.0         & 1.9 & 1.03  & 1.01 & 0.7 & 1.46 \\
                                               &                                    & 7.0         & 1.9 & 0.98  & 0.97 & 0.59 & 1.13 \\
                                               &                                    & 10.0        & 1.93 & 0.94  & 0.92 & 0.37 & 0.76 \\
                                               &                                    & 15.0        & 2.0 & 0.93  & 0.91 & 0.26 & 0.48 \\ \cline{2-8}
                                               & 0.5                                & 6.0         & 1.9 & 1.2  & 1.18 & 0.92 & 1.76 \\
                                               &                                    & 7.0         & 1.86 & 1.09  & 1.08 & 0.7 & 1.3 \\
                                               &                                    & 10.0        & 1.84 & 0.97  & 0.95 & 0.46 & 0.75 \\
                                               &                                    & 15.0        & 1.86 & 0.92  & 0.91 & 0.28 & 0.47 \\ \cline{2-8}
                                               & 1.0                                & 6.0         & 2.07 & 1.62  & 1.59 & 1.22 & 2.4 \\
                                               &                                    & 7.0         & 1.95 & 1.41  & 1.38 & 0.92 & 1.83 \\
                                               &                                    & 10.0        & 1.8 & 1.09  & 1.07 & 0.54 & 0.93 \\
                                               &                                    & 15.0        & 1.76 & 0.94  & 0.92 & 0.36 & 0.68 \\ \hline
1.0                                            & 0.25                               & 6.0         & 1.77 & 1.04  & 0.98 & 0.73 & 1.69 \\
                                               &                                    & 7.0         & 1.74 & 0.98  & 0.92 & 0.59 & 1.38 \\
                                               &                                    & 10.0        & 1.75 & 0.92  & 0.86 & 0.41 & 0.8 \\
                                               &                                    & 15.0        & 1.78 & 0.9  & 0.84 & 0.3 & 0.63 \\ \cline{2-8}
                                               & 0.5                                & 6.0         & 1.81 & 1.24  & 1.18 & 0.96 & 1.99 \\
                                               &                                    & 7.0         & 1.74 & 1.13  & 1.07 & 0.78 & 1.51 \\
                                               &                                    & 10.0        & 1.71 & 0.96  & 0.9 & 0.46 & 0.82 \\
                                               &                                    & 15.0        & 1.69 & 0.9  & 0.84 & 0.32 & 0.63 \\ \cline{2-8}
                                               & 1.0                                & 6.0         & 2.06 & 1.67  & 1.64 & 1.29 & 2.72 \\
                                               &                                    & 7.0         & 1.9 & 1.46  & 1.43 & 1.03 & 2.05 \\
                                               &                                    & 10.0        & 1.71 & 1.1  & 1.06 & 0.55 & 1.0 \\
                                               &                                    & 15.0        & 1.65 & 0.92  & 0.88 & 0.35 & 0.76 \\ \hline
2.0                                            & 0.25                               & 6.0         & 1.42 & 1.09  & 0.97 & 0.83 & 2.13 \\
                                               &                                    & 7.0         & 1.39 & 1.0  & 0.87 & 0.68 & 1.58 \\
                                               &                                    & 10.0        & 1.29 & 0.88  & 0.73 & 0.46 & 0.86 \\
                                               &                                    & 15.0        & 1.21 & 0.82  & 0.67 & 0.36 & 0.73 \\ \cline{2-8}
                                               & 0.5                                & 6.0         & 1.61 & 1.42  & 1.33 & 1.08 & 2.63 \\
                                               &                                    & 7.0         & 1.51 & 1.23  & 1.13 & 0.85 & 1.92 \\
                                               &                                    & 10.0        & 1.36 & 0.98  & 0.85 & 0.48 & 1.03 \\
                                               &                                    & 15.0        & 1.24 & 0.86  & 0.71 & 0.34 & 0.7 \\ \cline{2-8}
                                               & 1.0                                & 6.0         & 2.03 & 2.0  & 1.95 & 1.44 & 3.41 \\
                                               &                                    & 7.0         & 1.78 & 1.72  & 1.65 & 1.18 & 2.62 \\
                                               &                                    & 10.0        & 1.49 & 1.23  & 1.12 & 0.62 & 1.23 \\
                                               &                                    & 15.0        & 1.31 & 0.95  & 0.82 & 0.37 & 0.76 \\ \hline
\end{tabular}
\end{table}

\begin{table}
\caption{ Same as Tab.\ref{tab:real_3_thin_Shell} but displaying the details about the \GalacticWind\ geometry at redshift 5.7 . }
\label{tab:real_5_galactic_Wind}
\begin{tabular}{ccccccccc}
\multicolumn{3}{c}{ $z=5.7$ , Galactic Wind}                                                     & GM             & GM-F            & uGM-F     & NN:U           & NN:B   \\ \hline
$\rm W_g ^{Rest}$                              & $\rm \Delta \lambda _{pix}^{Rest}$ & $\rm S/N_{F}$ & $\sigma$       & $\sigma$        & $\sigma$  & $\sigma$       & $\sigma$    \\
$\;$[\AA{}]                                    & [\AA{}]                            &               & [\AA{}]        & [\AA{}]         & [\AA{}]   & [\AA{}]        & [\AA{}]        \\ \hline
0.5                                            & 0.25                               & 6.0         & 2.92 & 0.87  & 0.86 & 0.54 & 1.06 \\
                                               &                                    & 7.0         & 2.95 & 0.85  & 0.85 & 0.44 & 0.84 \\
                                               &                                    & 10.0        & 3.03 & 0.85  & 0.84 & 0.32 & 0.67 \\
                                               &                                    & 15.0        & 3.11 & 0.85  & 0.84 & 0.25 & 0.43 \\ \cline{2-8}
                                               & 0.5                                & 6.0         & 2.77 & 0.93  & 0.92 & 0.65 & 1.26 \\
                                               &                                    & 7.0         & 2.78 & 0.88  & 0.88 & 0.53 & 0.9 \\
                                               &                                    & 10.0        & 2.84 & 0.84  & 0.84 & 0.36 & 0.68 \\
                                               &                                    & 15.0        & 2.93 & 0.83  & 0.83 & 0.28 & 0.5 \\ \cline{2-8}
                                               & 1.0                                & 6.0         & 2.71 & 1.19  & 1.16 & 0.86 & 1.52 \\
                                               &                                    & 7.0         & 2.7 & 1.09  & 1.06 & 0.68 & 1.26 \\
                                               &                                    & 10.0        & 2.74 & 0.94  & 0.91 & 0.39 & 0.87 \\
                                               &                                    & 15.0        & 2.8 & 0.88  & 0.86 & 0.28 & 0.65 \\ \hline
1.0                                            & 0.25                               & 6.0         & 2.8 & 0.86  & 0.85 & 0.57 & 1.65 \\
                                               &                                    & 7.0         & 2.81 & 0.84  & 0.83 & 0.49 & 1.01 \\
                                               &                                    & 10.0        & 2.89 & 0.83  & 0.82 & 0.37 & 0.8 \\
                                               &                                    & 15.0        & 2.97 & 0.83  & 0.82 & 0.3 & 0.59 \\ \cline{2-8}
                                               & 0.5                                & 6.0         & 2.65 & 0.95  & 0.93 & 0.69 & 1.34 \\
                                               &                                    & 7.0         & 2.66 & 0.89  & 0.87 & 0.55 & 1.12 \\
                                               &                                    & 10.0        & 2.72 & 0.83  & 0.82 & 0.38 & 0.87 \\
                                               &                                    & 15.0        & 2.81 & 0.82  & 0.81 & 0.3 & 0.61 \\ \cline{2-8}
                                               & 1.0                                & 6.0         & 2.58 & 1.22  & 1.22 & 0.98 & 1.87 \\
                                               &                                    & 7.0         & 2.58 & 1.07  & 1.08 & 0.72 & 1.4 \\
                                               &                                    & 10.0        & 2.58 & 0.88  & 0.89 & 0.4 & 1.12 \\
                                               &                                    & 15.0        & 2.65 & 0.82  & 0.83 & 0.3 & 0.84 \\ \hline
2.0                                            & 0.25                               & 6.0         & 2.22 & 0.95  & 0.89 & 0.64 & 2.0 \\
                                               &                                    & 7.0         & 2.23 & 0.91  & 0.85 & 0.57 & 1.51 \\
                                               &                                    & 10.0        & 2.25 & 0.85  & 0.8 & 0.38 & 1.06 \\
                                               &                                    & 15.0        & 2.29 & 0.82  & 0.79 & 0.36 & 1.32 \\ \cline{2-8}
                                               & 0.5                                & 6.0         & 2.21 & 1.13  & 1.09 & 0.89 & 2.19 \\
                                               &                                    & 7.0         & 2.19 & 1.02  & 0.98 & 0.67 & 1.44 \\
                                               &                                    & 10.0        & 2.19 & 0.89  & 0.84 & 0.4 & 0.99 \\
                                               &                                    & 15.0        & 2.2 & 0.83  & 0.79 & 0.32 & 1.12 \\ \cline{2-8}
                                               & 1.0                                & 6.0         & 2.26 & 1.56  & 1.54 & 1.31 & 2.68 \\
                                               &                                    & 7.0         & 2.21 & 1.36  & 1.33 & 0.97 & 2.06 \\
                                               &                                    & 10.0        & 2.14 & 1.02  & 0.98 & 0.51 & 1.07 \\
                                               &                                    & 15.0        & 2.13 & 0.86  & 0.83 & 0.32 & 1.01 \\ \hline
\end{tabular}
\end{table}

\section*{Acknowledgements}

 { We acknowledge and thank the great labor of the scientific referee. Their comments  improved greatly the quality, clearance and content of this work.} This work has made used of CEFCA's Scientific High Performance Computing system which has been funded by the Governments  of  Spain  and  Aragón  through the  Fondo  de  Inversiones  de  Teruel, and the Spanish Ministry of Economy and Competitivenes (MINECO-FEDER, grant AYA2012-30789). 
 The authors acknowledge the support of the Spanish Ministerio de Economia y Competividad project No. AYA2015-66211-C2-P-2. 
 We acknowledge also STFC Consolidated Grants ST/L00075X/1 and ST/P000451/1 at Durham University. 
 This work used the DiRAC@Durham facility managed by the Institute for Computational Cosmology on behalf of the STFC DiRAC HPC Facility (www.dirac.ac.uk). The equipment was funded by BEIS capital funding via STFC capital grants ST/P002293/1, ST/R002371/1 and ST/S002502/1, Durham University and STFC operations grant ST/R000832/1. DiRAC is part of the National e-Infrastructure.
 SS was supported in part by JSPS KAKENHI Grant Number JP15H05896, and by World Premier International Research Center Initiative (WPI Initiative), MEXT, Japan. 
 SS was also supported in part by the Munich Institute for Astro- and Particle Physics (MIAPP) which is funded by the Deutsche Forschungsgemeinschaft (DFG, German Research Foundation) under Germany's Excellence Strategy (EXC-2094-390783311).
 
\section*{Data Availability}

The data underlying this article will be shared on reasonable request to the corresponding author.

%%%%%%%%%%%%%%%%%%%%%%%%%%%%%%%%%%%%%%%%%%%%%%%%%%



\bibliographystyle{mnras}
\bibliography{ref} % if your bibtex file is called example.bib

%\begin{thebibliography}{99}
%\end{thebibliography}

%%%%%%%%%%%%%%%%%%%%%%%%%%%%%%%%%%%%%%%%%%%%%%%%%%

%%%%%%%%%%%%%%%%% APPENDICES %%%%%%%%%%%%%%%%%%%%%

\appendix

 %%%%%%%%%%%%%%%%%%%%%%%%%%%%%%%%%%%%%%%%%%%%%%%%%%%%%%%%%%%%%%%%%%%%%%%%%%%%%%%%%%%%%%%%%%%%%%%%
\begin{figure*} 
\includegraphics[width=5.5in]{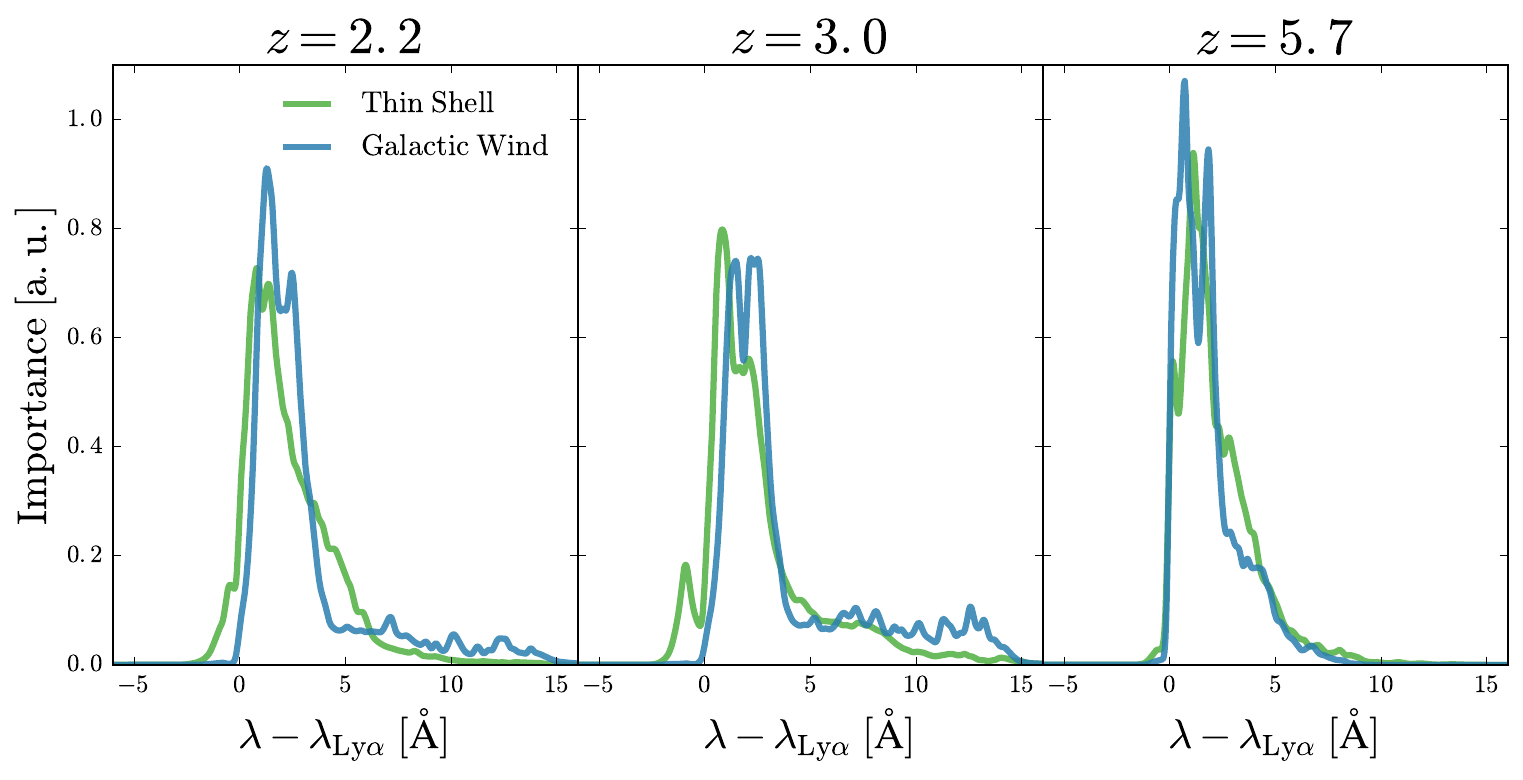}%
\caption{ Importance of each wavelength bin used in the \NNU\ algorithm  at redshift 2.2, 3.0 and 5.7 from left to right. The models using the \ThinShell\ outflow geometry are displayed in \colorThin, while their \GalacticWind\ counterparts are shown in \colorWind . }
\label{fig:importance}
\end{figure*}
%%%%%%%%%%%%%%%%%%%%%%%%%%%%%%%%%%%%%%%%%%%%%%%%%%%%%%%%%%%%%%%%%%%%%%%%%%%%%%%%%%%%%%%%%%%%%%%%

\section{Wavelength importance}

    In this section we quantify which wavelength of the line profile contribute the most to the determination of the \wlyaO\ in the neural networks in this work. With this goal in mind, for a given wavelength bin centered at the wavelength $\rm \lambda_{pix}$ we define its importance as \begin{equation}
    \label{eq:importance}
    \rm
    I(\lambda_{pix}) = \sigma( \lambda_{pix}) \;  \displaystyle{ \omega( \lambda_{pix}) },
    \end{equation} 
    where $\rm \omega( \lambda_{pix})$ is each weight for the pixel centered at $\rm  \lambda_{pix}$ and $\rm  \sigma( \lambda_{pix})$ is the standard deviation of that pixel across the training set. 
    
    In Fig.~\ref{fig:importance} we show the importance as a function of wavelength for the \NNU\ at different redshifts and for the two outflow geometries. In general, we find that the wavelength range with a significant importance is wider than the \lya\ stacked line profiles at $z=2.2$ and 3.0, while at $z=5.7$ both are similar. On one hand, at $z=5.7$ the IGM absorbs mostly all the flux bluewards \lya\ \cite{zheng10,laursen11,GurungLopez_2020}. Therefore, \wlya\ is extracted mainly from the information closer than 2\AA{} to \wlya. Meanwhile, the regions with $\lambda<\lambda_{\rm Ly\alpha}$ and $\lambda>\lambda_{\rm Ly\alpha}+5$\AA{} contain little information in comparison. On the other hand, at lower redshifts ($z=2.2$ and 3.0), the IGM absorption is not that strong and the neural networks need the information from a broader wavelength range. In fact, at these redshifts, most of the importance is located between \wlya\ and 5\AA{}  redwards.  Additionally, the spectral region with a significant importance spawns from $\sim 1$\AA{} bluewards \wlya\ to $\sim 15$\AA{} redwards \wlya .  This broad wavelength range exhibits the large variety of line profiles in our LAE population, as the presence of very broad lines (FWHM$\sim 10$\AA{}) extend the importance range up to $\sim 15$\AA{}.

\section{Comparison of the accuracy between models.}

In this section we speculate about why a given methodology works better for a certain model (combination of redshift and outflow geometry) than for other. Sometimes this can be intuitive. For example, \VER\ sometimes fails at z=5.7, as discussed above, due to the large IGM absorption. However, there are cases in which it is not trivial to find an answer. For example, in the ideal case at redshift 3.0 (Fig.\ref{fig:accuracy_NN}), in which \NNU\ reaches a better performance for the \ThinShell\ than for the \GalacticWind\ for large enough training sets. 

We think that the cause of the different performances is the 'diversity' of lines of each model, i.e., the amount of different line profiles. For example, if a population exhibits lines with one, two and three peaks, most likely, it would have a larger line diversity than another population containing only line with one peak. Other factor that matter for this would be the width of the distributions of the global maximum or of the FWHM of the line, among others.

In principle, the larger the variety of line profiles, the larger the dispersion of \NNU\ might be for a fixed training set size. As explained above, our models are build so that they reproduce the LAE LF, which makes that every model has a unique distribution outflow expansion velocity $\rm V_{exp}$, neutral hydrogen column density $\rm N_H$, dust optical depth $\tau_a$ and IGM transmission curves $T(\lambda)$. 

It is challenging to quantitatively characterise the diversity of lines. For example, given a fix outflow geometry, one could think that the bigger the covered volume in the 4-D space ($\rm V_{exp}$, $\rm N_H$, $\tau_a$ and $T(\lambda)$), the larger would be the diversity of lines, thus, the larger would be $\sigma(\Delta\lambda)$ for a fix methodology. However, this is not the case: in \cite{GurungLopez_2020} (Fig.A1), we showed the \{$\rm V_{exp}$, $\rm N_H$, $\tau_a$\} region covered by the \ThinShell\ and the \GalacticWind\ at z=2.2, 3.0 and 5.7. For both geometries the volume occupied at z=5.7 is larger than at z=2.2 and z=3.0. This would explain that the \NNU\ performance in the thin shell is a 25\% lower at z=5.7 than at the other redshifts. However, the \NNU\ in the \GalacticWind\ has greater accuracy at z=5.7 than at z=2.2 and z=3.0 in the ideal case, which does not support the above, apparently, naive reasoning. This can be explained because there are degeneracies between the different parameters. For example between $\rm V_{exp}$ and $\rm N_H$. 

Then, the comparison between outflow geometries is even more complicated. Even if two models cover exactly the \{$\rm V_{exp}$, $\rm N_H$, $\tau_a$, $T(\lambda)$\} region, the diversity of line profiles might be different, as most probably, the mapping between the outflow properties and line profile would be different.

Also, it is not always the case in which the accuracy of methodology is better in the \ThinShell. For example, \NNU\ at z=5.7 in the ideal case is more accurate in the \GalacticWind . At the same time \VER\ at z=2.2 is more accurate for the galactic wind (as opposite to \NNU\ at z=2.2).

%%%%%%%%%%%%%%%%%%%%%%%%%%%%%%%%%%%%%%%%%%%%%%%%%%%%%%%%%%%%%%%%%%%
%%%%%%%%%%%%%%%%%%%%%%%%%%%%%%%%%%%%%%%%%%%%%%%%%%%%%%%%%%%%%%%%%%%
%%%%%%%%%%%%%%%%%%%%%%%%%%%%%%%%%%%%%%%%%%%%%%%%%%%%%%%%%%%%%%%%%%%

% Don't change these lines
\bsp	% typesetting comment
\label{lastpage}
\end{document}